\documentclass[preprint,12pt]{elsarticle}

\makeatletter
\def\ps@pprintTitle{%
 \let\@oddhead\@empty
 \let\@evenhead\@empty
 \def\@oddfoot{}%
 \let\@evenfoot\@oddfoot}
\makeatother

\usepackage{amssymb}

\usepackage{float}
\usepackage{subfigure}
\usepackage{amsmath}
\usepackage{afterpage}
\usepackage{longtable}

\usepackage{graphicx,subfigmat,etoolbox,amssymb,float}

\usepackage{multirow}
\usepackage{booktabs}

\usepackage{url}
\usepackage{hyperref}

\newcounter{subfigcount}
\begin{document}

\begin{frontmatter}



\title{An Experimental Study of Noise Reduction in Wind Turbine Airfoils with Serrated Trailing Edges}

 \author[label1,label3]{Weicheng Xue}
 \affiliation[label1]{organization={Pengcheng Laboratory},
             city={Shenzhen, Guangdong},
             postcode={518055},
             country={China}}

 \author[label5]{Shaohong Jia}
  \affiliation[label5]{organization={Xiangxuelanxi Community},
             city={Beijing},
             postcode={101102},
             country={China}}

  \affiliation[label2]{organization={Goldwind Sci \& Tech Co.,Ltd.},
             city={Beijing},
             postcode={100176},
             country={China}} 

 \author[label4]{Hongyu Wang}
 \affiliation[label4]{organization={Internet Based Engineering},
             city={Beijing},
             postcode={100083},
             country={China}}

 \author[label3]{Zhe Chen}          

 \author[label2,label3]{Bing Yang*\footnote{*Bing Yang, corresponding author, bingyang1215@126.com}} 
 \affiliation[label3]{organization={Institute of Engineering Thermophysics, Chinese Academy of Sciences},
             city={Beijing},
             postcode={100083},
             country={China}}             


\begin{abstract}
This study explores the noise reduction achieved by airfoils with serrated trailing edges in a low turbulence wind tunnel, focusing on acoustic spectral characteristics and wake flow field measurements. We analyze the effects of various factors, including Reynolds number, angle of attack, serration parameters, and model type, on sound power levels and far-field radiation patterns. Our findings reveal that serrated trailing edges significantly reduce blunt vortex shedding noise and laminar separation bubble noise across a broader frequency range, particularly in the mid-to-high frequency range, with reductions delineated by two boundaries. Interestingly, the serration geometry exhibits minimal impact on noise reduction, which varies with the angle of attack and airfoil profile across all tested conditions. Additionally, while serrations effectively lower noise levels, especially at higher frequencies, they do not significantly alter the airfoil's acoustic directivity patterns. Measurements of wake flow velocity spectra demonstrate a clear correlation between reduced wake turbulence and noise reduction, as serrated edges decrease the power spectral density of turbulent velocity fluctuations, effectively disrupting larger vortex structures responsible for noise generation. These valuable insights contribute to understanding the acoustic benefits of serrated trailing edges.
\end{abstract}


\begin{highlights}
\item This study shows that serrated trailing edges can effectively reduce noise over a broader frequency range than previously observed, with notable reductions delineated by two critical boundaries.
\item The noise reduction performance of serrated trailing edges is influenced by the angle of attack and the airfoil's geometry, leading to unique noise reduction profiles in both frequency and intensity.
\item The impact of serration geometry on noise attenuation was found to be minimal.
\item This study indicate a strong link between reduced wake turbulence and noise suppression, suggesting that serrations disrupt larger vortices, which are key contributors to noise generation.
\end{highlights}


\end{frontmatter}


\section{Introduction}

In the pursuit of enhancing wind turbine efficiency and productivity, there has been a noticeable trend towards increasing the size of wind turbines and extending blade lengths~\cite{kaewniam2022recent,hoen2023effects}. As a result, the tip speed ratio of wind turbine rotors has been steadily rising in order to reduce the load and cost~\cite{kosasih2016influence}. However, aerodynamic noise generated by wind turbine blades increases with the fifth power of inflow speed~\cite{SORENSEN2012225}. While higher tip speeds can improve wind turbine performance, they also lead to increased noise emissions. For instance, a 15\% increase in the tip speed of wind turbine blades can result in a 3 dB rise in noise, which may adversely affect the living conditions of nearby residents~\cite{barone2011survey}.

The primary sources of noise in wind turbines are aerodynamic noise from the blades and mechanical noise from the nacelle~\cite{deshmukh2019wind}. Mechanical noise, stemming from the operation of mechanical components, is characterized by tonal qualities~\cite{pinder1992mechanical}. Conversely, aerodynamic noise arises from the interaction between air and the wind turbine blade and can be categorized into inflow turbulence noise and airfoil self-noise~\cite{oerlemans2011wind}. Within airfoil self-noise, turbulent and laminar boundary layer trailing edge noise are significant due to their origin from boundary layer interaction with the sharp trailing edge~\cite{brooks1989airfoil}. Stall separation noise and blunt trailing edge vortex shedding noise are less influential because wind turbines usually operate below stall conditions, and sharp trailing edges can mitigate vortex shedding noise.

Effective reduction of trailing edge noise is essential for optimizing wind turbine blade design. Several trailing edge modifications have been proposed to reduce noise while maintaining aerodynamic performance. These include brush trailing edges~\cite{herr2005experimental,herr2007design,finez2010broadband,suryadi2023identifying}, perforated trailing edges~\cite{suryadi2023identifying,geyer2010measurement,zhang2020experimental}, and serrated trailing edges~\cite{oerlemans2009reduction,moreau2011flat,moreau2013noise,avallone2018noise,celik2021aeroacoustic,pereira2022aeroacoustics}. Among these, serrated trailing edges have become particularly prominent due to their ability to effectively disrupt vortex formation and reduce noise levels, as well as their adaptable design options.

The efficacy of serrated trailing edges in mitigating aerodynamic noise has been partially investigated~\cite{gruber2010experimental, gruber2011mechanisms, oerlemans2009reduction}. Gruber et al.\cite{gruber2010experimental, gruber2011mechanisms} reported noise reductions of up to 5 dB across a broad frequency spectrum, while Oerlemans et al.\cite{oerlemans2009reduction} conducted experiments on a full-scale 2.3 MW wind turbine, observing reductions of up to 3.2 dB below 1000 Hz. These studies demonstrate that serrated trailing edges effectively reduce wind turbine noise, particularly at low to mid frequencies. However, their effectiveness diminishes at higher frequencies, where they can occasionally increase noise above critical levels, underscoring the need for further investigation.

Celik et al.~\cite{celik2021aeroacoustic} conducted experiments on a flat plate with a serrated trailing edge, finding that larger serrations significantly reduce noise. They noted that low-frequency analysis may need high-fidelity simulations for accuracy. However, discrepancies between current numerical and experimental results highlight the need for further investigation and a deeper understanding of the underlying noise reduction mechanisms.

Zhou et al.~\cite{zhou2020study} investigated flexible trailing edge serrations on airfoils, finding that they achieve greater high-frequency noise reduction compared to rigid serrations. Additionally, they observed that flexible serrations align better with the flow, reducing crossflow intensity near the serration roots. However, selecting the appropriate flexibility depends on specific working conditions to ensure good stability of the serrations.

The geometry of serrated trailing edges plays a critical role in their noise reduction capabilities. Dassen et al.\cite{dassen1996results} found that serrated trailing edges on flat plates were more effective at noise reduction than those on airfoils. Finez et al.\cite{finez2011broadband} explored various geometric parameters of serrated trailing edges on a cascade of seven NACA 651210 airfoils, discovering no significant impact of the cascade effect on noise reduction in the low to mid-frequency range. Singh et al.\cite{singh2023control} examined non-uniform sinusoidal serrations, revealing that increased spanwise decoherence and vortex pairing resulted in greater noise reduction compared to uniform serrations. Gruber et al.\cite{gruber2011mechanisms} proposed design guidelines, emphasizing key parameters that influence noise reduction efficiency.

Significant advancements have been achieved in reducing tonal noise using serrated and porous trailing edges. Chong et al.\cite{chong2013experimental} investigated the effects of serrated trailing edges on unsteady pure tone noise, particularly focusing on Tollmien-Schlichting waves and separation bubbles on a NACA 0012 airfoil at moderate Reynolds numbers ($10^5 < Re_c < 10^6$). They found that serrations reduced instability noise by modifying the length scales of separation bubbles and suppressing their amplification. Additionally, the three-dimensional flow induced by serrations weakened normal fluctuations and adverse pressure gradients, further contributing to noise reduction. Subsequent studies\cite{chong2013airfoil, chong2011noise} demonstrated that non-planar serrated trailing edges accelerated the laminar-to-turbulent transition, further reducing noise levels. The trailing edge noise followed a power law of velocity, underscoring its primary role as a noise source. Zhang and Chong~\cite{zhang2020experimental} also examined the impact of porous parameters on tonal noise, identifying optimal settings for broadband noise reduction.

Despite significant advancements, the fundamental mechanisms behind noise reduction through modified trailing edges remain unclear. This study aims to enhance understanding by investigating the noise reduction mechanisms of wind turbine blade airfoils with serrated trailing edges. Experiments were conducted in an open-circuit anechoic low turbulence wind tunnel using various airfoil models, including two wind turbine blade airfoils and a reference flat plate. The findings from this study will aid in optimizing trailing edge designs for noise reduction and developing low-noise wind turbine blades. Looking ahead, drones and low altitude aircraft may gradually enter daily life for express delivery and human transportation. Therefore, rotor noise also needs careful attention.

\section{Experimental Apparatus and Procedure}

\subsection{Anechoic Wind Tunnel}

The experiments were conducted in a small open-circuit anechoic low turbulence wind tunnel located at the Fluid and Acoustic Engineering Laboratory, Beihang University. The anechoic chamber has internal dimensions of $8.9\  \text{m} \times 6.8\ \text{m} \times 4.65\ \text{m}$, with a cutoff frequency of 200 Hz, ensuring minimal acoustic reflections during measurements. The wind tunnel, which extends from the outside of the chamber to its interior, features a circular outlet with a diameter of 150 mm and can achieve a maximum wind speed of 50 m/s. Table~\ref{velocity_turbulence} provides the characteristics of the turbulence measured using a hot wire anemometer at the center of the outlet, indicating a low intensity of the turbulence, which is critical for ensuring accurate flow and noise measurements in the enrironment of this work. The experiments were carried out at wind speeds of 15 m/s, 25 m/s, and 35 m/s, with noise and flow field measurements performed at each speed.


\begin{table}[H]
    \centering
    \caption{Measured Wind Velocity and Corresponding Turbulence Intensity at the Wind Tunnel Outlet}
    \renewcommand{\arraystretch}{1.5} 
    \setlength{\tabcolsep}{12pt} 
    \begin{tabular}{ccccc}
        \hline
        \textbf{Wind Velocity} (m/s) & 15.19 & 25.12 & 35.45 & 45.69 \\
        \hline
        \textbf{Turbulence Intensity} (\%) & 0.036 & 0.080 & 0.101 & 0.144 \\
        \hline
    \end{tabular}
    \label{velocity_turbulence}
\end{table}

Table~\ref{velocity_turbulence} shows the measured wind velocities and their corresponding turbulence intensities at the wind tunnel outlet. The turbulence intensity remains relatively low across all wind speeds, with values ranging from 0.036\% to 0.144\%. This indicates that the flow remains stable and well controlled, even at higher velocities, ensuring minimal turbulence interference. These low turbulence levels are particularly important for the experimental setup, as the study aims to investigate noise reduction under conditions of low inflow turbulence, which is critical for isolating the effects of trailing edge modifications on acoustic emissions.

\begin{figure}[H]
	\centering 
	\subfigure[V = 25.12 m/s]{ 
		\label{V25}
		\includegraphics[width=.45\textwidth,trim=7 7 3 7,clip]{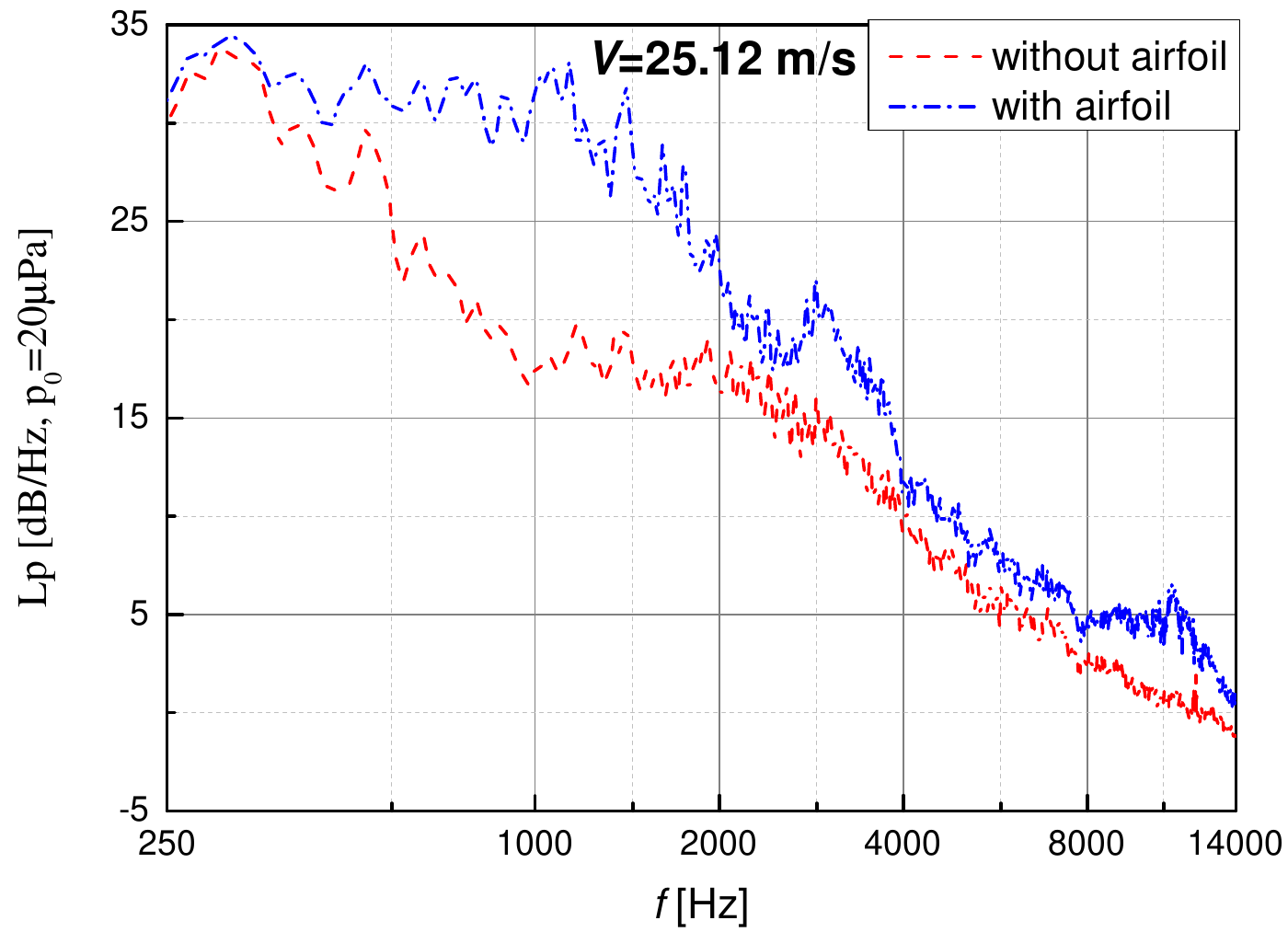} 
	} 
	\subfigure[V = 35.45 m/s]{ 
		\label{V35}
		\includegraphics[width=.45\textwidth,trim=7 7 3 7,clip]{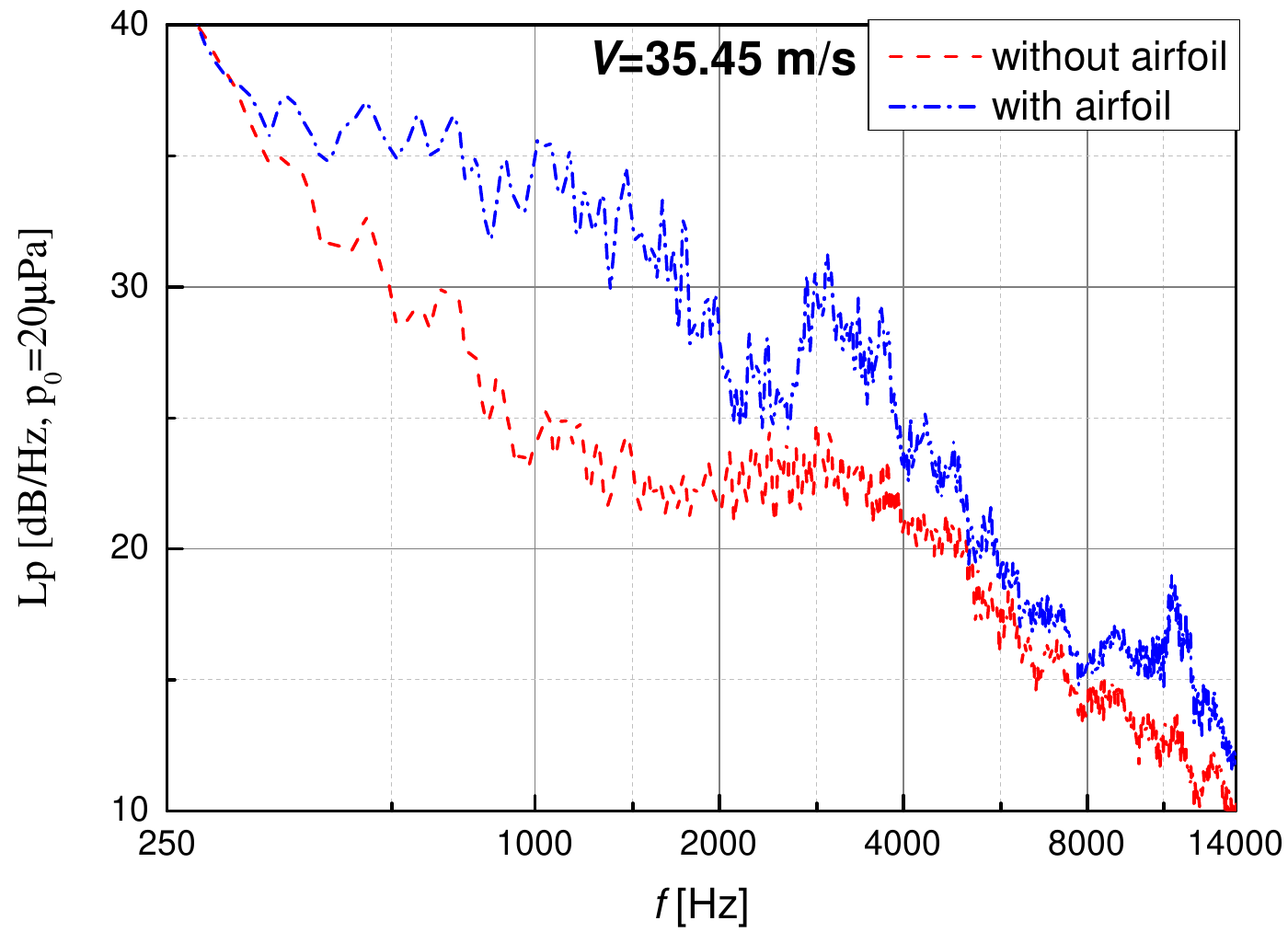} 
	} 
	\caption{Spectral comparison of background and airfoil noise} 
	\label{background_airfoil}
\end{figure}

Fig.~\ref{background_airfoil} compares the sound pressure level of the sound pressure signal, denoted as \( L_p \), in the anechoic chamber with and without a NACA 63(4)-421 airfoil at wind speeds of 25.12 m/s and 35.45 m/s. The sound pressure level \( L_p(f) \) at a frequency \( f \) is derived from the sound pressure signal captured by a microphone and is calculated using the formula shown in Eq.~\ref{eq:lp}:

\begin{equation}
L_p(f) = 20 \log_{10}\left(\frac{|P(f)|}{p_0}\right)
\label{eq:lp}
\end{equation}
where \( P(f) \) is the frequency domain representation of the sound pressure signal, obtained by applying the Fast Fourier Transform (FFT) to the time-domain signal \( p(t) \), measured by the microphone. Specifically, \( P(f) = \text{FFT}(p(t)) \), and the magnitude \( |P(f)| \) represents the amplitude of the sound pressure at frequency \( f \). The reference sound pressure \( p_0 \) is typically set to \( 20 \ \mu\text{Pa} \). The sound pressure level \( L_p(f) \), expressed in decibels (dB), is then calculated by comparing \( |P(f)| \) to \( p_0 \), using a logarithmic scale to reflect the human auditory system's sensitivity to sound.

As shown in Fig.~\ref{background_airfoil}, below 400 Hz, the airfoil has a negligible effect on noise levels. However, between 400 and 4000 Hz, the airfoil significantly increases noise by more than 10 dB at both wind speeds, suggesting a strong influence of the airfoil on mid-frequency noise. In the 4000 to 8000 Hz range, the increase is more moderate ($1\sim3$ dB), while from 8000 to 14000 Hz, it rises by 5 to 6 dB. These measurement results indicate that the airfoil mainly affects mid to high frequencies, with noise levels rising with wind speed. The maximum difference observed was 13.8 dB at higher wind speeds. The noise pattern, characterized by narrowband tonal components over broadband noise, highlights the complexity of the airfoil's noise generation. More research is needed to ensure that serrated trailing edges effectively reduce both broadband and tonal noise.

\subsection{Experimental Airfoils}

The experimental setup featured two wind turbine airfoil models, NACA 63(3)-418 and NACA 63(4)-421, each with a uniform chord length of 74 mm and a span of 160 mm. The NACA 63(3)-418 and NACA 63(4)-421 airfoils are low-drag designs optimized for wind turbine applications, with the 63(3)-418 being thinner (18\% thickness) for higher aerodynamic efficiency, and the 63(4)-421 slightly thicker (21\% thickness) for improved structural strength, especially in large turbine blades. The trailing edges, designed with a blunt profile and a thickness of $t=1$ mm, included a slot of $s=0.6$ mm to allow for the attachment of serrations of various sizes. This design enabled a detailed investigation into noise reduction techniques using different serrated trailing edges.



\subsection{Serrated Trailing Edges}

The tested half-height values ($h$) range from 4 mm to 6 mm, and the wavelength-to-half-height ratio ($\lambda/h$) varies from 0.2 to 0.8, offering a range of serration geometries to evaluate their effects on noise reduction. Aspect ratio is also shown in the table. Models labeled with "S" correspond to serrated trailing edges, while those labeled with "F" represent straight bars at half the serration height. The "0-0" model serves as the baseline configuration, featuring the original airfoil without any modifications to the trailing edge. These parameters are detailed in Table~\ref{serration_bar_parameters}.

\begin{table}[H]
    \centering
    \renewcommand{\arraystretch}{1.3}
    \setlength{\tabcolsep}{12pt}
    \caption{Serration and Straight Bar Parameters}
    \begin{tabular}{lccc}
        \toprule
        \textbf{Model} & \textbf{Half-height, $h$ (mm)} & \textbf{Wavelength, $\lambda$ (mm)} & \textbf{$\lambda/h$} \\
        \midrule
        4-1S & 4 & 3.2 & 0.8 \\
        4-2S & 4 & 1.6 & 0.4 \\
        4-0F & 4 & - & - \\
        \midrule
        5-1S & 5 & 4.0 & 0.8 \\
        5-2S & 5 & 2.0 & 0.4 \\
        5-3S & 5 & 1.0 & 0.2 \\
        5-0F & 5 & - & - \\
        \midrule
        6-1S & 6 & 4.8 & 0.8 \\
        6-2S & 6 & 2.4 & 0.4 \\
        6-3S & 6 & 1.2 & 0.2 \\
        6-0F & 6 & - & - \\
        \midrule
        0-0 & 0 & 0 & 0 \\
        \bottomrule
    \end{tabular}
    \label{serration_bar_parameters}
\end{table}

\subsection{Noise Data Acquisition and Processing System}

The noise measurement setup utilized BSWA MPA 416 pressure transducer microphones (1/4-inch radius, 20 to 20,000 Hz frequency range). Data acquisition was performed with a National Instruments (NI) system, comprising a PXIe-1071 chassis, a PXIe-8102 controller, and multiple PXIe-4496 data acquisition cards. Each PXIe-4496 card supported synchronous sampling across 16 channels with a rate of 204.8 kS/s, ensuring compliance with the frequency requirements for accurate noise measurement.


The sound signals from the microphones were processed and stored in \emph{.tdms} format using a LabVIEW-based data acquisition program. For analysis, these \emph{.tdms} files were imported into a LabVIEW-based FFT (Fast Fourier Transform) analysis tool, which allowed for parameter customization such as the starting point of the time segment, number of segments, sliding window, and reference channels. The data acquisition sampling rate was set at 65536 samples/s, with a time segment length of 4096 samples and a 10-second recording duration per data acquisition session. The sampling rate of 65536 samples/s is sufficient to capture the human audible range (20$\sim$20000 Hz), while the short segment length ensures precise frequency resolution per the uncertainty principle. Averaging across multiple segments reduces random errors for more accurate results. A spectral overlap rate of 50\% was used to ensure smooth spectrum calculation and effective averaging. The LabVIEW program used for data acquisition and FFT analysis is available at \href{https://doi.org/10.5281/zenodo.13768804}{https://doi.org/10.5281/zenodo.13768804}.




\subsection{Microphone Arrangement}

\begin{figure}[H]
\centering
\includegraphics[width=0.9\textwidth]{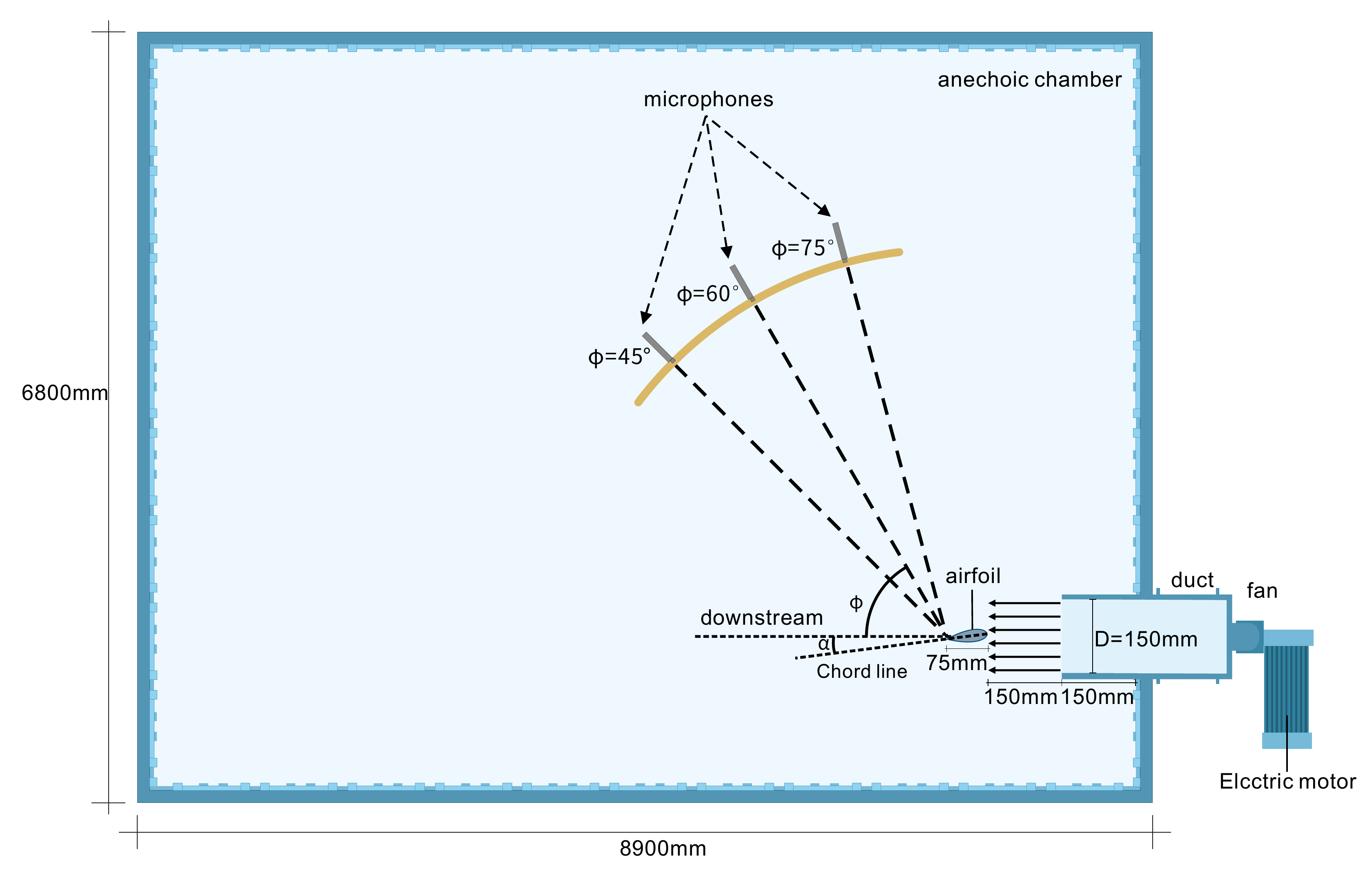}
\caption{Microphone placement for directivity measurement}
\label{directivity_microphones}
\end{figure}

The microphone arrangement in the experiment is shown in Fig.~\ref{directivity_microphones}. The angle between the line connecting the microphones and the blade airfoil trailing edge, and the direction of the airflow, is denoted as $\phi$. The angle between the airfoil chord and the downstream, denoted as $\alpha$ in Fig.~\ref{directivity_microphones}, is the opposite of the angle of attack. Therefore, the microphones are oriented towards the pressure side of the airfoil. Due to spatial constraints and equipment limitations within the wind tunnel, three microphones are positioned at a radius of 2.35 m from the model's trailing edge, corresponding to $\phi$ angles of 45\textdegree, 60\textdegree, and 75\textdegree, respectively. In this experiment, when directivity is not a primary concern, the microphone located at the 75\textdegree \ position is used for acoustic data acquisition and subsequent processing. No corrections have been made to the sound pressure level equations.

\subsection{Hot Wire Anemometry Measurement System}

\begin{figure}[H]
\centering
\includegraphics[width=0.8\textwidth]{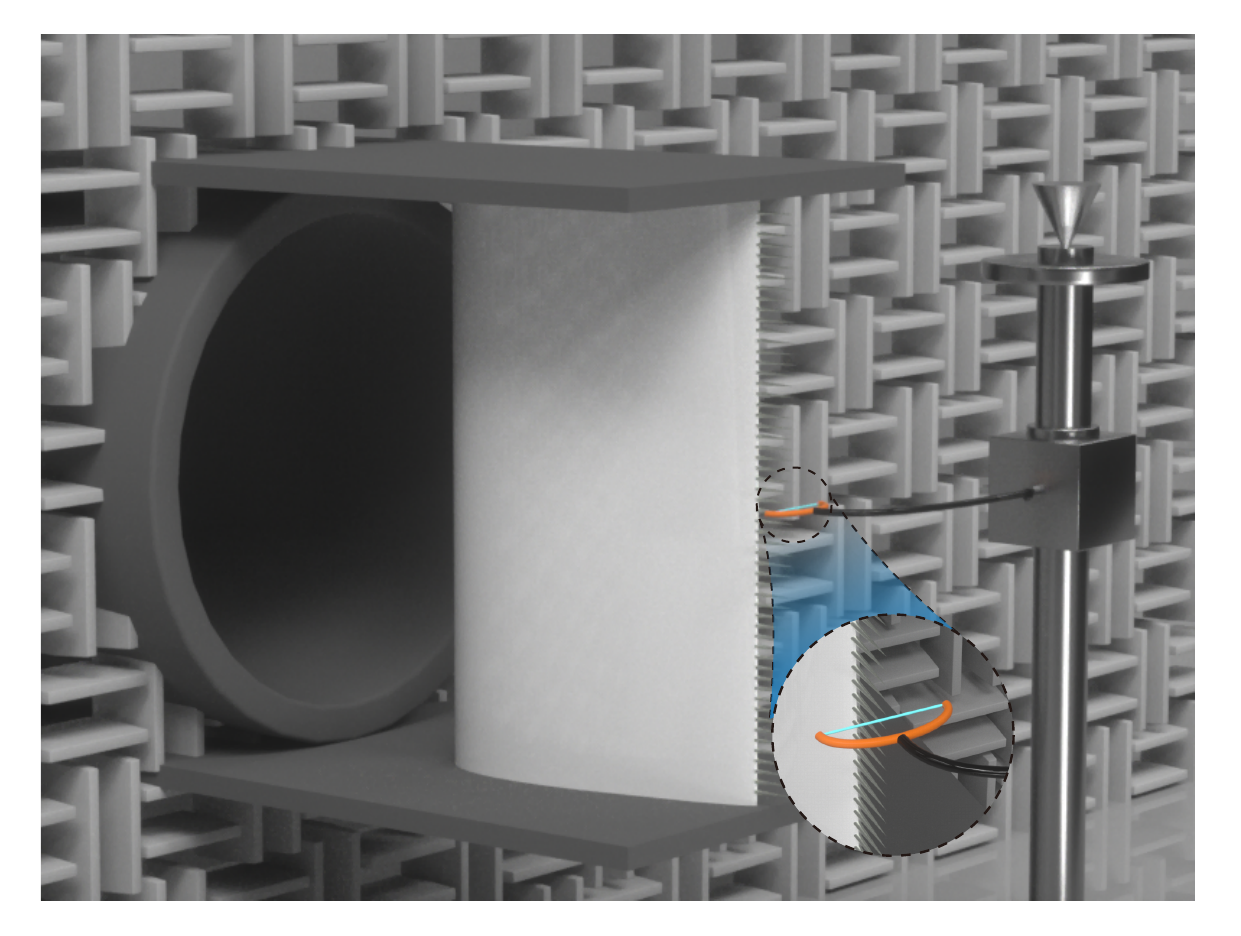}
\caption{Hot wire anemometry measurement setup}
\label{hot_wire}
\end{figure}

Fig.~\ref{hot_wire} diagrams the spatial arrangement of components within the hot wire anemometry measurement system, using Dantec Dynamics A/S products. The experiment utilized the 55P13 thermal wire probe, featuring a 90° bend, a hot wire diameter of 5 µm, and a length of 1.25 mm. This design minimizes interference with the flow field and has a maximum response frequency of 90 kHz. The three-dimensional displacement device employs a helical micro-displacement tool, providing high-resolution measurements with a precision of 0.05 mm.

\begin{figure}[H]
	\centering
	\includegraphics[width=0.8\textwidth,trim=175 123 163 140,clip]{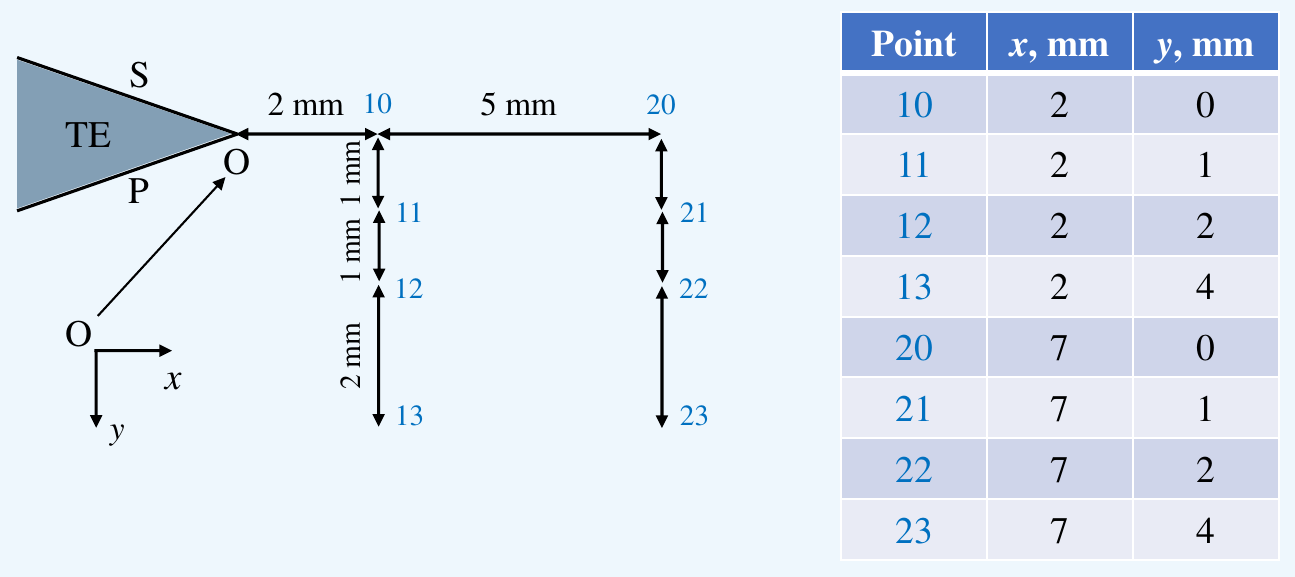}
	\caption{Measurement point locations}
	\label{measurement_locations}
\end{figure}

Velocity measurements were conducted on the wake flow of three different types: the original airfoil (0-0 model), the airfoil with a straight edge (5-0F model), and three airfoil models with serrated edges (5-XS, where X is 1, 2, or 3). The arrangement and numbering of measurement points are depicted in Fig.~\ref{measurement_locations}. For instance, measurement point "12" is located 2 mm downstream of the trailing edge in the pressure surface direction, with a 2 mm offset.

\section{Noise Measurement}

\subsection{Impact of Reynolds Number}

\begin{figure}[H]
\centering
\subfigure[$Re = 0.7 \times 10^5$]{
\includegraphics[width=.3\textwidth]{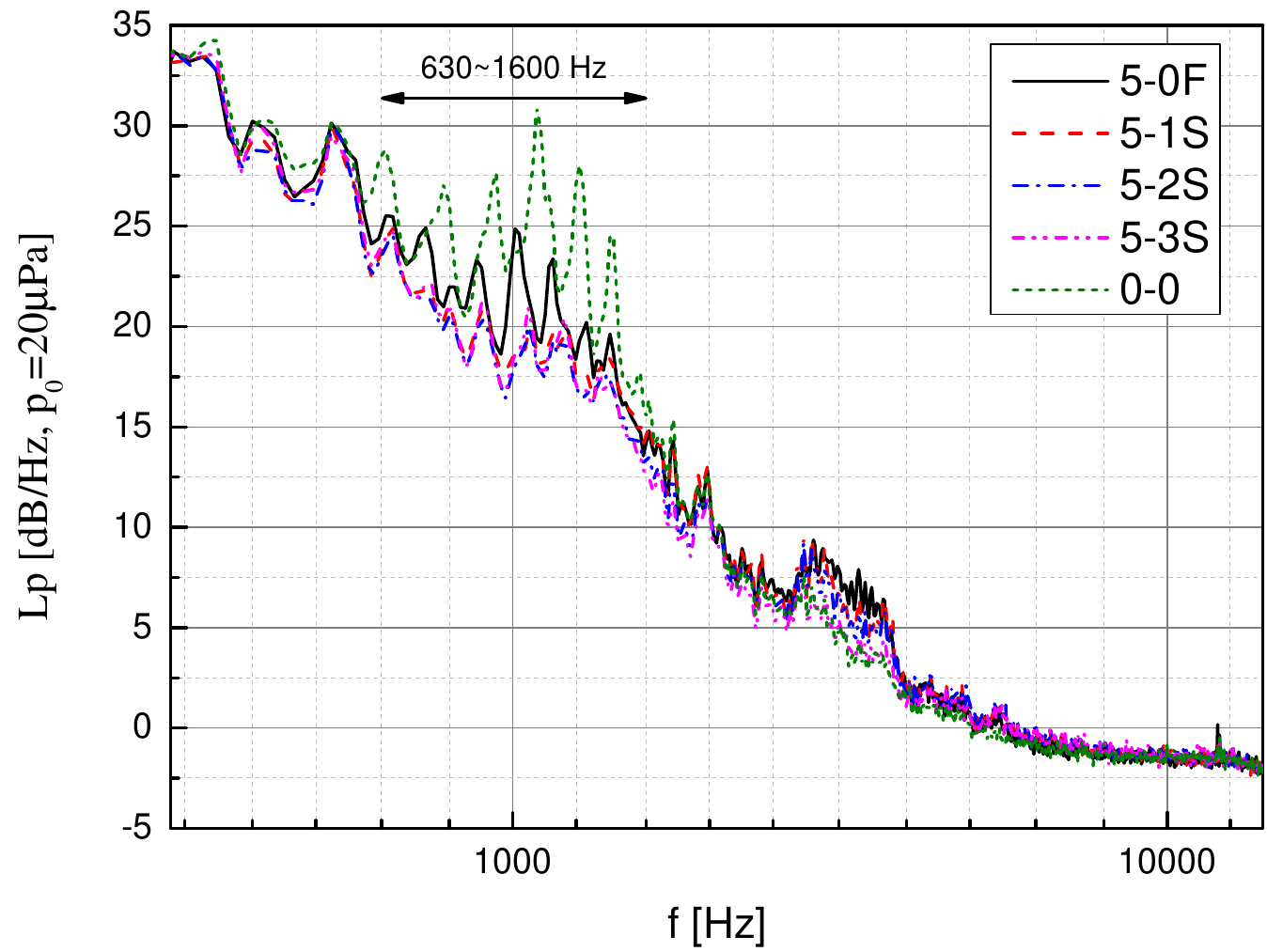}
}
\subfigure[$Re = 1.2 \times 10^5$]{
\includegraphics[width=.3\textwidth]{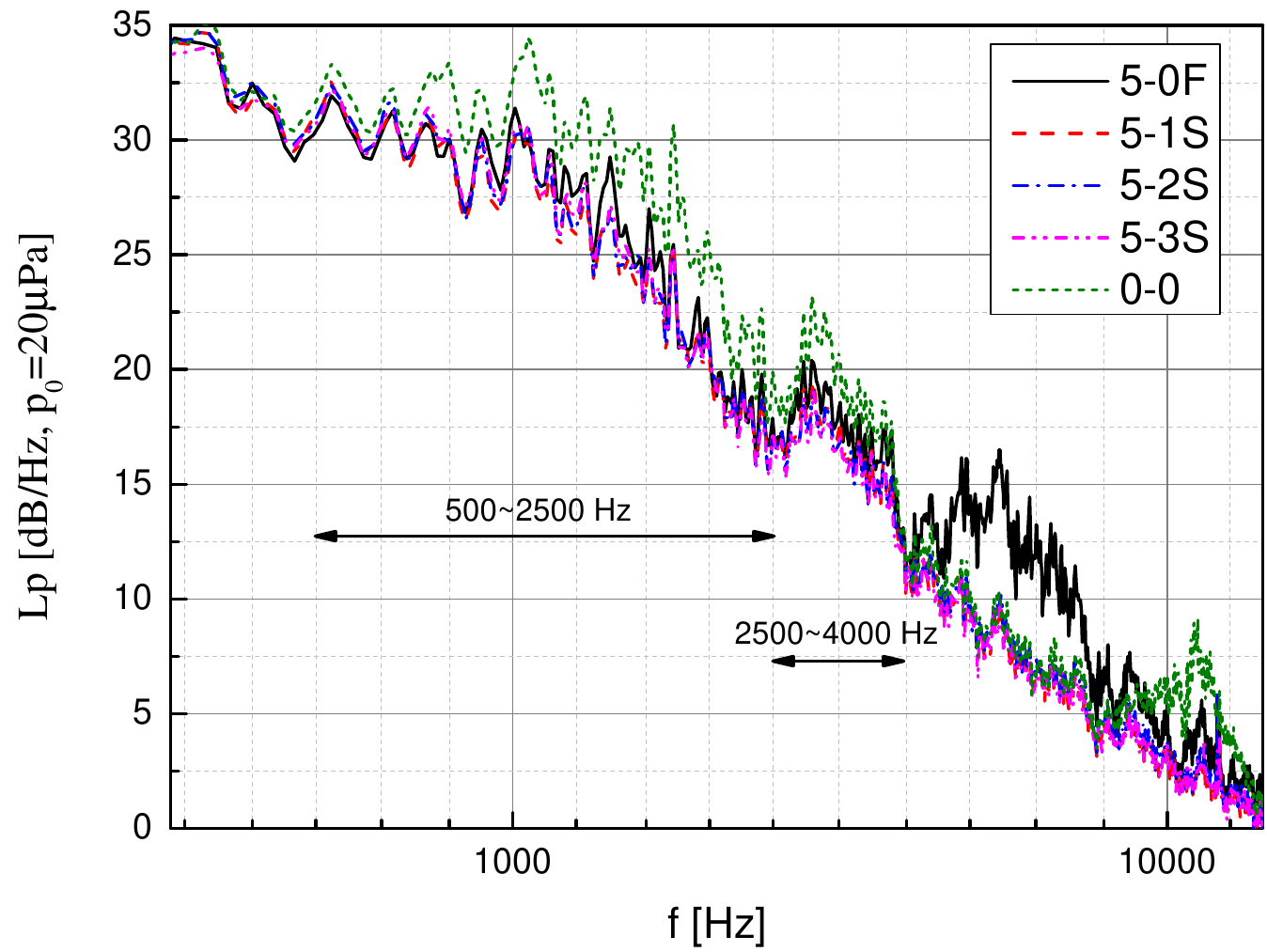}
}
\subfigure[$Re = 1.6 \times 10^5$]{
\includegraphics[width=.3\textwidth]{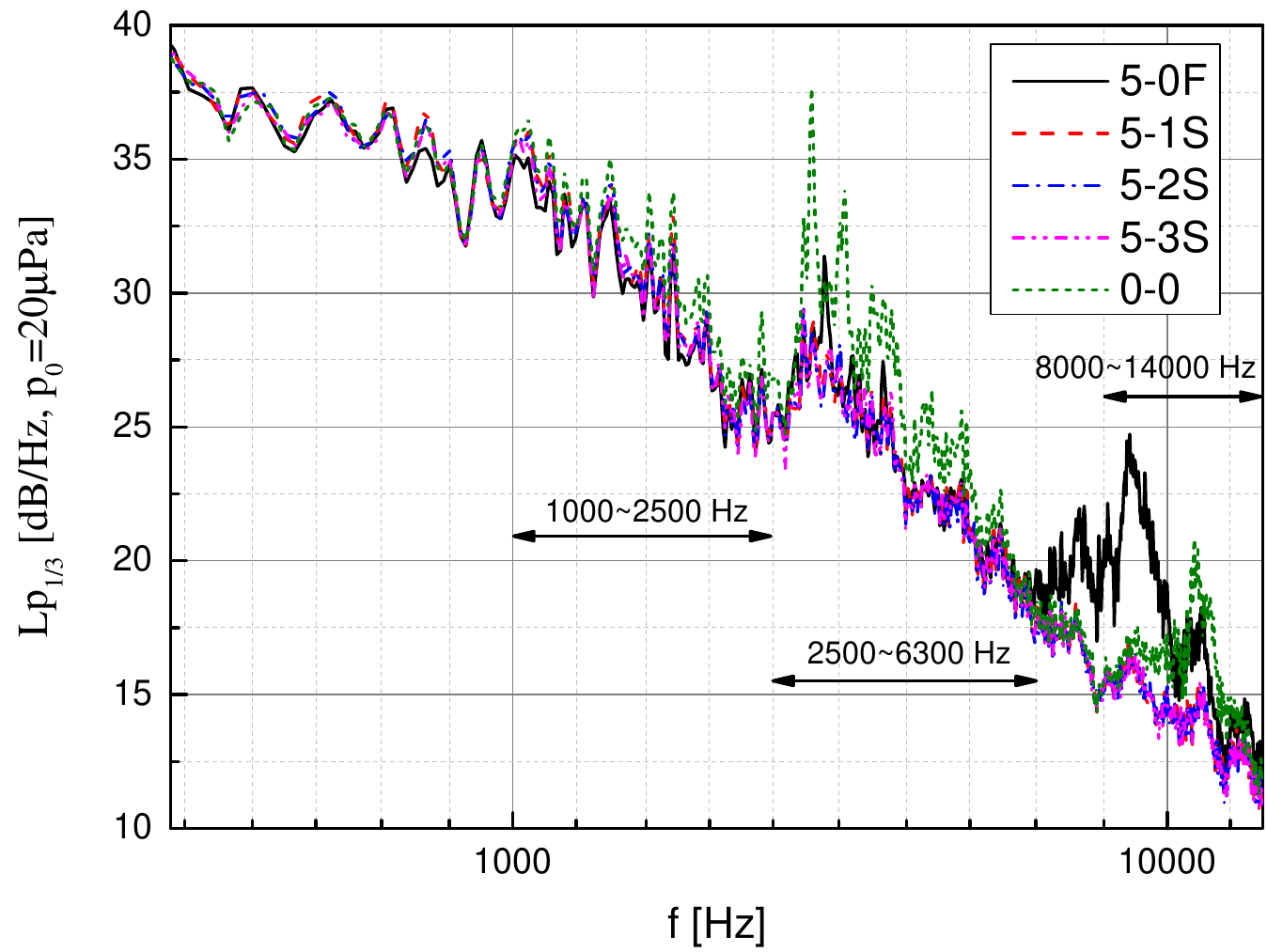}
}
\caption{Sound power level for NACA 63(3)-418 airfoil with trailing edge serrations} 
\label{Re_633418}
\end{figure}

\begin{figure}[H]
\centering
\subfigure[$Re = 0.7 \times 10^5$]{ 
	\includegraphics[width=.3\textwidth]{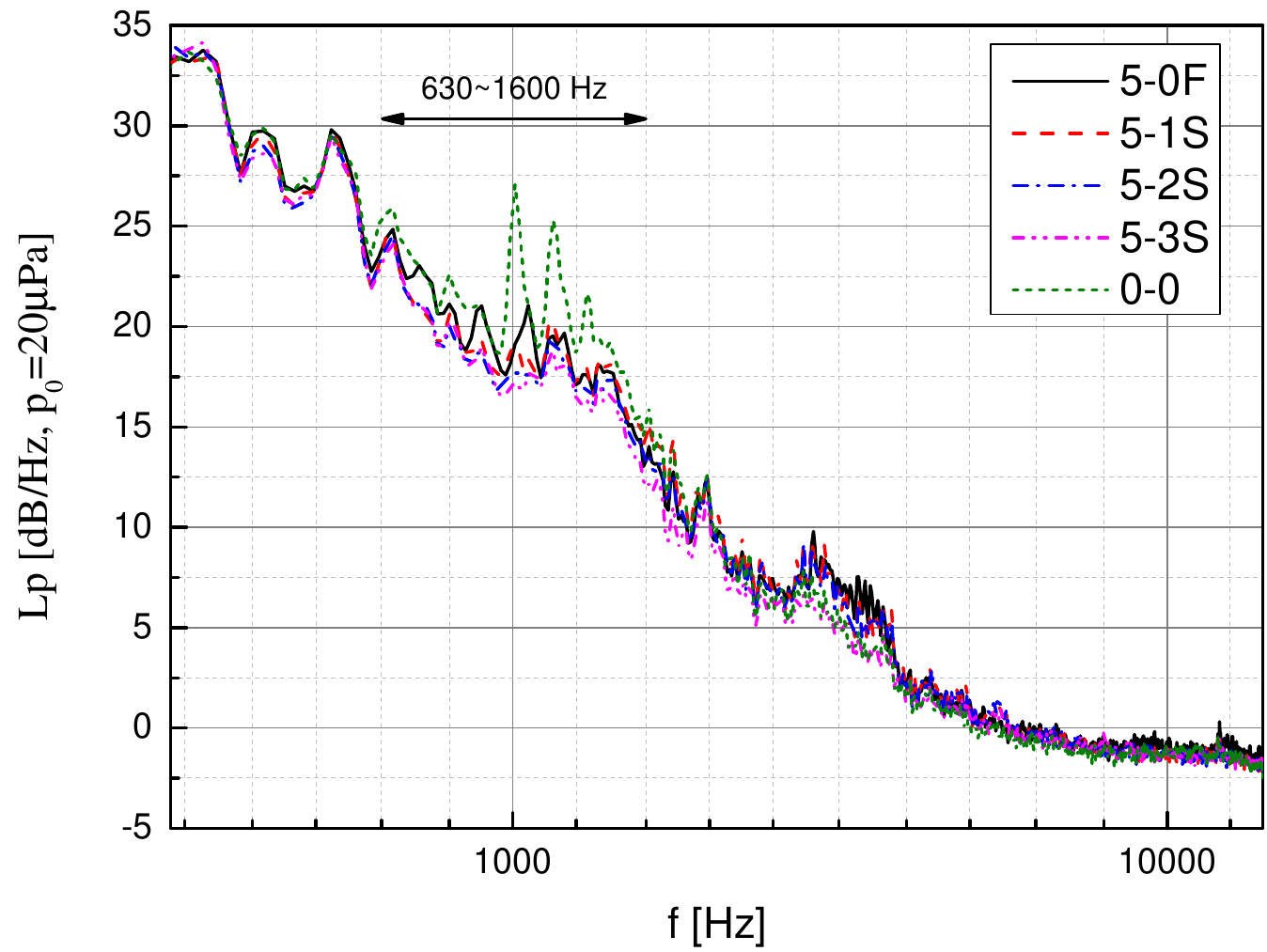} 
} 
\subfigure[$Re = 1.2 \times 10^5$]{ 
	\includegraphics[width=.3\textwidth]{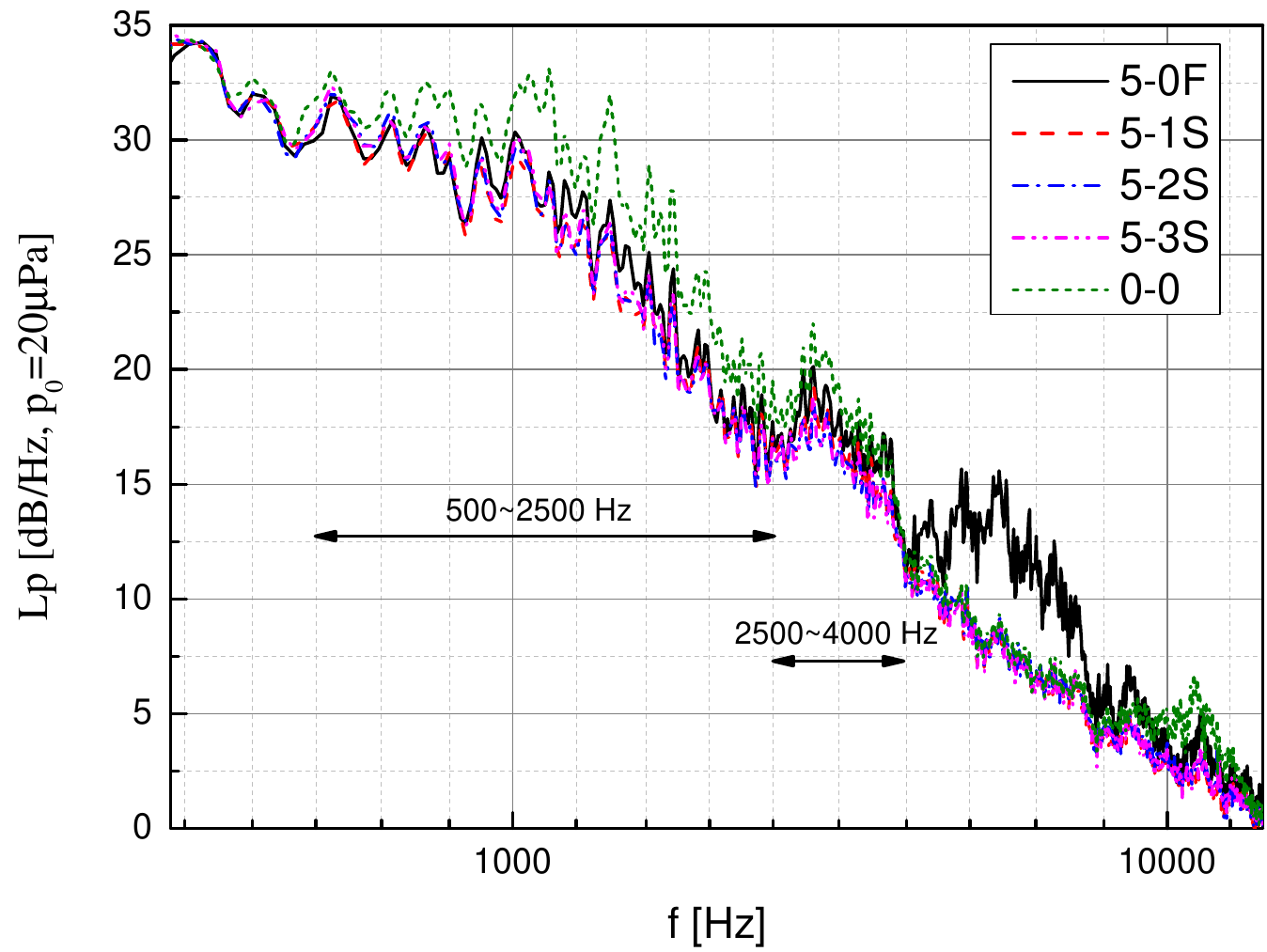} 
}
\subfigure[$Re = 1.6 \times 10^5$]{ 
	\includegraphics[width=.3\textwidth]{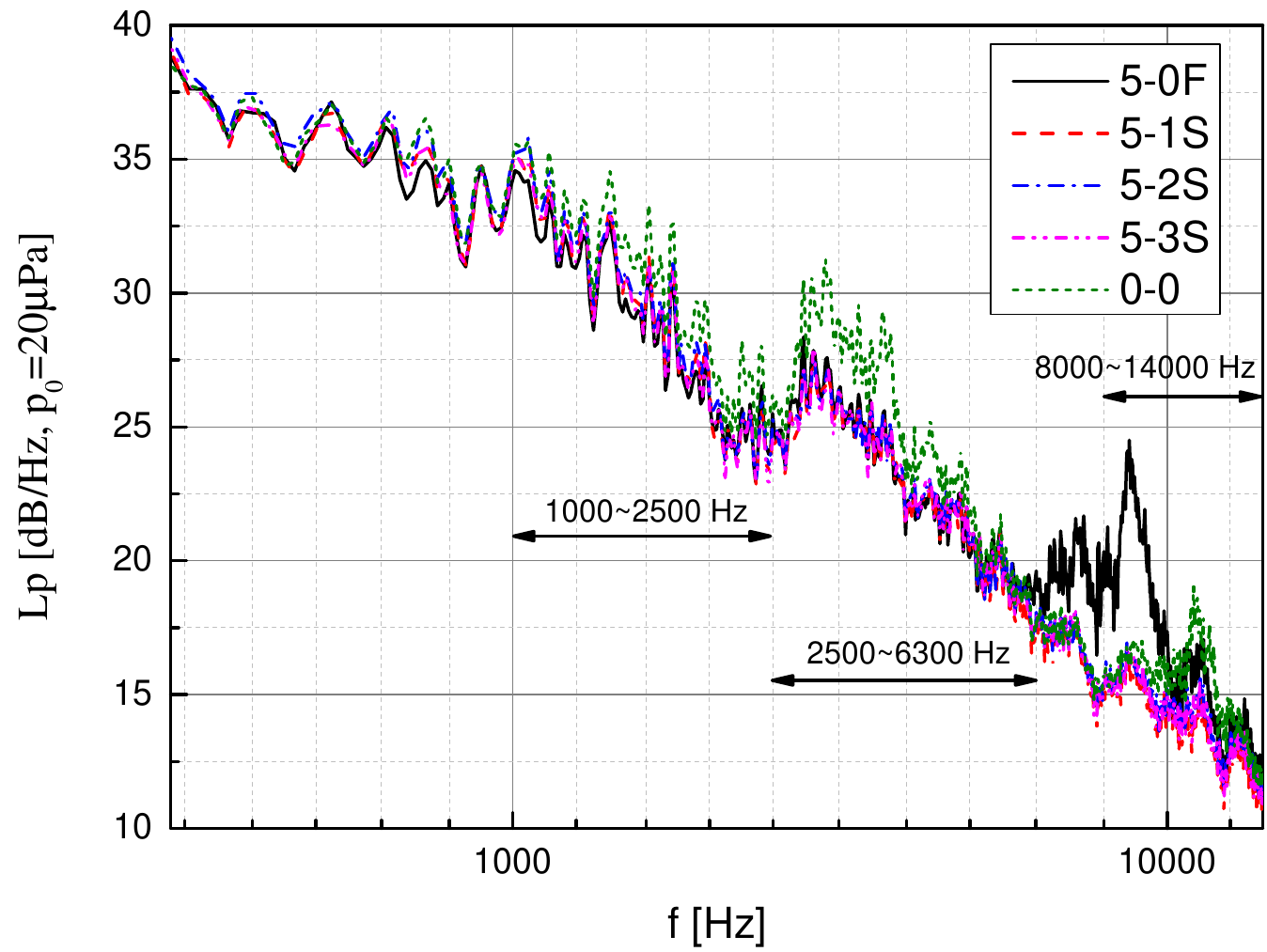} 
}  
\caption{Sound power level for NACA 63(4)-421 airfoil with trailing edge serrations} 
\label{Re_634421}
\end{figure}

\begin{figure}[H]
\centering
\subfigure[$Re = 0.7 \times 10^5$]{ 
	\includegraphics[width=.3\textwidth]{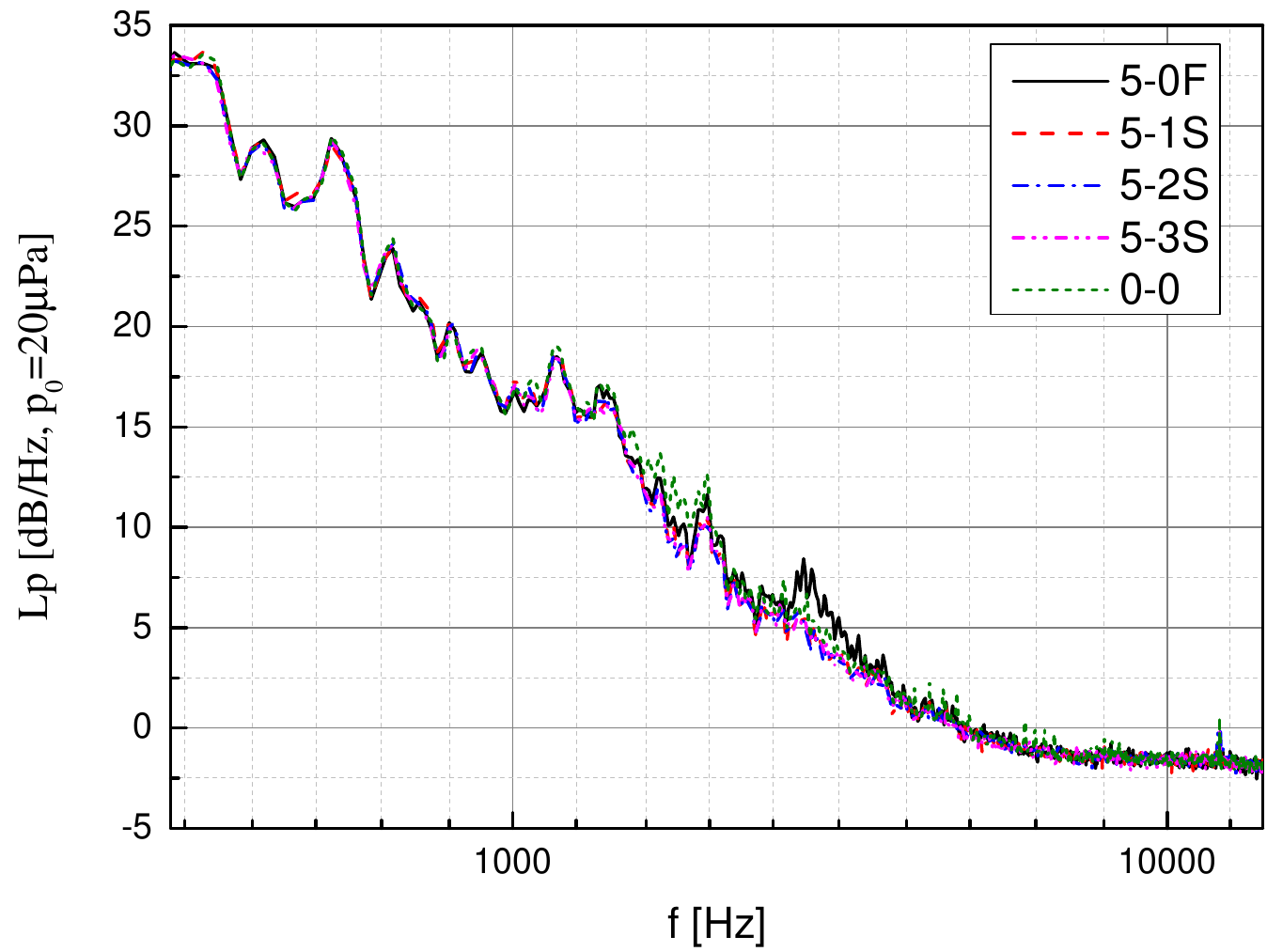} 
} 
\subfigure[$Re = 1.2 \times 10^5$]{ 
	\includegraphics[width=.3\textwidth]{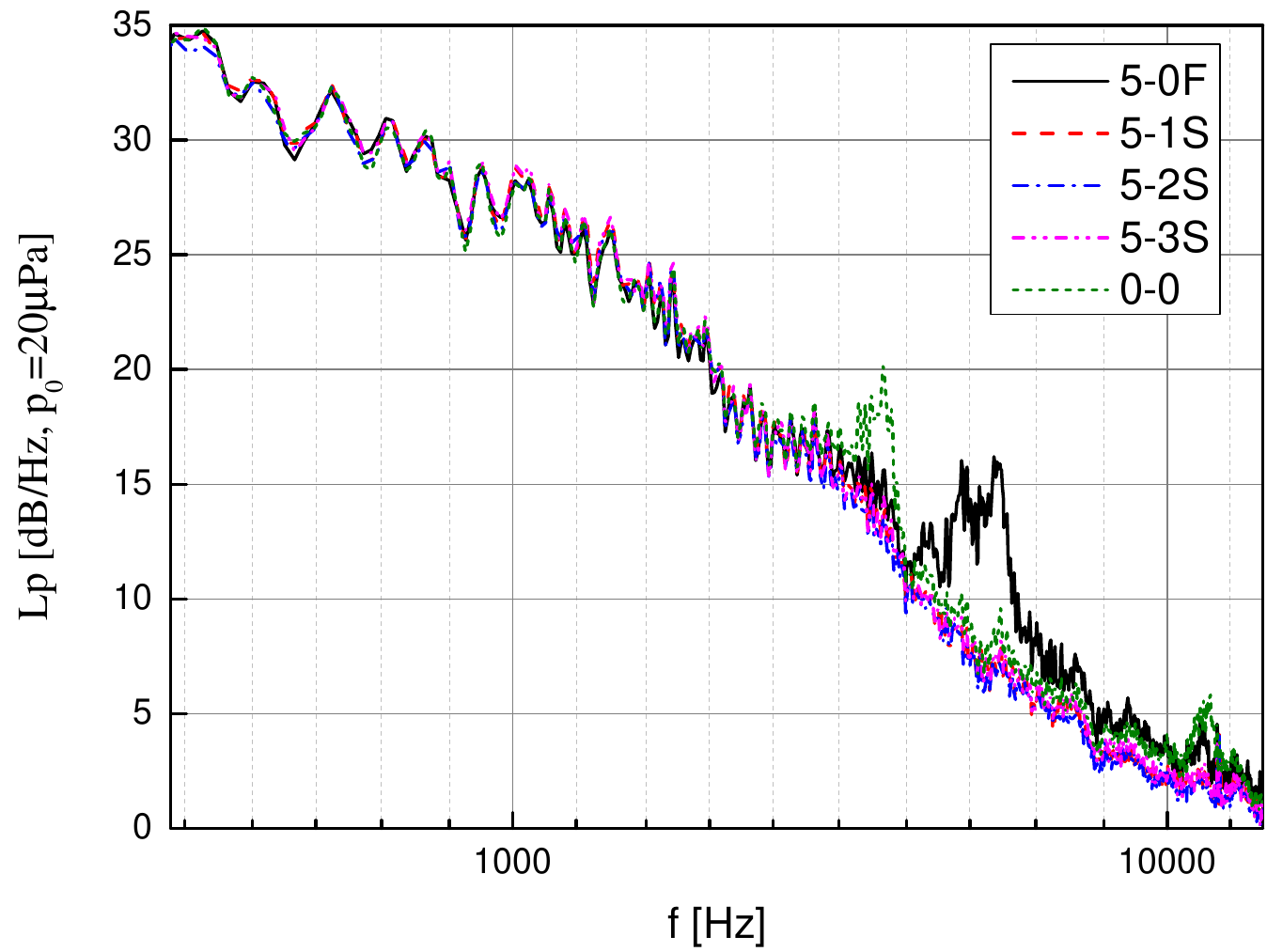} 
}  
\caption{Sound power level for a reference flat plate with trailing edge serrations} 
\label{Re_plate}
\end{figure}

The sound pressure levels for the NACA 63(3)-418, NACA 63(4)-421 airfoils were measured at Reynolds numbers of $0.7 \times 10^5$, $1.2 \times 10^5$, and $1.6 \times 10^5$. In the experiments testing the impact of Reynolds number, the angle of attack was consistently set to 0 degrees. Fig.~\ref{Re_633418} and Fig.~\ref{Re_634421} show the sound power levels for the NACA 63(3)-418 and NACA 63(4)-421 airfoils, including airfoils with trailing edge serrations, original airfoils, and those with added straight bars. The sound pressure levels for a reference flat plate with the same chord length are also presented in Fig.~\ref{Re_plate} at Reynolds numbers of $0.7 \times 10^5$ and $1.2 \times 10^5$.

The variation of noise reduction frequency ranges with Reynolds number ($Re$) reveals the effectiveness of trailing edge serrations. Trailing edge serrations consistently demonstrate their capacity to reduce noise levels. Compared to the original airfoil, serrations are effective in reducing noise across both low-to-medium and high frequencies. At lower Reynolds numbers, the NACA 63(3)-418 and NACA 63(4)-421 airfoils show primary noise reductions of 2$\sim$7 dB in the 630$\sim$1600 Hz range, featuring a broadband frequency range with superimposed tones. At moderate Reynolds numbers, the serrated trailing edge reduces noise by 2$\sim$5 dB in the 500$\sim$4000 Hz and 10000$\sim$12500 Hz ranges for both airfoil models. At higher Reynolds numbers, the serrated trailing edge is especially effective in reducing noise in the 2500$\sim$6300 Hz range. The noise reduction effect of serrated trailing edges on the flat plate is less significant than on the airfoil, with reductions primarily observed in the mid-to-high frequency range. The less significant noise reduction effect on the flat plate compared to the airfoil can be attributed to the absence of an aerodynamic profile. However, similar to the airfoil, the serrations on the flat plate show improved noise reduction at higher Reynolds numbers. This indicates that the effectiveness of the serrated trailing edge in reducing noise is both frequency dependent and highly sensitive to flow conditions.

\begin{figure}[H]
	\centering 
	\subfigure[NACA 63(3)-418]{ 
		\label{NACA633418_BL}
		\includegraphics[width=.45\textwidth]{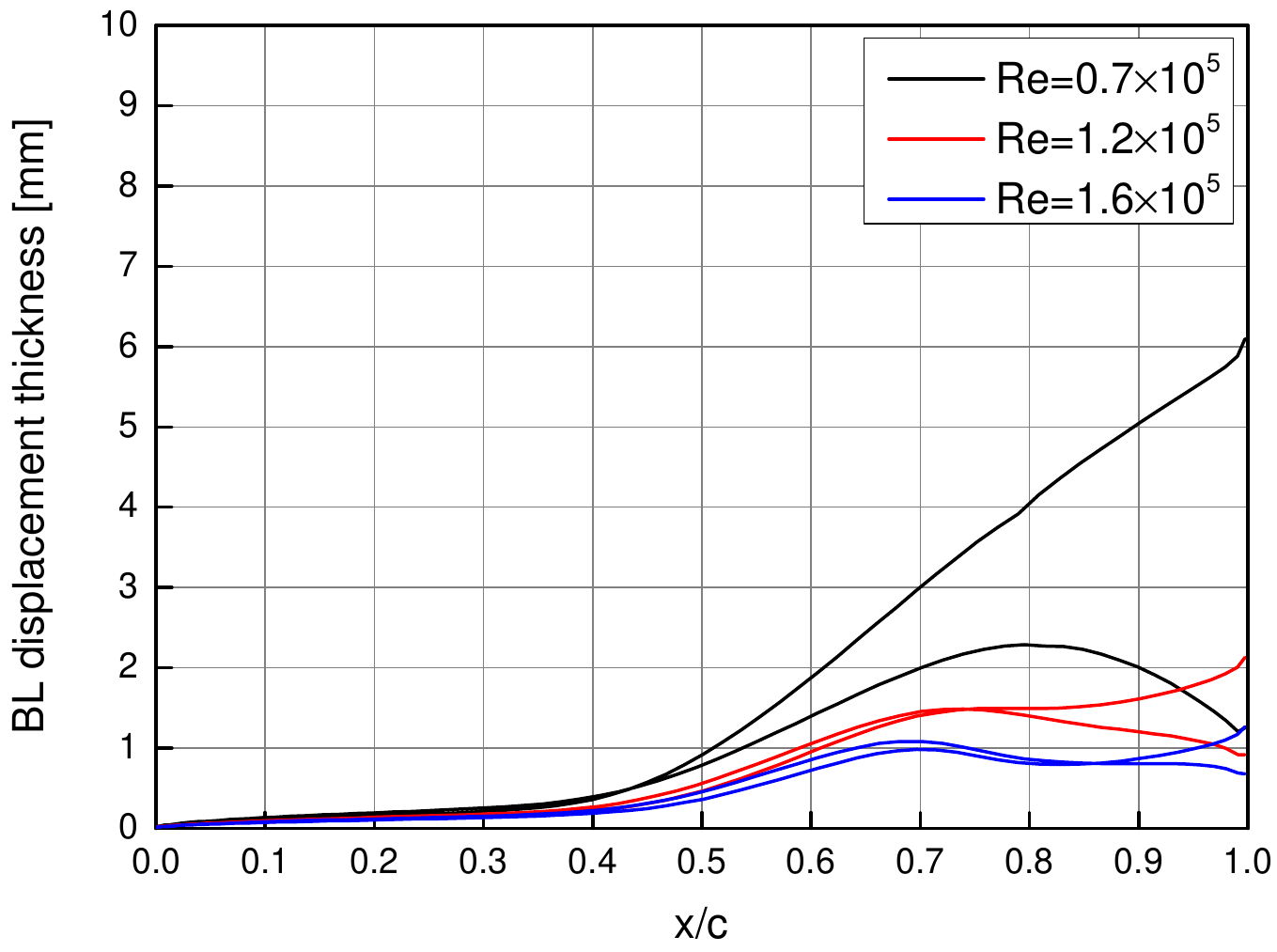} 
	} 
	\subfigure[NACA 63(4)-421]{ 
		\label{NACA634421_BL}
		\includegraphics[width=.45\textwidth]{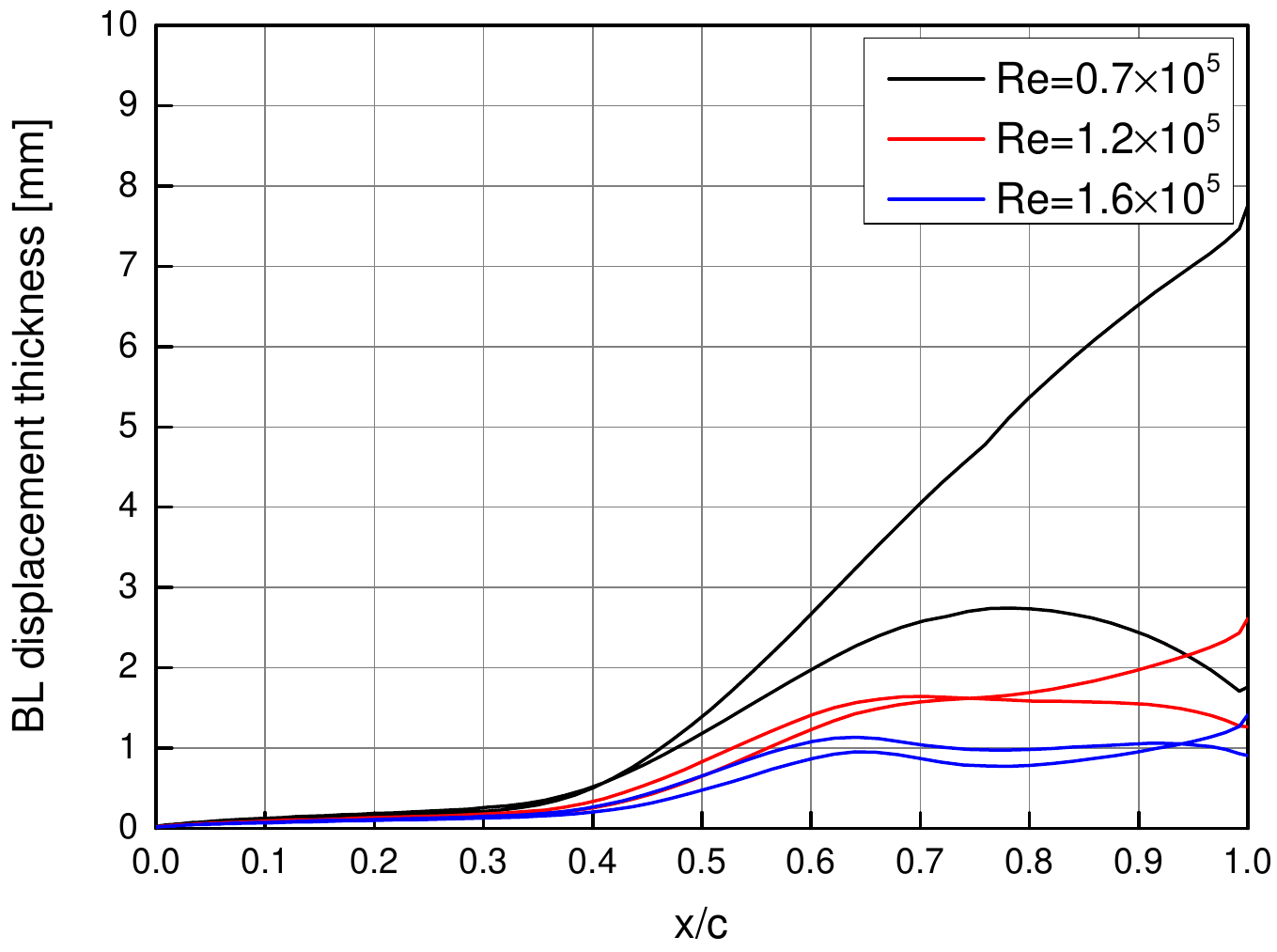} 
	}
	\caption{Boundary layer displacement thickness of NACA 63(3)-418 and NACA 63(4)-421 airfoils} 
	\label{airfoil_BL}
\end{figure}

Fig.~\ref{airfoil_BL} shows the boundary layer displacement thicknesses $\delta^*$ on both the suction and pressure sides for the NACA 63(3)-418 and NACA 63(4)-421 airfoils, calculated using XFoil\cite{drela1989xfoil}. Note that the boundary layer thickness is smaller on the pressure side and the microphones are positioned facing the pressure side of the airfoil in this work. According to Oerlemans~\cite{Oerlemans2018advances}, blunt vortex shedding noise occurs at frequencies around $f \sim \left(0.1 \sim 0.3\right) \times \frac{U}{t}$ when the ratio $\frac{t}{\delta^*}$ exceeds 0.3, where $t$ represents the blunt edge thickness. As shown in Fig.~\ref{Re_633418} (c) and Fig.~\ref{Re_634421} (c), the frequency range of 2500$\sim$6300 Hz contains multiple tonal components for the 0-0 model (the original airfoil with a 1 mm blunt trailing edge) at a Reynolds number of $Re = 1.6 \times 10^5$, consistent with Oerlemans' observations~\cite{Oerlemans2018advances}. A similar noise characteristic appears at $Re = 1.2 \times 10^5$ in the 2500$\sim$4000 Hz frequency range, though less pronounced. This suggests that for airfoils with blunt trailing edges, higher Reynolds numbers and reduced displacement thicknesses increase the likelihood of bluntness-induced vortex shedding.

\begin{figure}[H]
	\centering 
	\subfigure[NACA 63(3)-418]{ 
		\label{633418_Cp}
		\includegraphics[width=.45\textwidth]{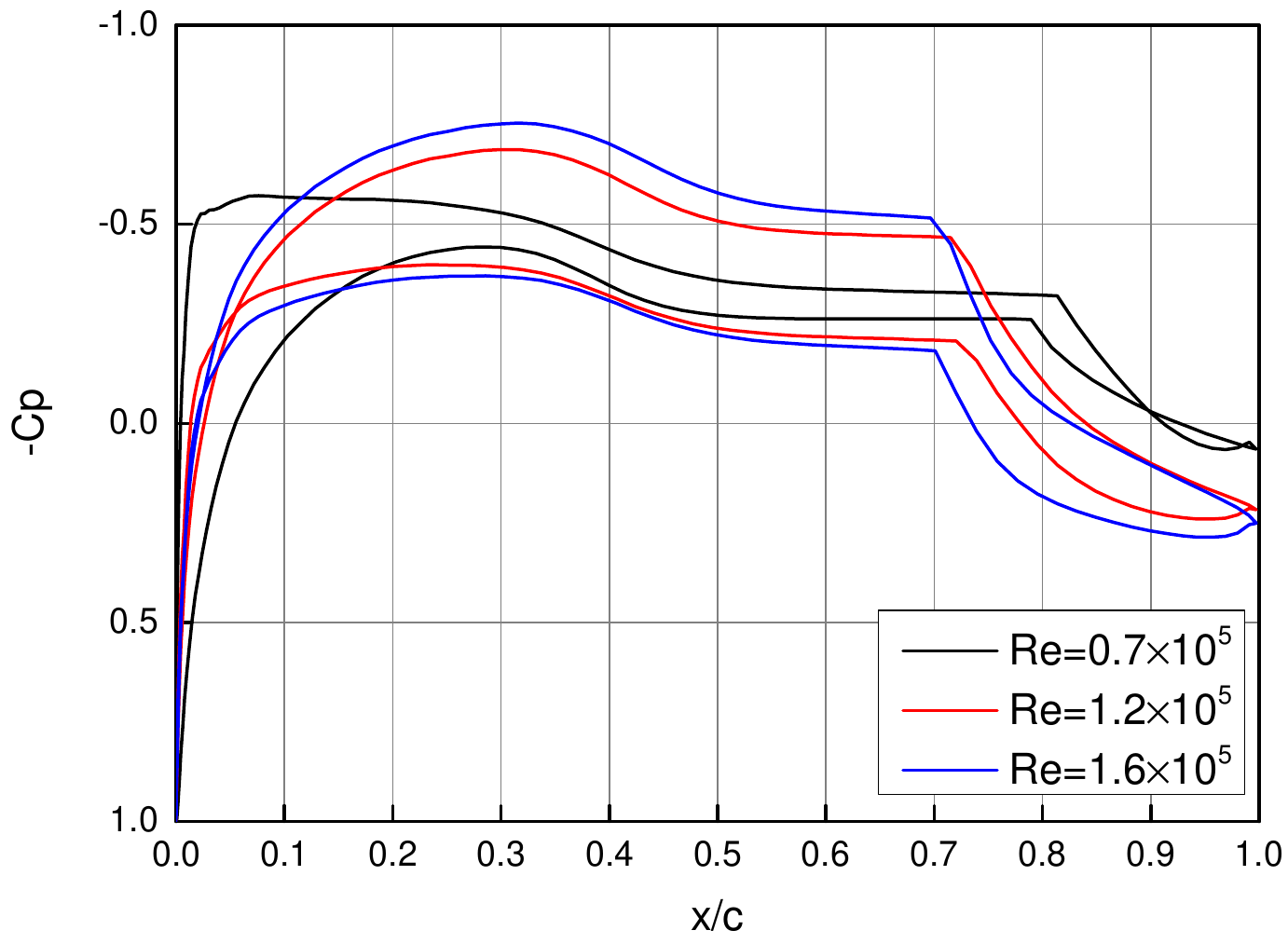} 
	} 
	\subfigure[NACA 63(4)-421]{ 
		\label{634421_Cp}
		\includegraphics[width=.45\textwidth]{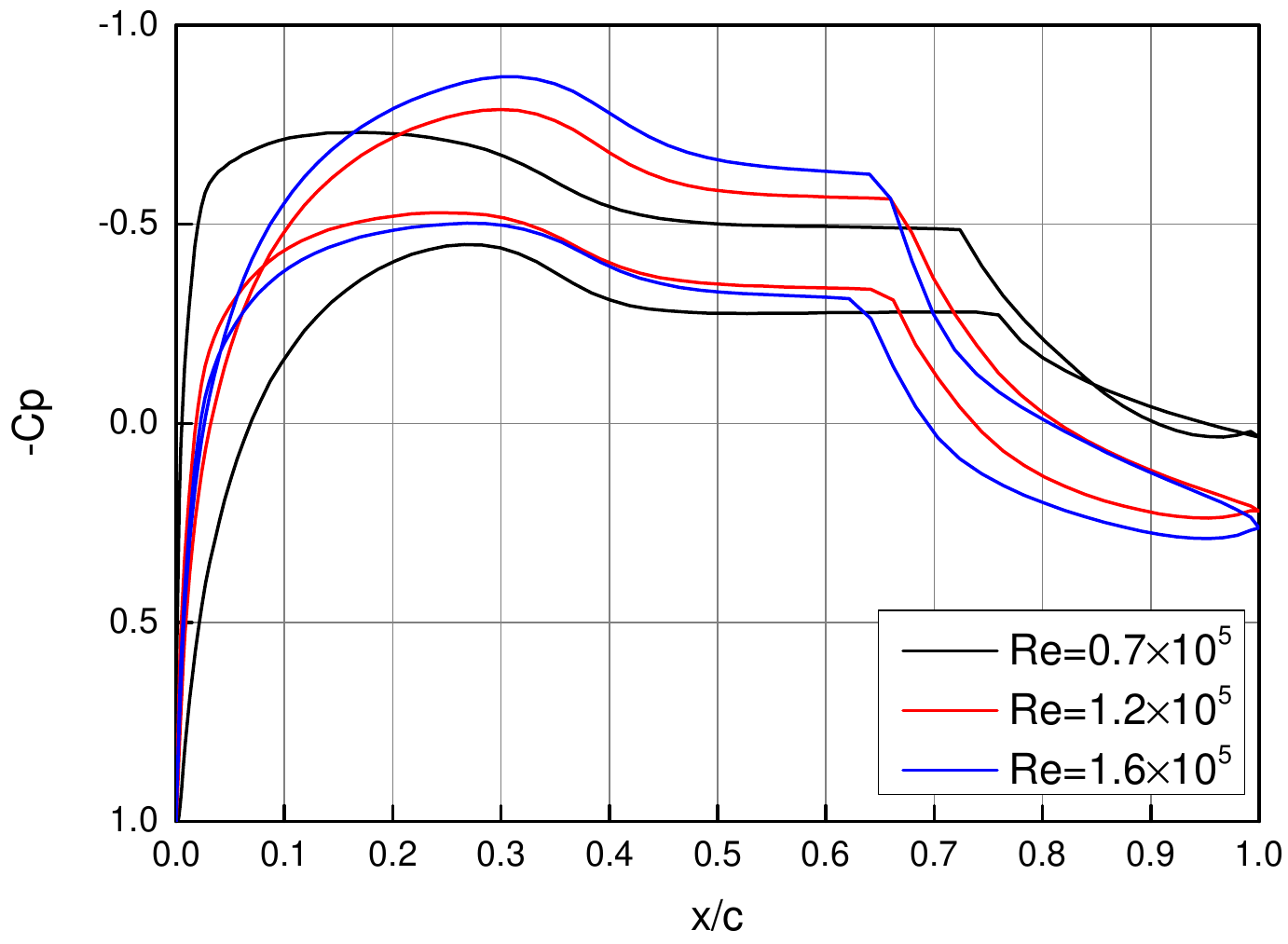} 
	}
	\caption{Pressure coefficient of NACA 63(3)-418 and NACA 63(4)-421 airfoils} 
	\label{airfoil_Cp}
\end{figure}

Fig.~\ref{airfoil_Cp} shows the pressure coefficient distributions for the NACA 63(3)-418 and NACA 63(4)-421 airfoils. For both airfoils, it can be observed that the pressure coefficient remains constant (pressure plateau) starting from the mid-chord location, indicating the occurrence of laminar separation bubble~\cite{drela2014flight}, and then increases sharply as the flow transitions to the turbulent region, particularly evident at the experimental Reynolds number of $Re = 0.7 \times 10^5$. At this Reynolds number, the laminar separation region on the pressure side of the NACA 63(4)-421 airfoil extends from 0.38 to 0.83 of the chord length. In contrast, at $Re = 1.6 \times 10^5$, the laminar separation on the pressure side extends from 0.42 to 0.68 of the chord length, which is smaller compared to the separation region observed at lower Reynolds numbers. The plateau arises because the fluid in the laminar section of the bubble remains nearly stagnant, preventing the development of significant pressure gradients. In contrast, the turbulent region exhibits intense mixing, allowing the flow to withstand substantial adverse pressure gradients.

\begin{figure}[H]
	\centering 
	\subfigure[NACA 63(3)-418]{ 
		\label{633418_H12}
		\includegraphics[width=.45\textwidth]{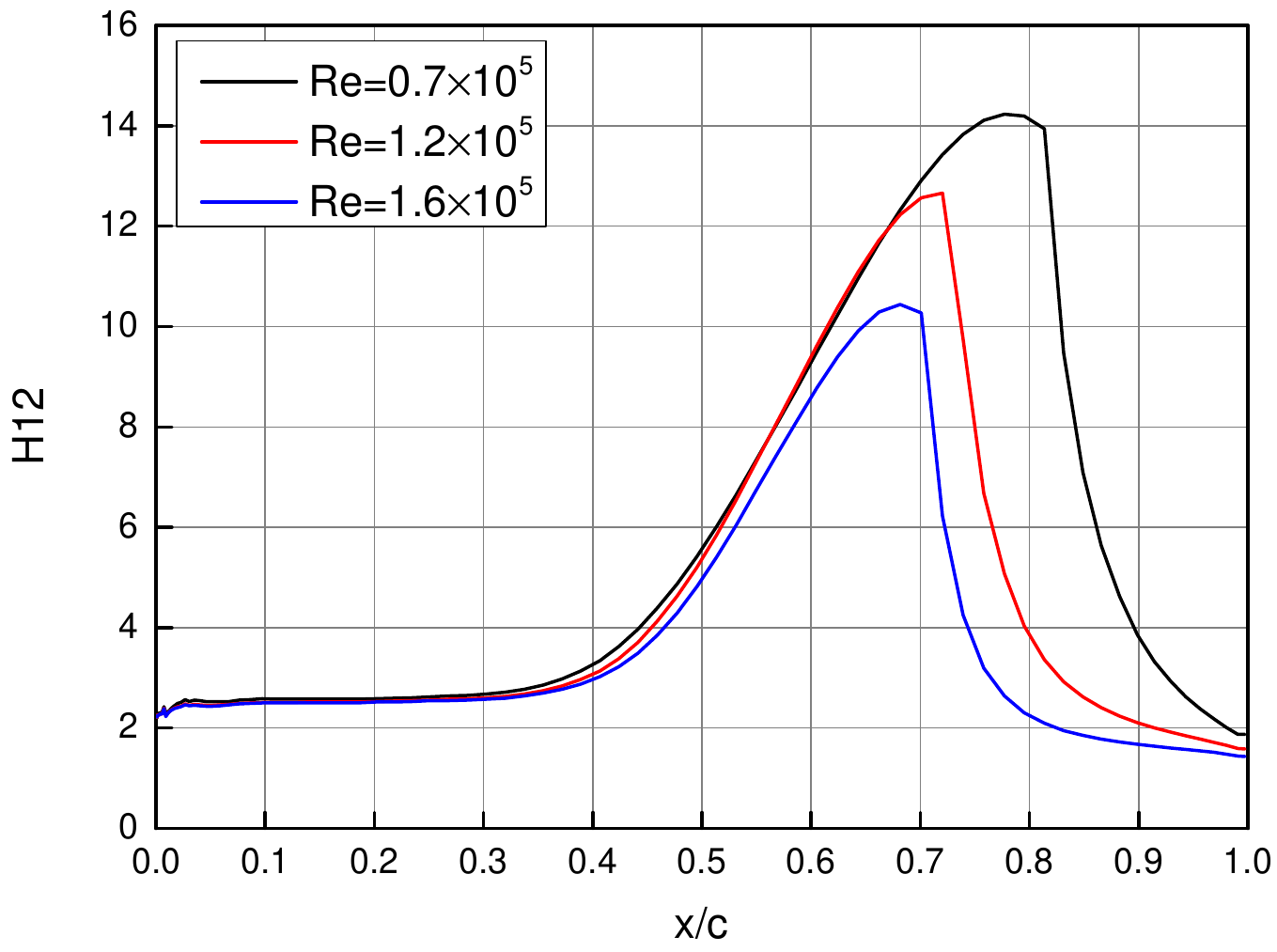} 
	} 
	\subfigure[NACA 63(4)-421]{ 
		\label{634421_H12}
		\includegraphics[width=.45\textwidth]{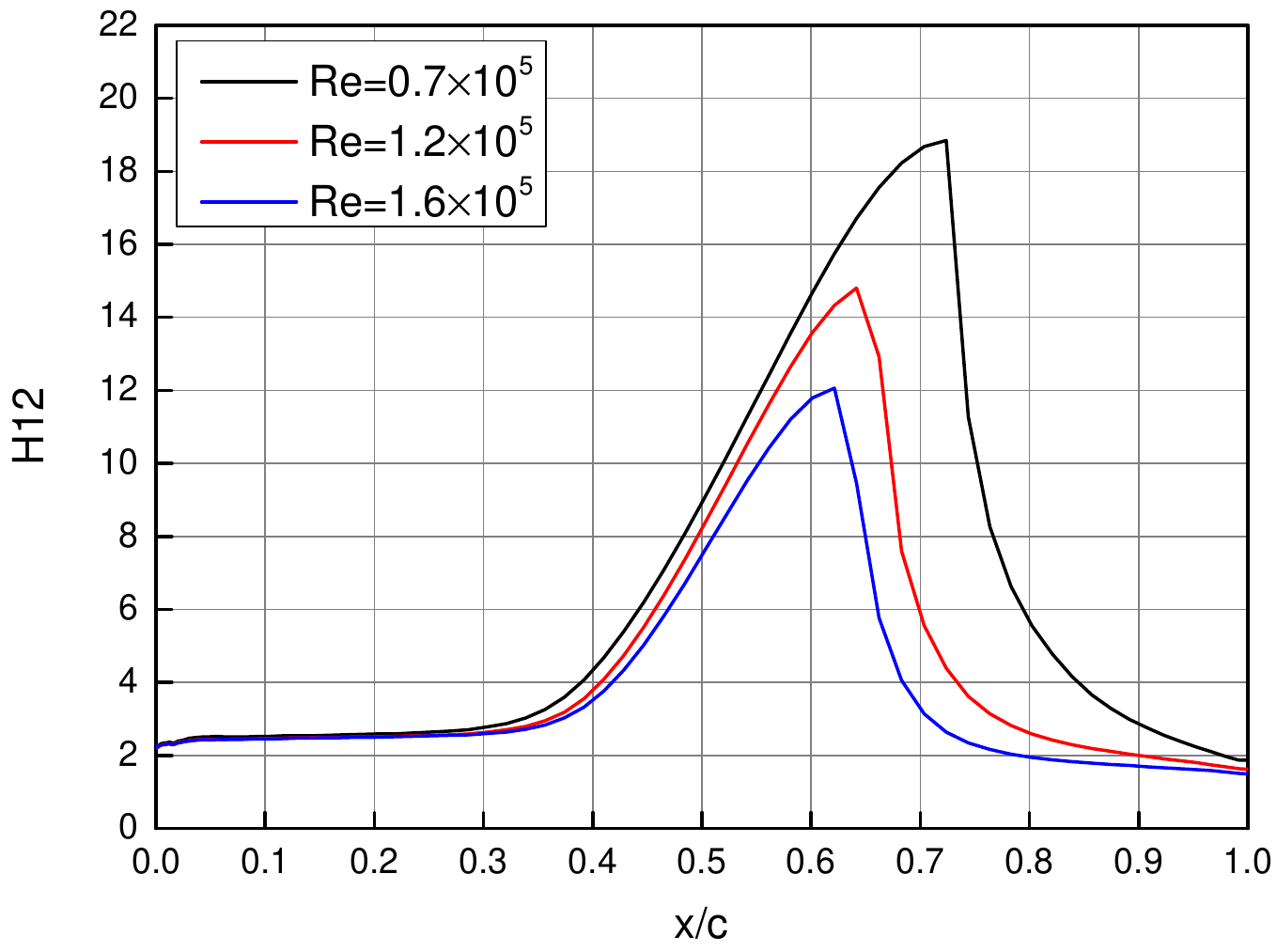} 
	}
	\caption{$H_{12}$ of NACA 63(3)-418 and NACA 63(4)-421 airfoils} 
	\label{airfoil_H12}
\end{figure}

Fig.~\ref{airfoil_H12} shows the shape factor \( H_{12} \), which is defined as the ratio of displacement thickness to momentum thickness in the boundary layer on the pressure side of the airfoil. The data shows that for both airfoils at different Reynolds numbers, the shape factor reaches its peak value at approximately 0.6 to 0.8 of the chord length. As the Reynolds number increases, the peak value of the shape factor (the position of transition onset~\cite{lobo2021investigation}) decreases, and the position of the peak shifts upstream. This suggests that higher Reynolds numbers promote earlier transition to turbulence, as the increased inertial forces at higher \(Re\) tend to destabilize the boundary layer, leading to a more rapid transition from laminar to turbulent flow. In contrast, at lower Reynolds numbers, the flow remains more stable, with a larger laminar separation bubble observed. The larger separation bubble at low \(Re\) is associated with a lower level of turbulence and a more significant region of laminar separation before the flow reattaches, consistent with the findings in Yang et al.~\cite{yang2022secondary}. This behavior highlights the interplay between Reynolds number, boundary layer stability, and the size of the laminar separation bubble in determining the transition characteristics of the flow. Additionally, a shape factor of around 2.0 is observed downstream of the turbulent reattachment~\cite{castillo2004separation}, particularly for \(Re = 1.2 \times 10^5\) and \(Re = 1.6 \times 10^5\). This shift suggests a faster turbulent reattachment with increasing Reynolds number. Additionally, the NACA 63(4)-421 airfoil exhibits a shape factor peak located further uptream compared to the NACA 63(3)-418 airfoil, pointing to differences in the boundary layer characteristics and development between the two airfoil configurations. These findings provide valuable insights into the influence of Reynolds number and airfoil geometry on boundary layer behavior.

\begin{figure}[H]
	\centering 
	\subfigure[Comparison with the original airfoil]{ 
		\label{with_original}
		\includegraphics[width=.45\textwidth]{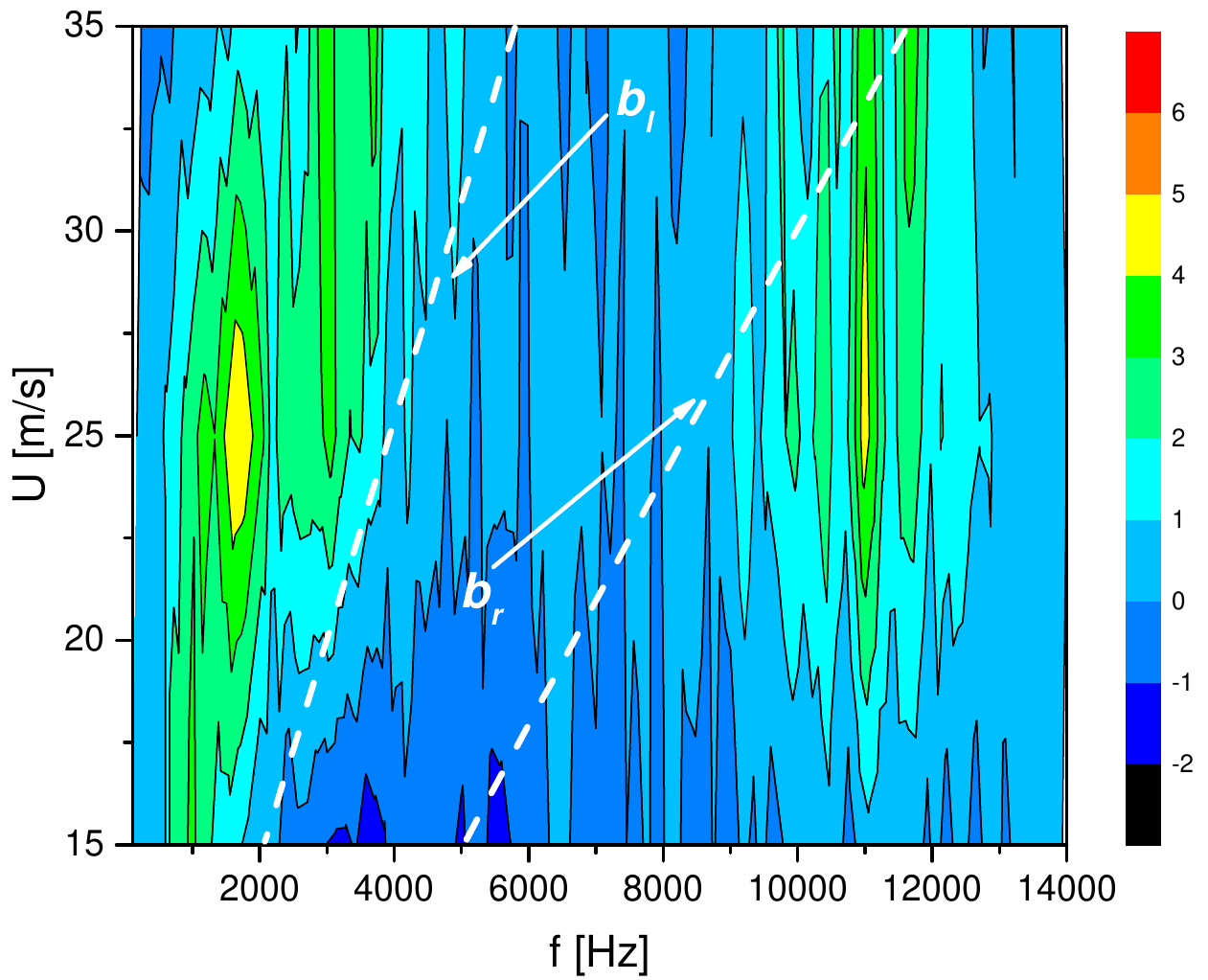} 
	} 
	\subfigure[Comparison with the airfoil with a bar trailing edge]{ 
		\label{with_bar}
		\includegraphics[width=.45\textwidth]{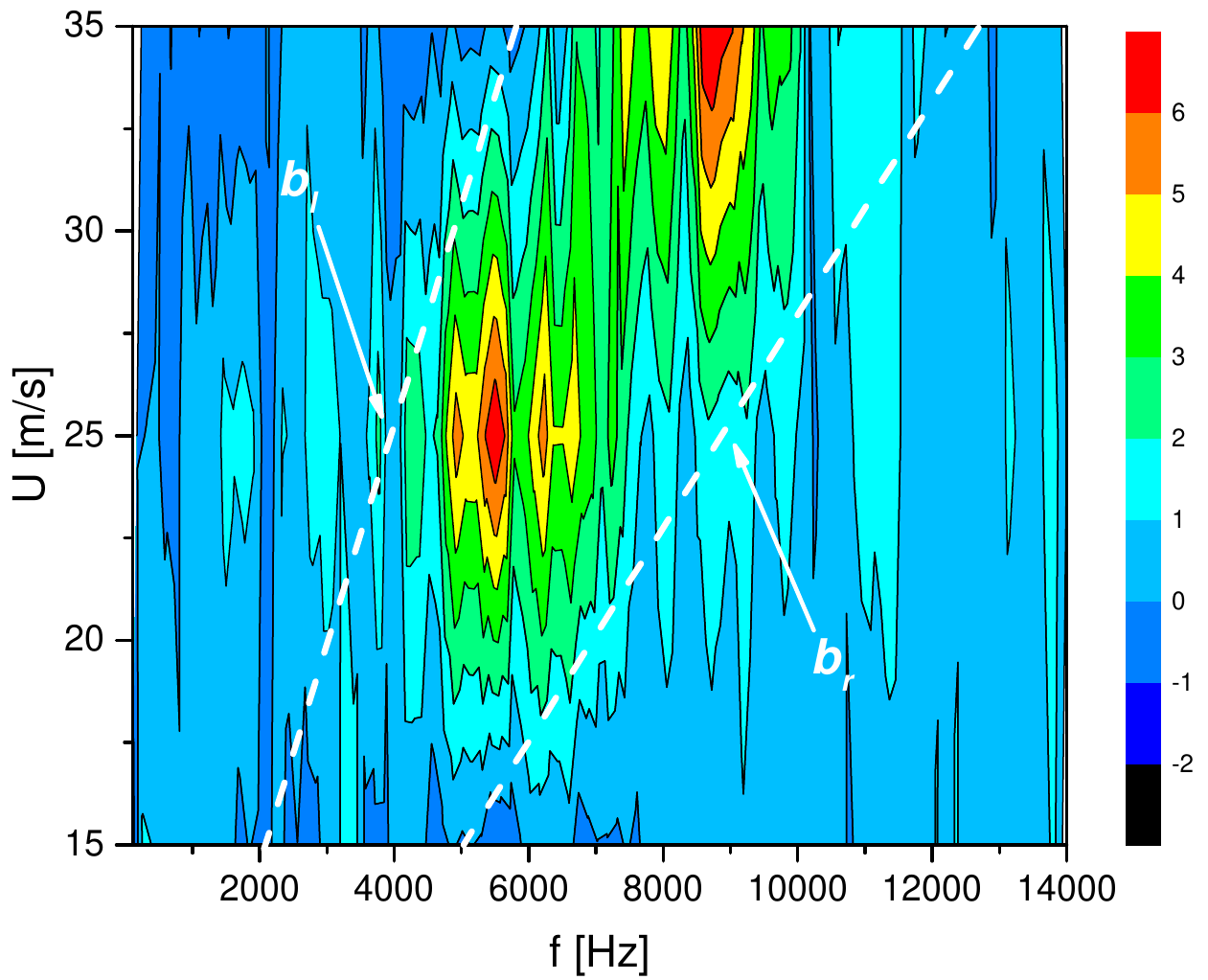} 
	} 
	\caption{Noise reduction with respect to frequency ($f$) and wind speed ($U$) for NACA 63(4)-421 airfoil} 
	\label{noise_reduction_f_U}
\end{figure}

\begin{figure}[H]
	\centering 
	\subfigure[Comparison with the original airfoil]{ 
		\label{633418_with_original}
		\includegraphics[width=.45\textwidth]{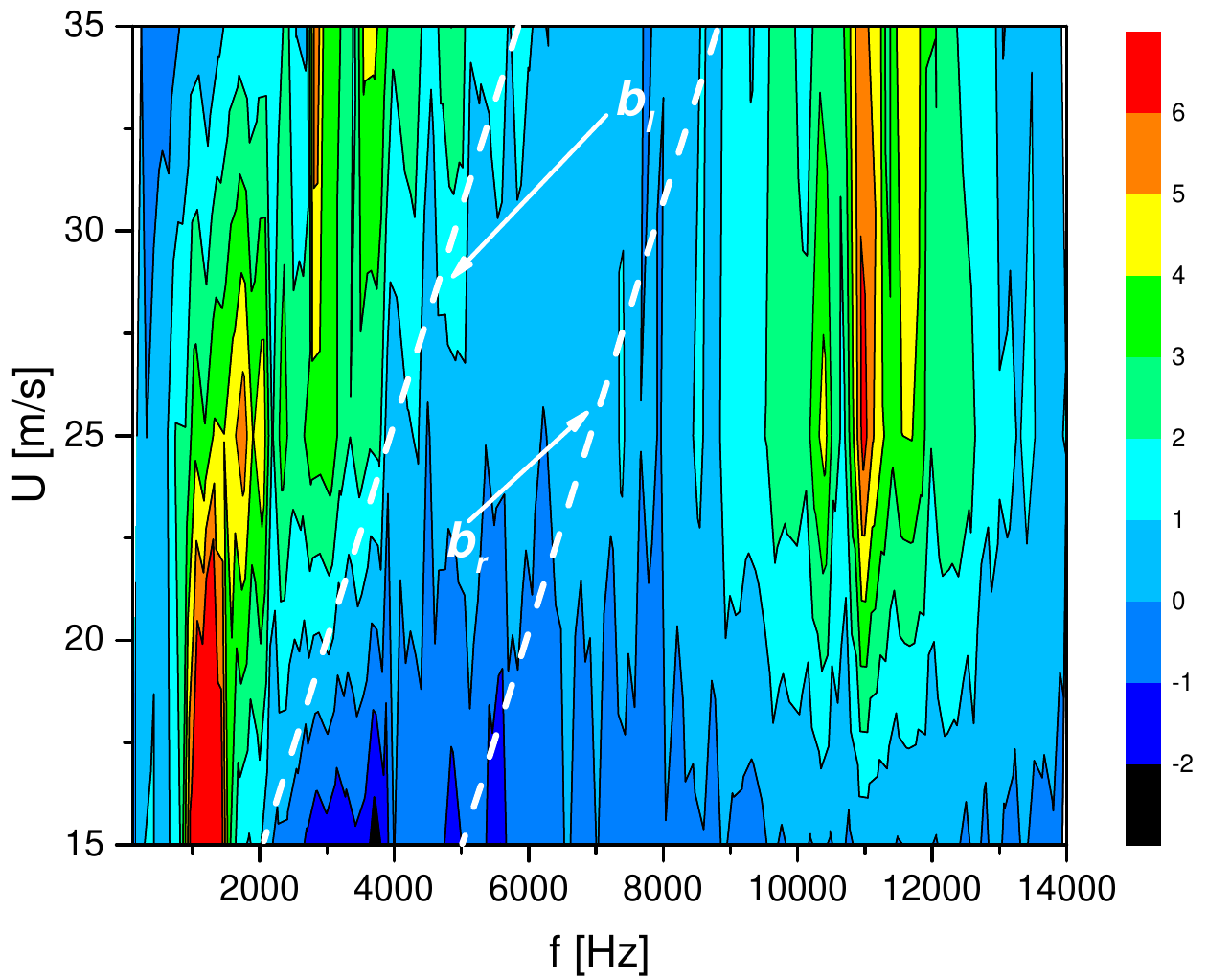} 
	} 
	\subfigure[Comparison with the airfoil with a bar trailing edge]{ 
		\label{633418_with_bar}
		\includegraphics[width=.45\textwidth]{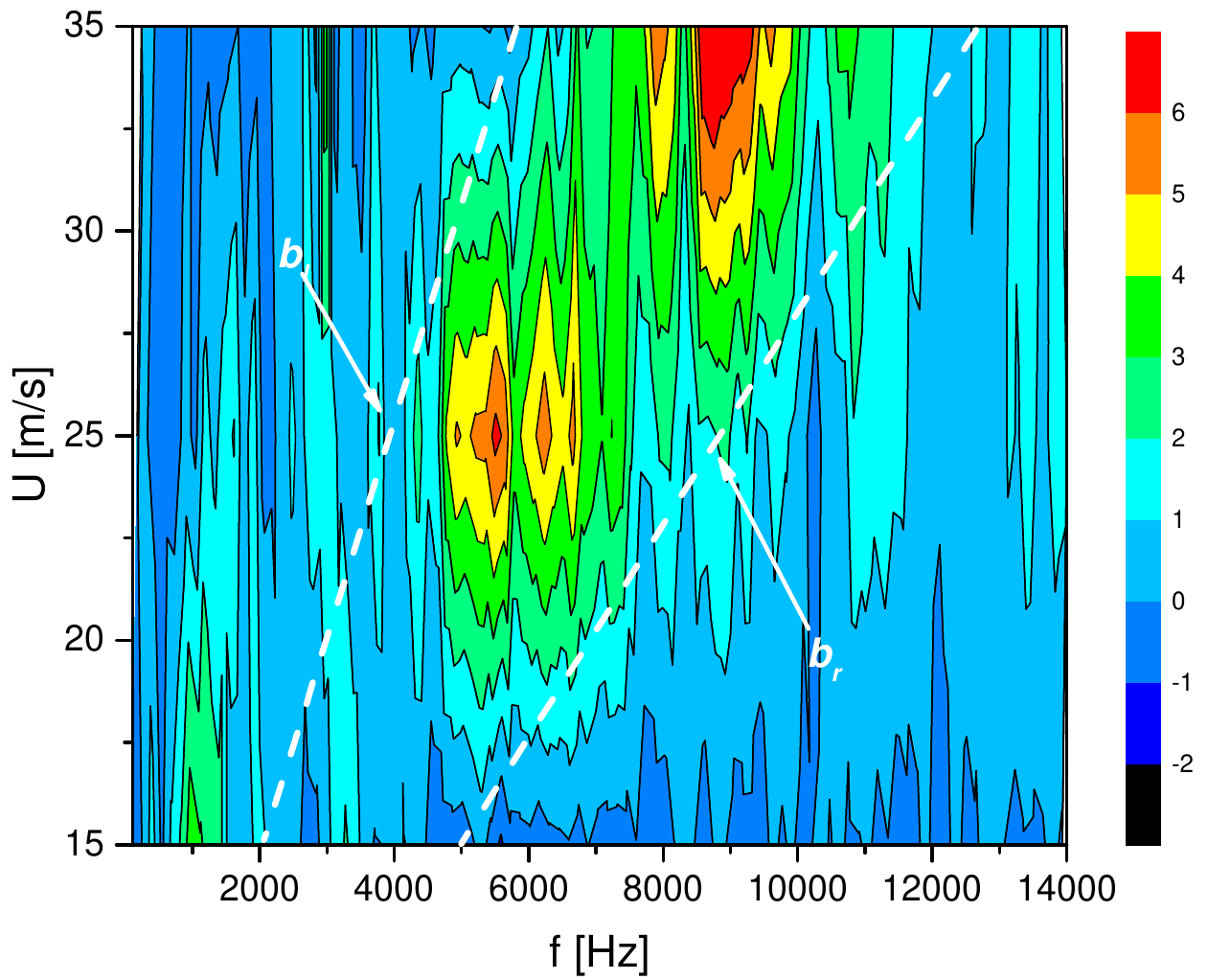} 
	} 
	\caption{Noise reduction with respect to frequency ($f$) and wind speed ($U$) for NACA 63(3)-418 airfoil} 
	\label{noise_reduction_f_U_633418}
\end{figure}

Fig.~\ref{noise_reduction_f_U} and Fig.~\ref{noise_reduction_f_U_633418} show the noise reduction as a function of inflow speed and frequency for the NACA 63(4)-421 and NACA 63(3)-418 airfoils, respectively. In both figures, the noise reduction regions are delineated by two boundaries, with contour colors indicating the magnitude of the noise reduction.

The plots in Fig.~\ref{noise_reduction_f_U} (a) and (b) or Fig.~\ref{noise_reduction_f_U_633418} (a) and (b) exhibit similarly patterns of noise reduction, where the airfoils with serrated trailing edges consistently outperform those with or without straight bars across various frequency ranges. The noise reduction zones, approximately bounded by two lines, $b_l$ and $b_r$, suggest frequency ranges where the serrations are particularly effective as wind speed changes. Moreover, the wider noise reduction zone observed at higher Reynolds numbers indicates that the serrations' effectiveness tends to increase with rising wind speeds. This comparison between the original airfoil and the one with a straight bar provides new insights not widely addressed in previous studies.

Sound power level is sometimes conducted within specific frequency bands, either as doubling frequency bands with a 2:1 ratio between cutoff frequencies or as 1/3-octave bands, which divide these bands into three equal parts. This study also adopts the 1/3-octave band approach since it provides more detailed analysis of sound characteristics. For a band centered at frequency $f$, the upper and lower cutoff frequencies are \(\sqrt[6]{2} f\) and \( f/\sqrt[6]{2} \), respectively. This method enhances noise data precision. A high-pass filter with a cutoff frequency of 100 Hz is used, and the analysis spans 17 frequency bands, starting with a center frequency of 315 Hz and ending with a center frequency of 12500 Hz.

The 1/3-octave band sound pressure level, \( L_p \), is computed by summing the sound pressure levels within each frequency band. The energy summation in decibels (dB) for each 1/3-octave band is performed using Eq.~\ref{eq:third_octave_lp}:

\begin{equation}
L_p = 10 \log_{10} \left( \sum_{f_l \leq f_i \leq f_u} 10^{L_p(f_i) / 10} \right)
\label{eq:third_octave_lp}
\end{equation}

In Eq.~\ref{eq:third_octave_lp}, \( L_p(f_i) \) represents the sound pressure level at frequency \( f_i \), and the summation is performed over all frequencies within the lower and upper cutoff frequencies, \( f_l \) and \( f_u \), of the 1/3-octave band. The exponential term converts the sound pressure level from decibels back to a linear scale, and the logarithmic operation returns the summed energy back into decibels. This summation provides an accurate representation of the total energy within the 1/3-octave band.

\begin{figure}[H]
\centering
\subfigure[$Re = 0.7 \times 10^5$]{ 
	\includegraphics[width=.3\textwidth]{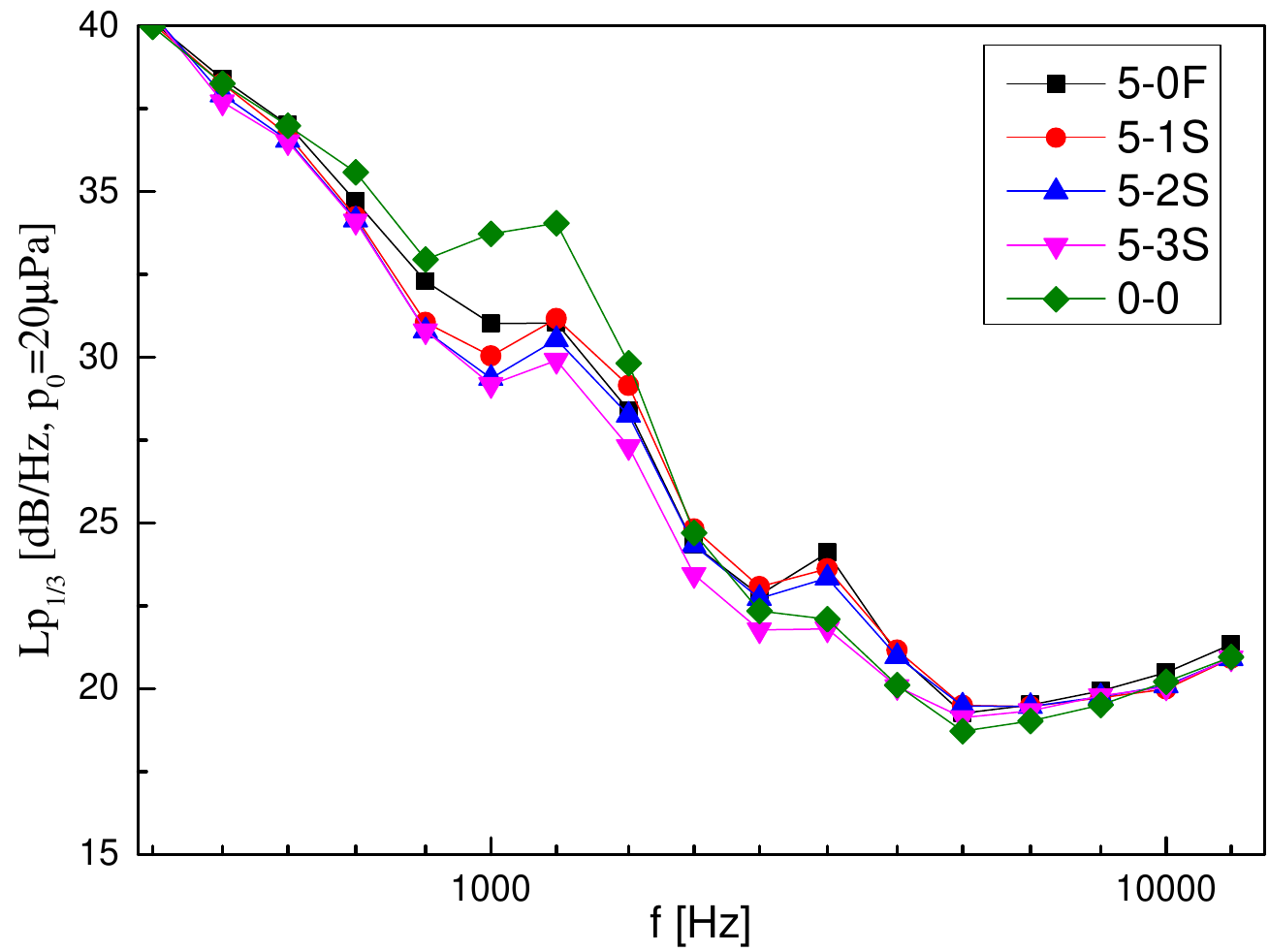} 
} 
\subfigure[$Re = 1.2 \times 10^5$]{ 
	\includegraphics[width=.3\textwidth]{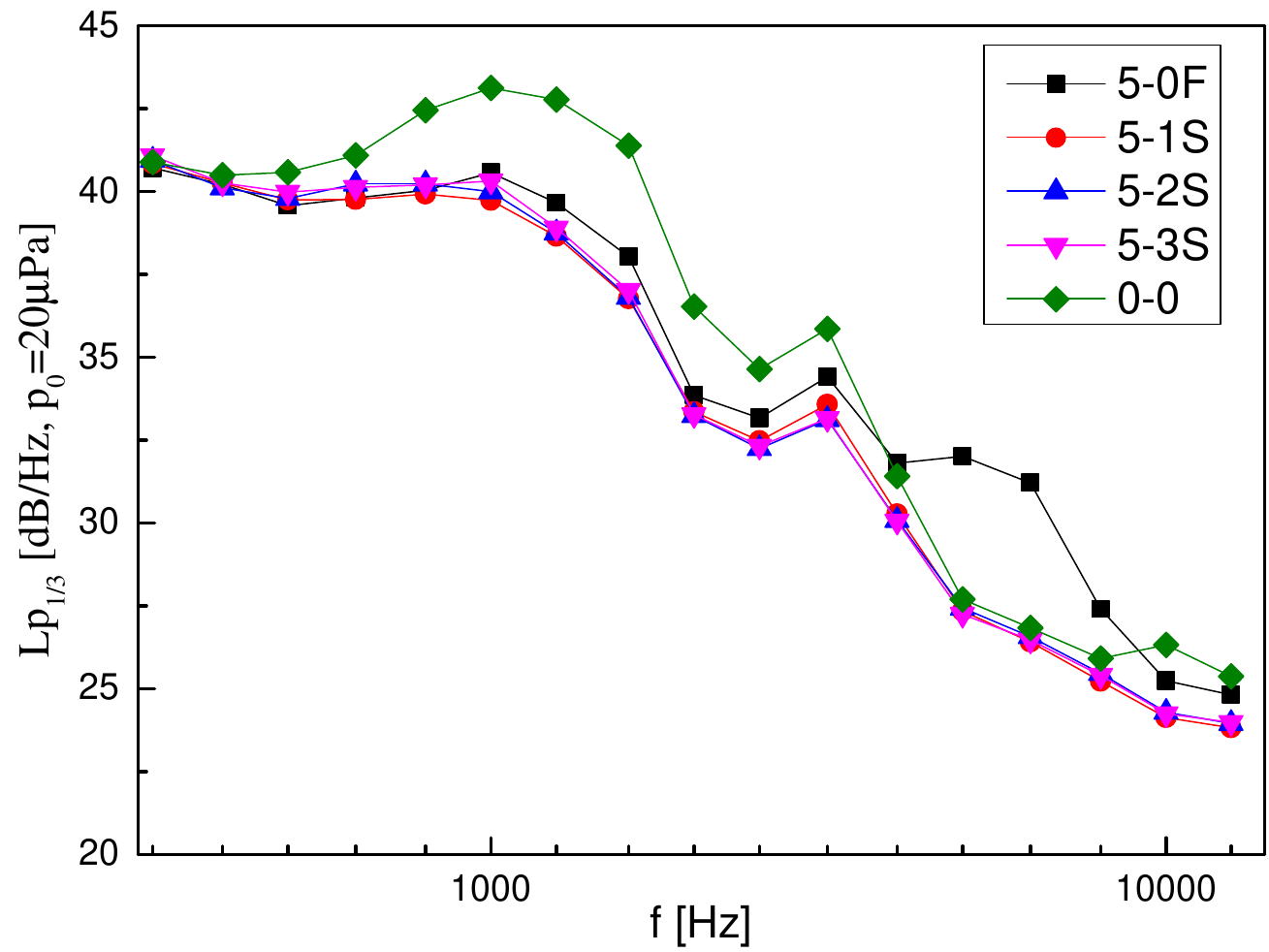} 
}
\subfigure[$Re = 1.6 \times 10^5$]{ 
	\includegraphics[width=.3\textwidth]{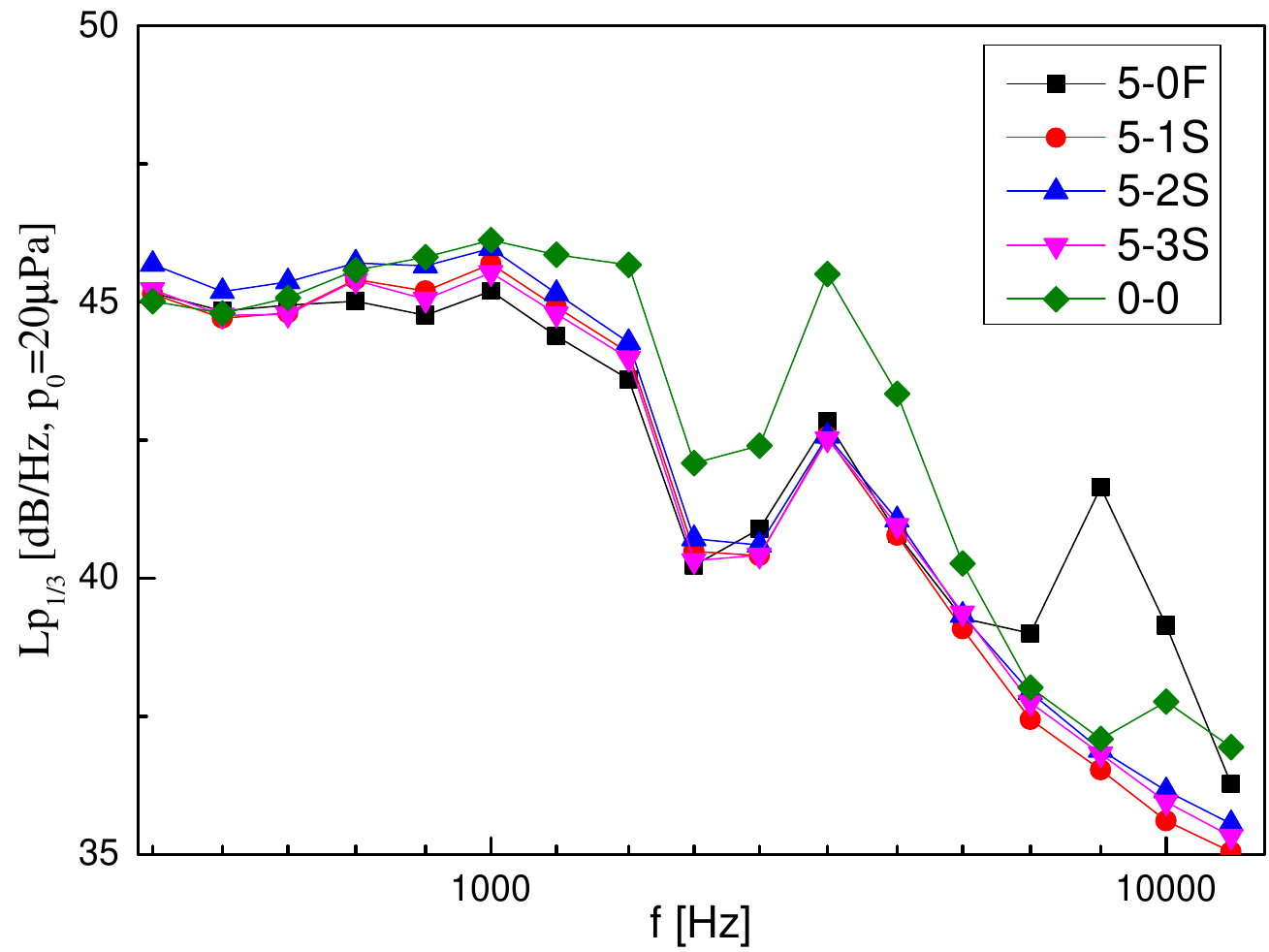} 
}  
\caption{1/3 Octave sound power level for NACA 63(4)-421 airfoil with trailing edge serrations} 
\label{Re_634421_13octave}
\end{figure}

Fig.~\ref{Re_634421_13octave} shows the 1/3-octave sound pressure level for the NACA 63(4)-421 airfoil. It clearly demonstrates that the serrated trailing edge reduces noise at two distinct frequency regions compared to the original airfoil: the broad-band noise range and tonal noise. The broad-band noise is related to the laminar separation bubble on the pressure side of the airfoil, while the tonal noise is associated with the effective thickness of the blunt trailing edge of the original airfoil. Under the three Reynolds number conditions tested in this study, the serrated trailing edge effectively mitigates both types of noise.

When compared to an airfoil with straight trailing edges, serrations exhibit greater noise reduction mainly in the mid-to-high frequency range. The influence of serration wavelength ($\lambda$) or wavelength-to-half-height ratio ($\lambda/h$) on sound power level is minimal, contrary to the observation that larger wavelength serrations provided better noise reduction performance compared to smaller wavelength serrations in Moreau et al.~\cite{moreau2011flat,moreau2013noise} and Celik et al.~\cite{celik2021aeroacoustic}. This may be due to the smaller serration height used in the wind tunnel measurement.

\subsection{Impact of Angle of Attack}

The angle of attack significantly impacts the performance of airfoil trailing edge modifications. Changes in the angle of attack alter the flow characteristics over the airfoil, affecting noise generation mechanisms. Investigating how trailing edge modifications interact with different angles of attack offers valuable insights into their effectiveness under various operating conditions.

\begin{figure}[H]
	\centering 
	\subfigure[Airfoil with serrated trailing edge vs. original airfoil]{ 
		\label{634421AOA_0-0}
		\includegraphics[width=.45\textwidth]{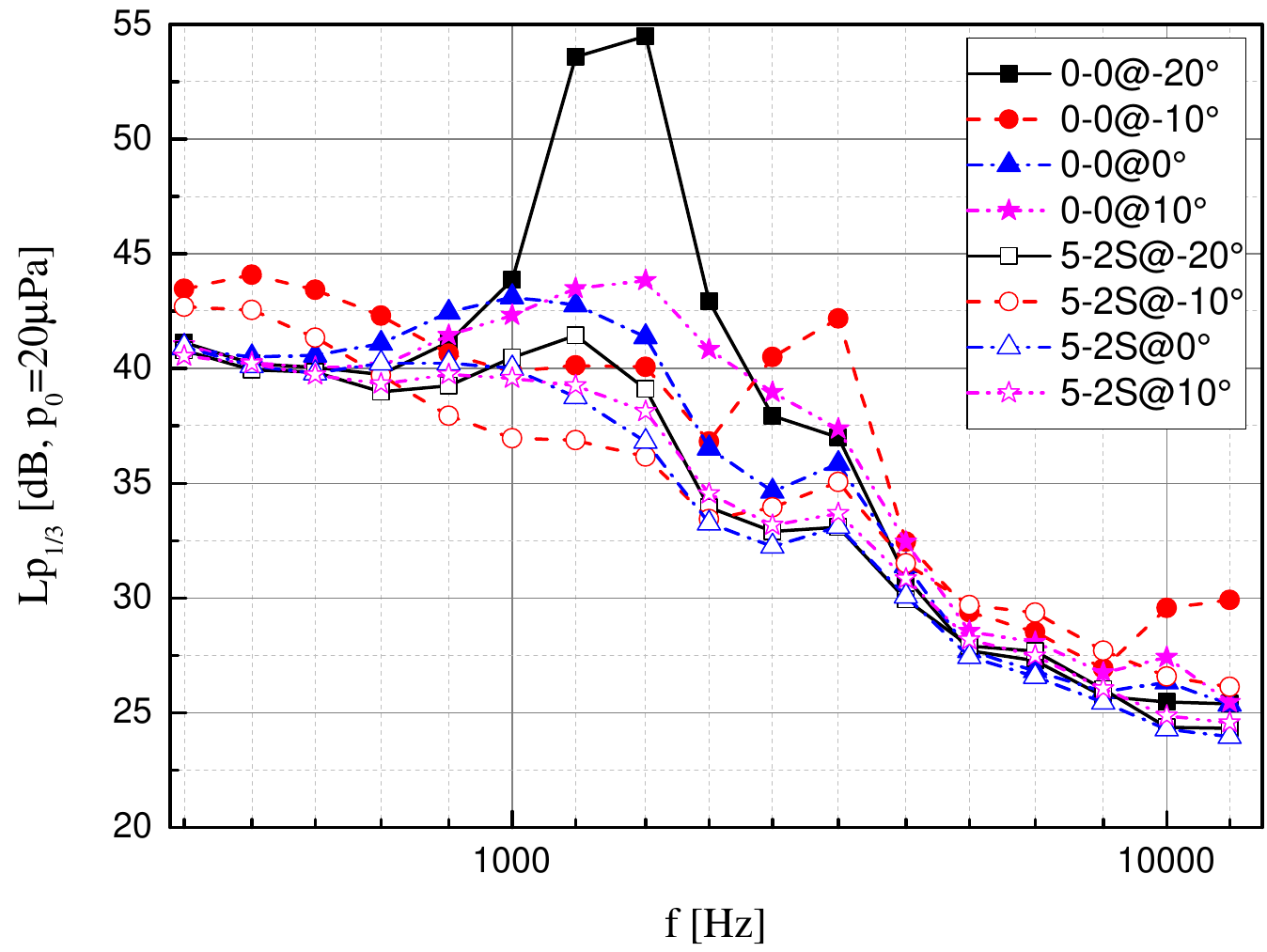} 
	} 
	\subfigure[Airfoil with serrated trailing edge vs. bar trailing edge]{ 
		\label{634421AOA}
		\includegraphics[width=.45\textwidth]{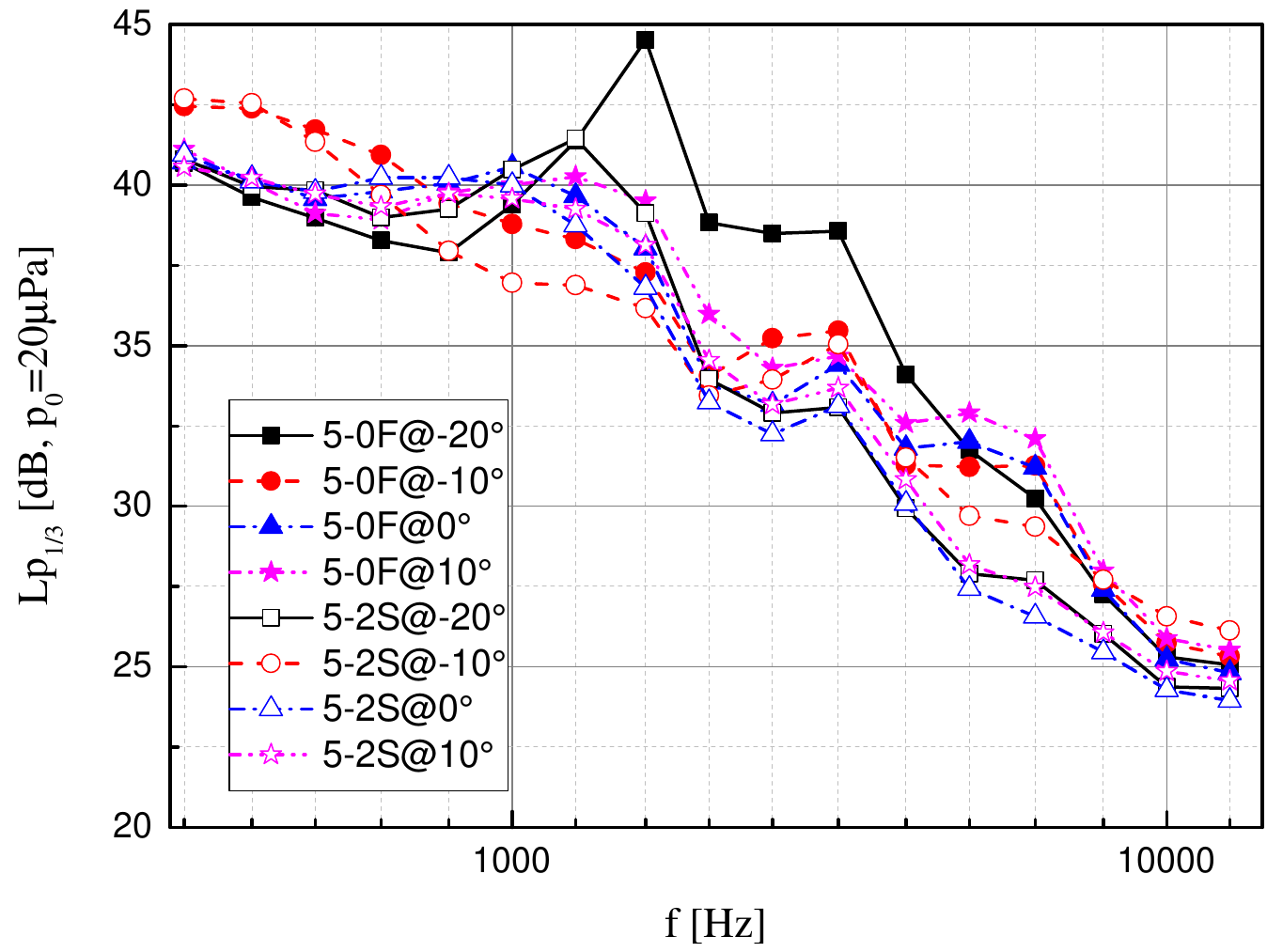} 
	} 
	\caption{Sound pressure level ($L_P$) as a function of angle of attack for the NACA 63(4)-421 airfoil with and without serrated trailing edges} 
	\label{AOA_634421}
\end{figure}

Fig.~\ref{AOA_634421} compares the noise levels between the NACA 63(4)-421 airfoil with and without serrated trailing edges under a Reynolds number of \( Re = 1.2 \times 10^5 \), with the 5-2S configuration selected as the experimental group. The airfoil without a trailing edge exhibits noticeable tonal noise, approximately 10 dB, within the frequency range of $1250 \, \mathrm{Hz} \leq f \leq 1600 \, \mathrm{Hz}$ at a high negative angle of attack ($-20^\circ$), which is primarily due to flow separation on the pressure side. The serrated trailing edge effectively mitigates this tonal noise. Similarly, the airfoil with a bar trailing edge also generates prominent tonal noise around $f = 1600 \, \mathrm{Hz}$ at a high negative angle of attack ($-20^\circ$), which the serrated trailing edge successfully eliminates. Across the four different angles of attack tested, the serrated trailing edge consistently demonstrates significant noise reduction within a certain frequency range, both in comparison to the airfoil without a trailing edge and the one with a bar trailing edge.

\begin{figure}[H]
	\centering 
	\subfigure[Comparison with the original airfoil]{ 
		\label{633418_with_original_alpha}
		\includegraphics[width=.45\textwidth]{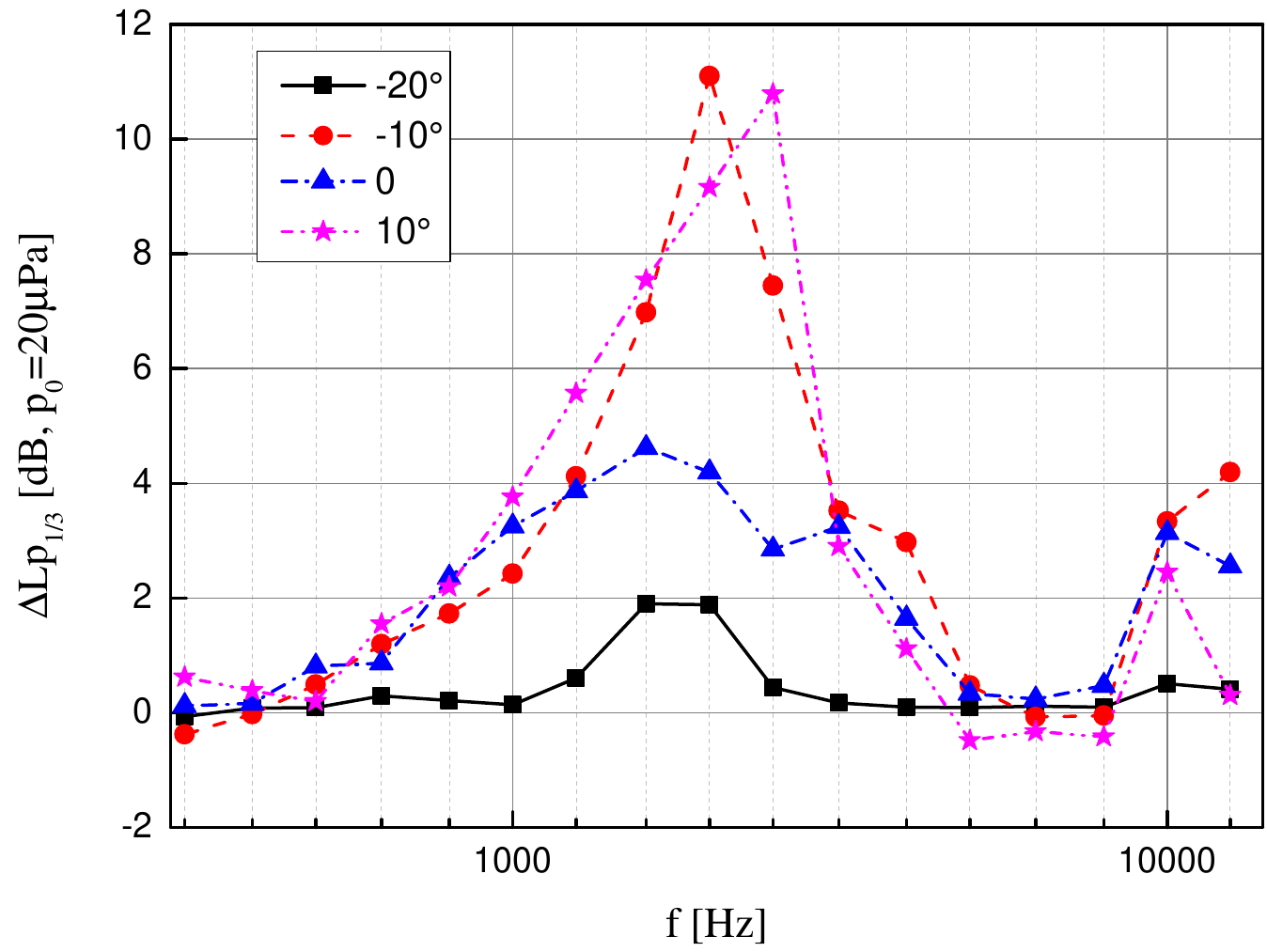} 
	} 
	\subfigure[Comparison with the airfoil with a bar trailing edge]{ 
		\label{633418_with_bar_alpha}
		\includegraphics[width=.45\textwidth]{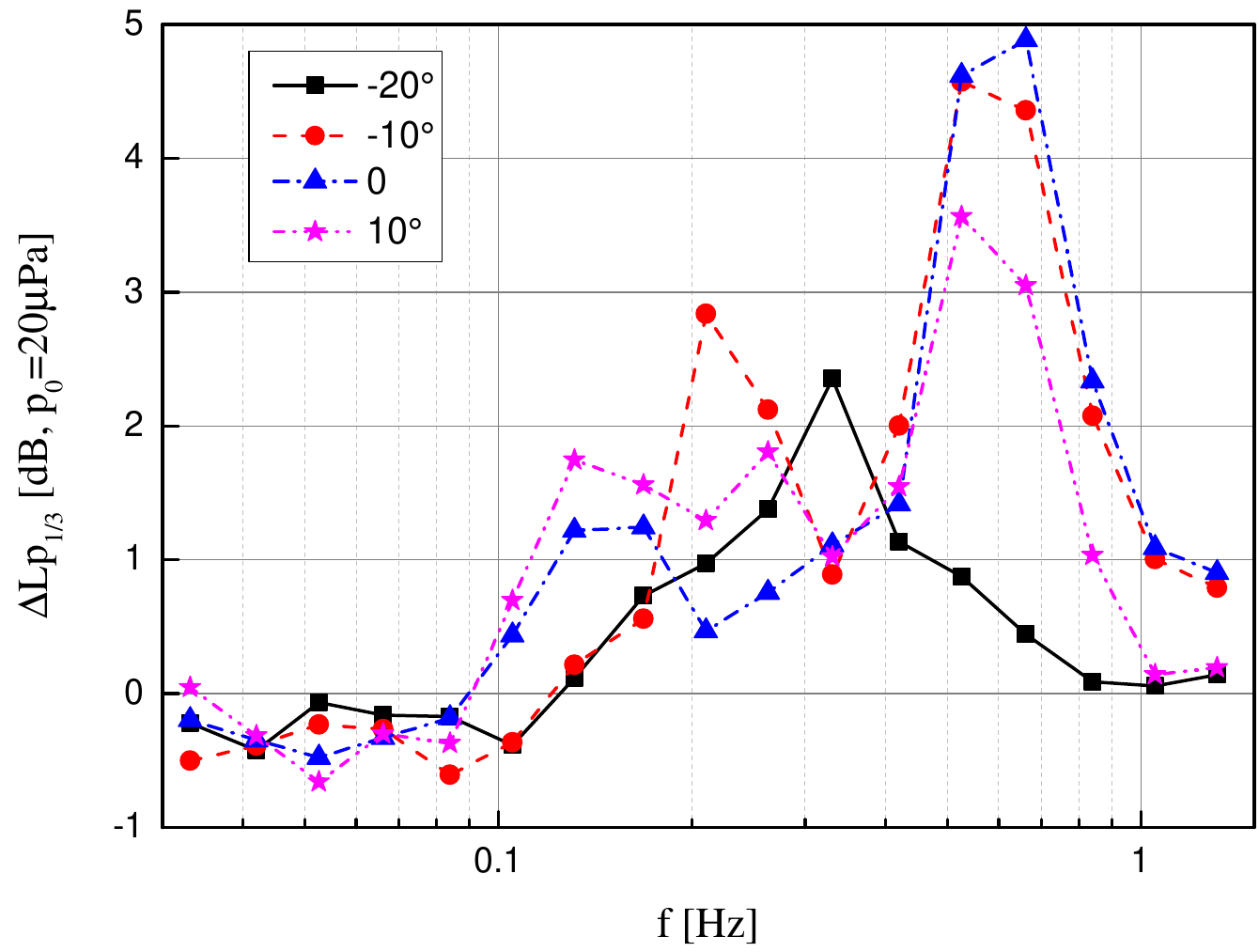} 
	} 
	\caption{Noise reduction ($\Delta L_P$) relative to the baselines as a function of angle of attack for the NACA 63(3)-418 airfoil with serrated trailing edges} 
	\label{noise_alpha_633418}
\end{figure}

\begin{figure}[H]
	\centering 
	\subfigure[Comparison with the original airfoil]{ 
		\label{634421_with_original_alpha}
		\includegraphics[width=.45\textwidth]{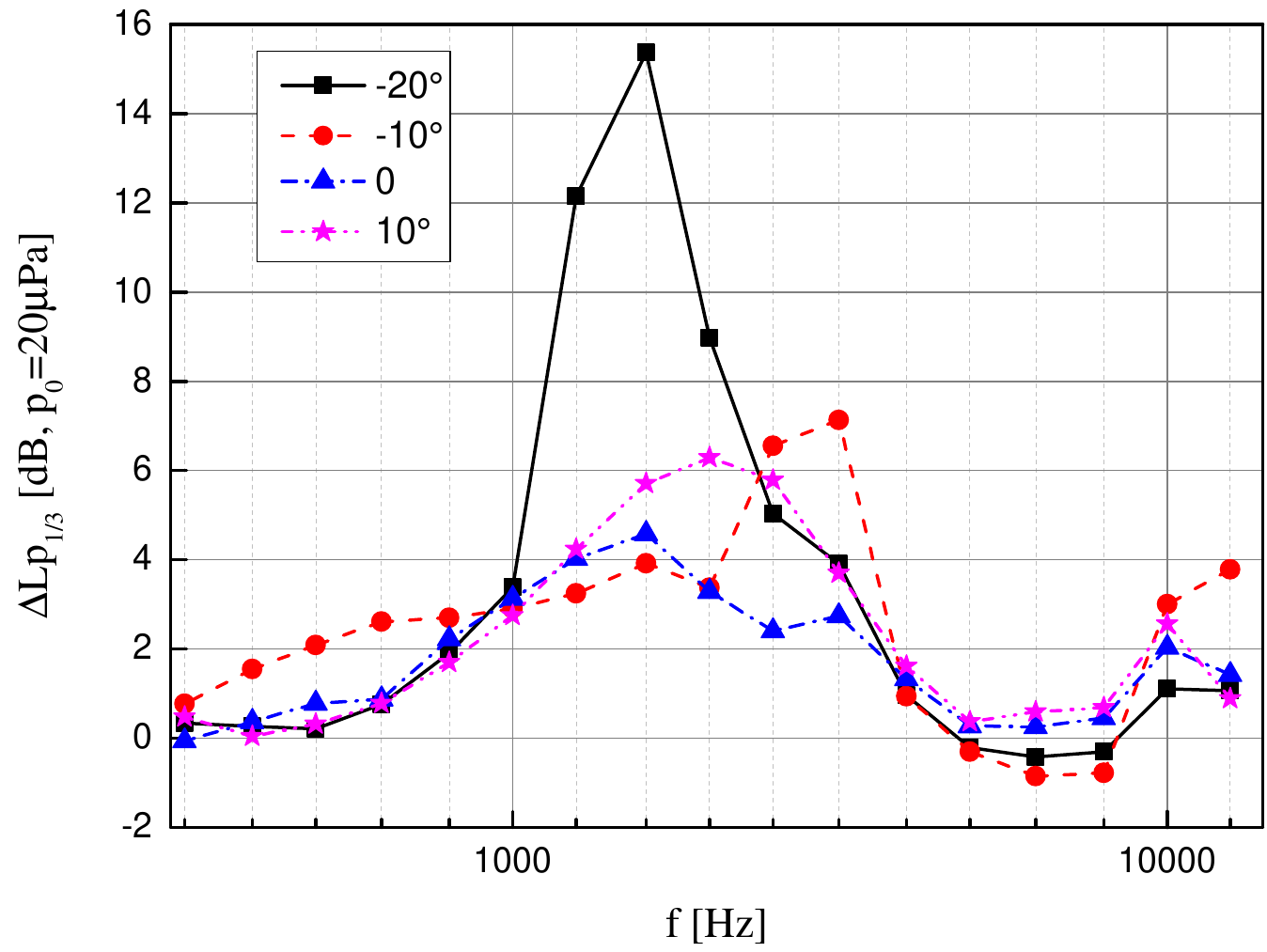} 
	} 
	\subfigure[Comparison with the airfoil with a bar trailing edge]{ 
		\label{634421_with_bar_alpha}
		\includegraphics[width=.45\textwidth]{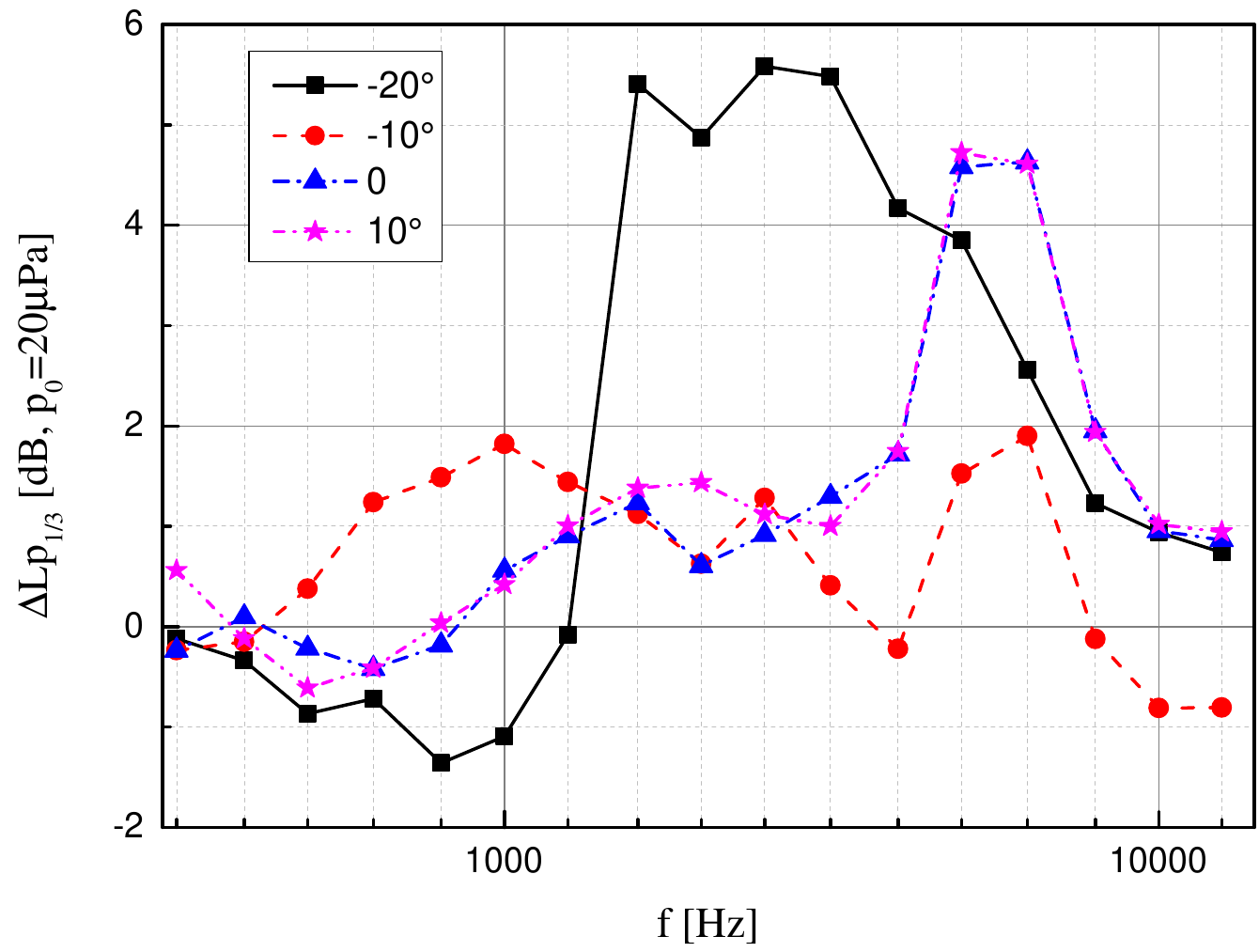} 
	} 
	\caption{Noise reduction ($\Delta L_P$) relative to the baseline airfoil as a function of angle of attack for the NACA 63(4)-421 airfoil with serrated trailing edges} 
	\label{noise_alpha_634421}
\end{figure}

Fig.~\ref{noise_alpha_633418} and Fig.~\ref{noise_alpha_634421} show the effect of the angle of attack on noise reduction for the NACA 63(3)-418 and NACA 63(4)-421 airfoils, respectively. Noise reduction is calculated as the difference in sound power level between the 5-2S serrated trailing edge and the reference cases.

The effectiveness of serrated trailing edges in reducing noise varies with the angle of attack for different airfoil profiles, leading to distinct noise reduction patterns in terms of frequency range and amplitude. For the NACA 63(3)-418 airfoil, the serrated trailing edge performs best at angles of $-10^\circ$ and $10^\circ$, achieving noise reductions over 4 dB across a broad frequency range from $f = 1000 \, \mathrm{Hz}$ to $f = 3150 \, \mathrm{Hz}$. This is likely due to the thinner profile of the NACA 63(3)-418, which optimizes its aerodynamic performance at moderate angles of attack, effectively reducing vortex shedding and turbulence over a wide frequency range. In contrast, the NACA 63(4)-421 airfoil achieves the most significant noise reduction at an angle of $-20^\circ$, with reductions exceeding 15 dB. The thicker profile of the NACA 63(4)-421 is optimized for higher angles of attack, significantly reducing noise by minimizing flow separation and reattachment regions.

Across different angles of attack, the bar trailing edge consistently shows inferior noise reduction performance compared to the serrated trailing edge for both airfoils tested. The maximum noise reduction difference between the serrated and bar trailing edges can reach up to 5 dB, with the serrated trailing edge being more effective across most frequency ranges. These conclusions provide valuable insights for the application of serrations in wind turbines, drones, and low altitude transport aircraft.

\subsection{Impact of Serration Size}

\begin{figure}[H]
	\centering 
	\subfigure[$Re = 0.7 \times 10^5$]{ 
		\label{633418_1000_13o_h}
		\includegraphics[width=.3\textwidth]{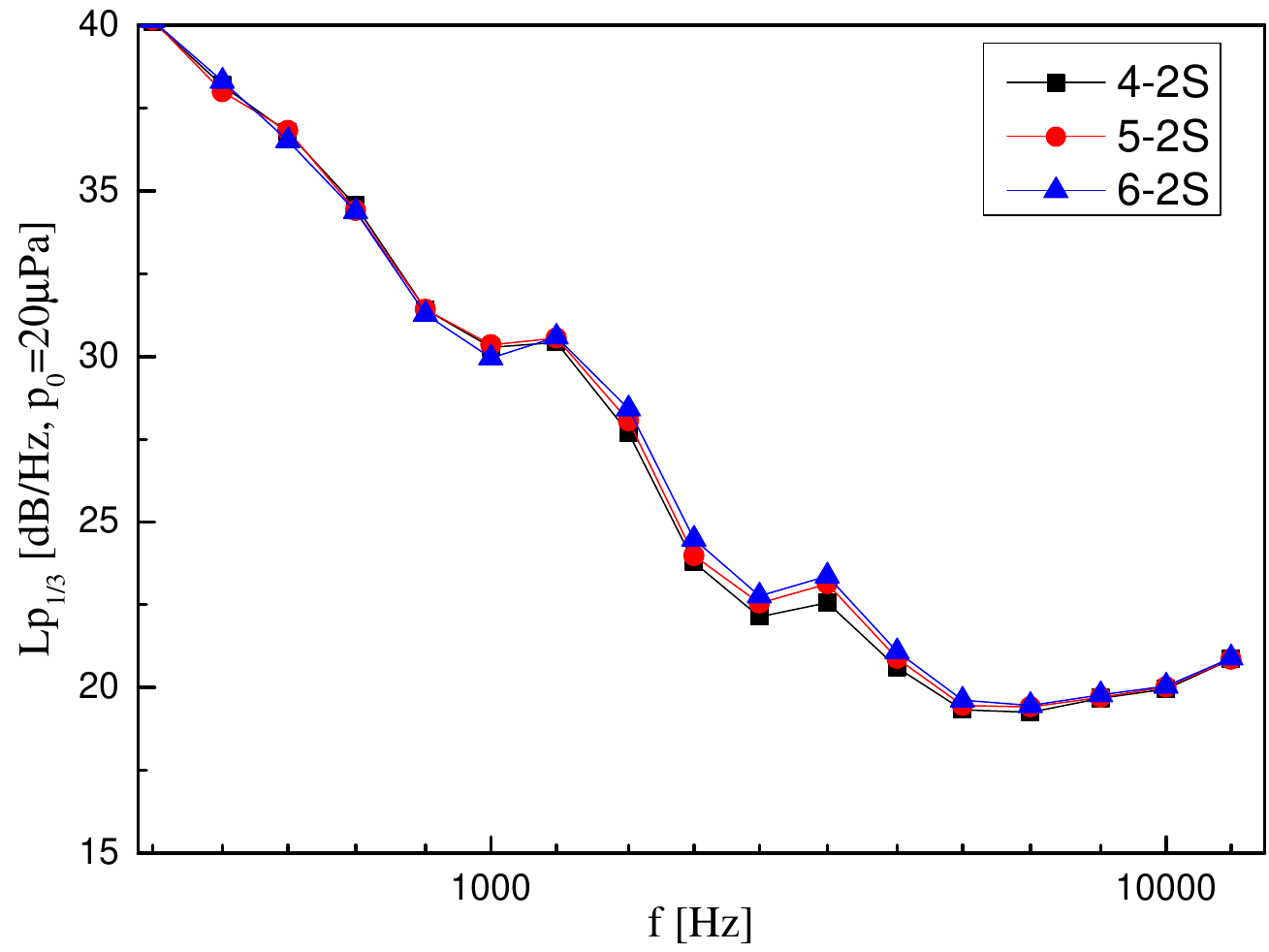} 
	} 
	\subfigure[$Re = 1.2 \times 10^5$]{ 
		\label{633418_1500_13o_h}
		\includegraphics[width=.3\textwidth]{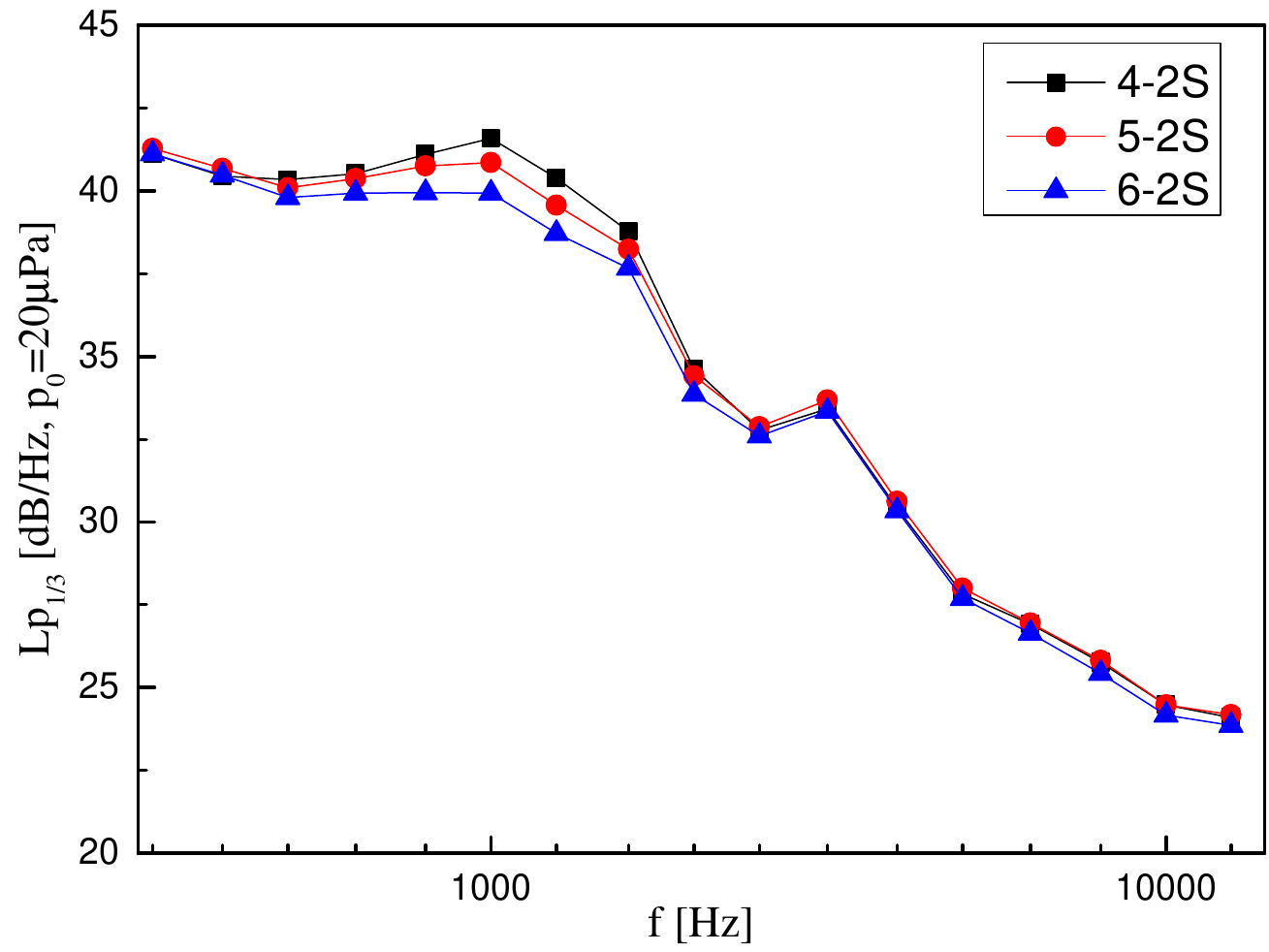} 
	}
	\subfigure[$Re = 1.6 \times 10^5$]{ 
		\label{633418_2000_13o_h}
		\includegraphics[width=.3\textwidth]{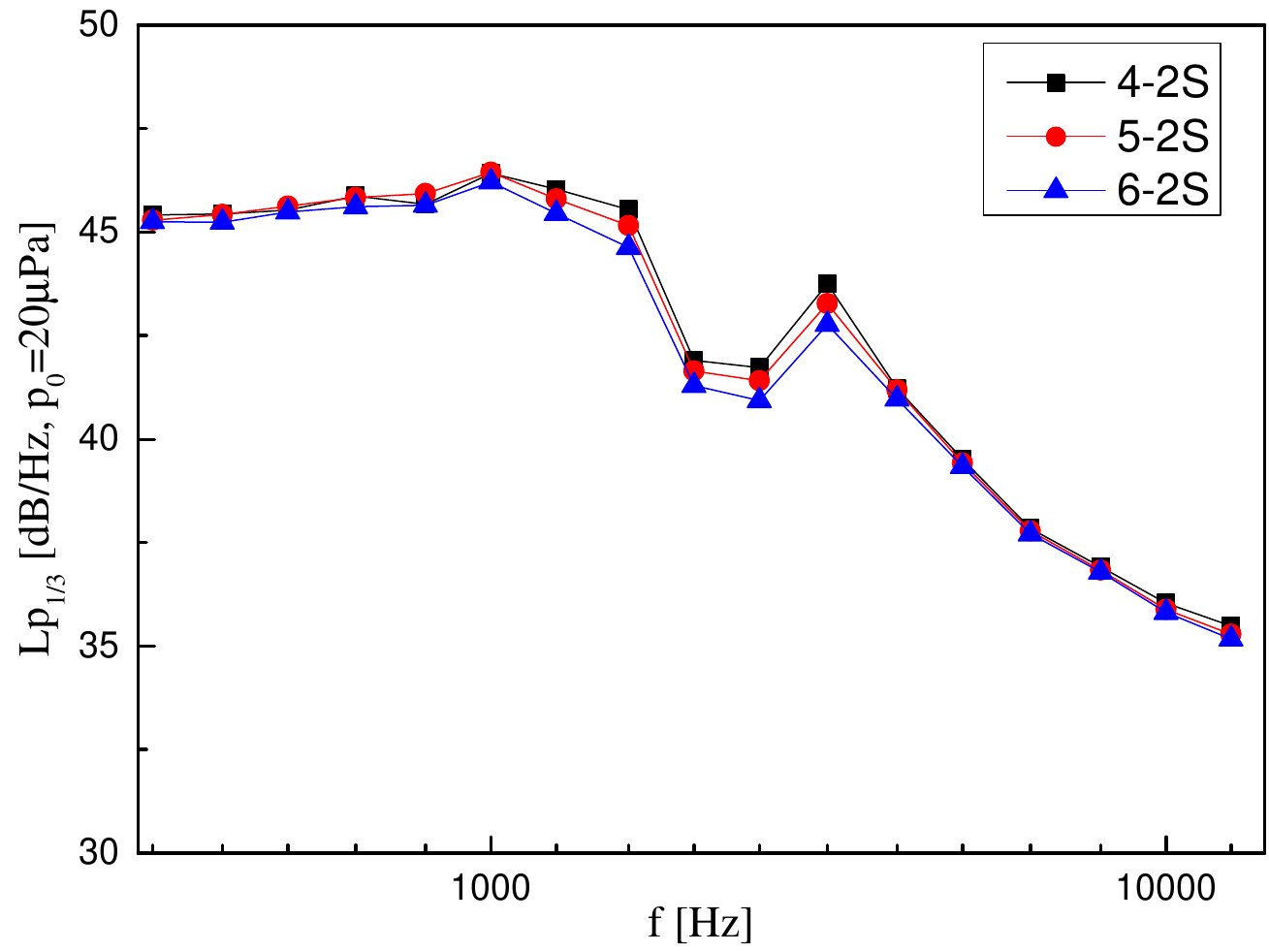} 
	} 
	\caption{Sound pressure level ($L_P$) as a function of serration size for the NACA 63(3)-418 airfoil with serrated trailing edges} 
	\label{633418_13o_h}
\end{figure}

\begin{figure}[H]
	\centering 
	\subfigure[$Re = 0.7 \times 10^5$]{ 
		\label{634421_1000_13o_h}
		\includegraphics[width=.3\textwidth]{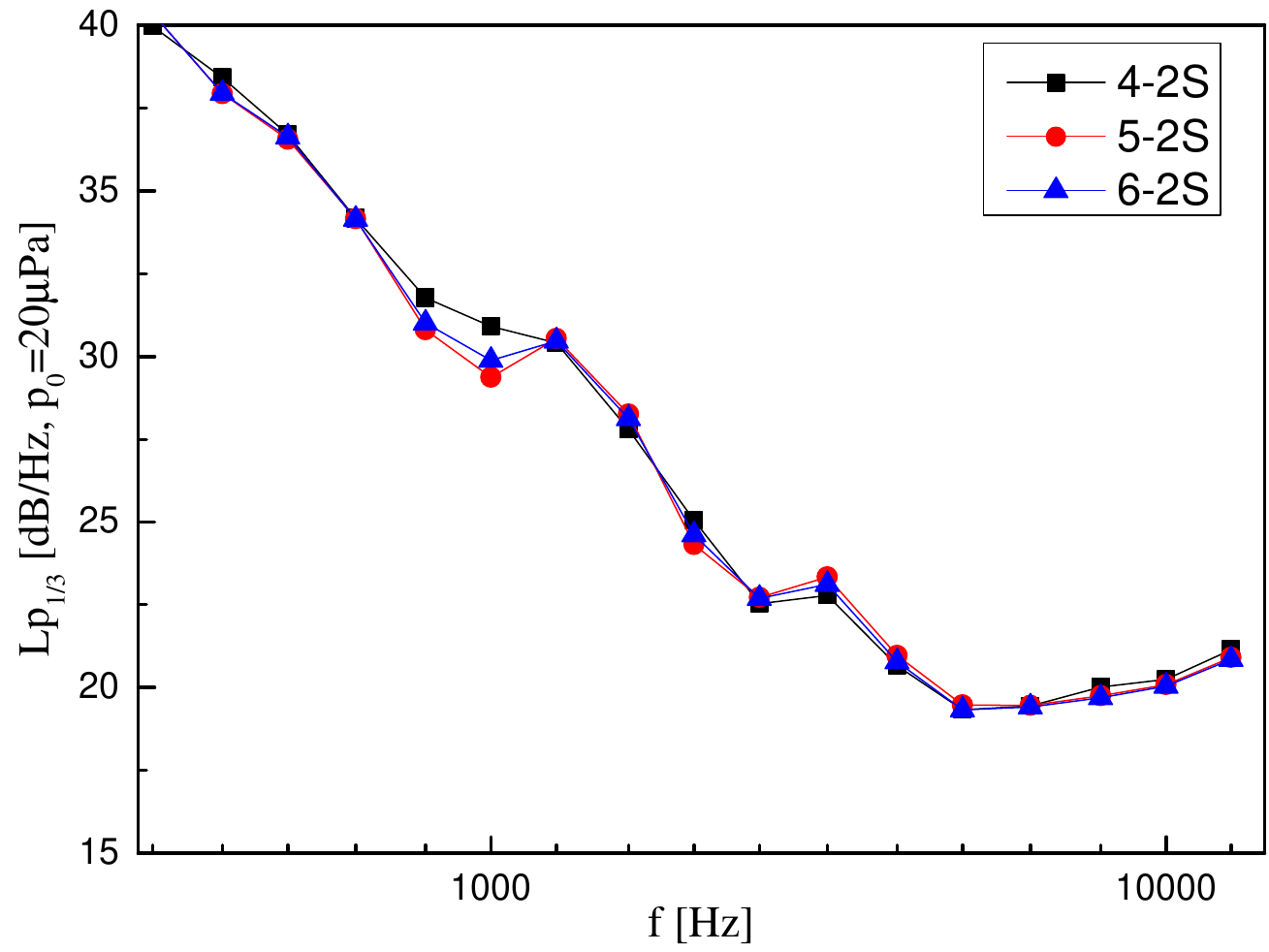} 
	} 
	\subfigure[$Re = 1.2 \times 10^5$]{ 
		\label{634421_1500_13o_h}
		\includegraphics[width=.3\textwidth]{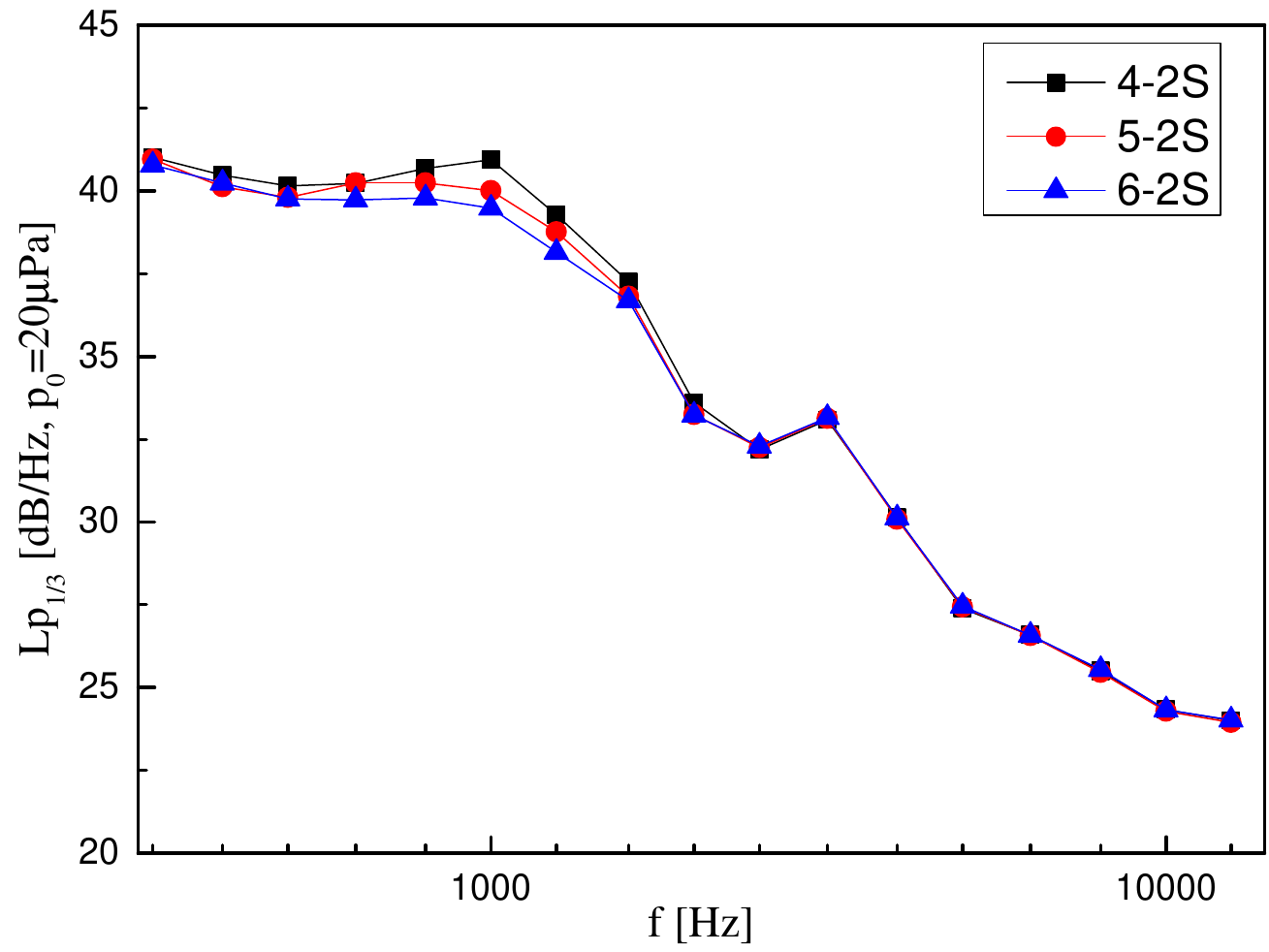} 
	}
	\subfigure[$Re = 1.6 \times 10^5$]{ 
		\label{634421_2000_13o_h}
		\includegraphics[width=.3\textwidth]{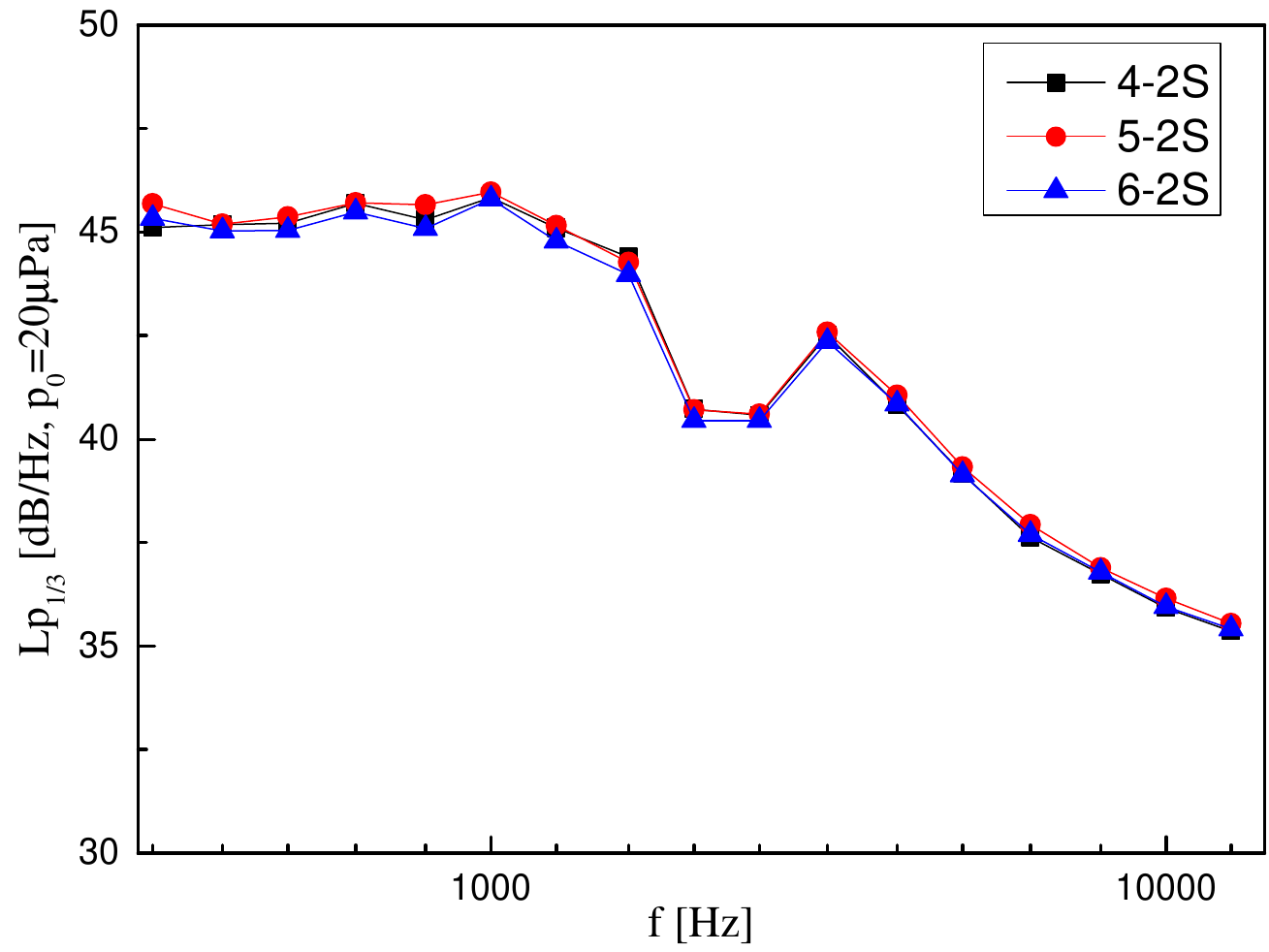} 
	} 
	\caption{Sound pressure level ($L_P$) as a function of serration size for the NACA 63(4)-421 airfoil with serrated trailing edges} 
	\label{634421_13o_h}
\end{figure}

Figs.~\ref{633418_13o_h} and \ref{634421_13o_h} demonstrate the impact of serration size on noise reduction for the NACA 63(3)-418 and NACA 63(4)-421 airfoils at various Reynolds numbers, with a fixed wavelength-to-half-height ratio of 0.4. While the effect of serration size on sound pressure level appears modest at lower Reynolds numbers, its impact becomes more pronounced at higher Reynolds numbers. The limited influence at lower Reynolds numbers may be attributed to smaller serration height compared to the airfoil boundary layer displacement thickness near the trailing edge, so the mismatch between serration height and airfoil boundary layer thickness result in minimal noise attenuation. As the Reynolds number increases, the flow becomes more turbulent and the boundary layer displacement thickness becomes smaller, so the interaction between the serrations and the turbulent eddies enhances noise reduction. 

In Fig.~\ref{airfoil_BL} which shows the boundary layer displacement thicknesses, at \( Re = 0.7 \times 10^5 \), the displacement thickness reaches 7.76 mm, which exceeds the maximum half-height of the serrations used in this study, suggesting that the serrations have a weak effect on disrupting coherent vortices at this Reynolds number. In contrast, at \( Re = 1.2 \times 10^5 \) and \( Re = 1.6 \times 10^5 \), the displacement thickness reduces to 1$\sim$3 mm, which is smaller than the minimum half-height of the serrations. Moreover, the displacement thickness for the NACA 63(3)-418 airfoil is consistently smaller than that for the NACA 63(4)-421 airfoil, indicating that at higher Reynolds numbers, the serrations are more effective in disrupting coherent vortices. Further experimental investigation is required to fully understand the mechanisms behind this effect.

While this study shows that longer and greater serrations generally provide better noise reduction, consistent with Howe's findings \cite{howe1991aerodynamic,howe1991noise}, it is important to note that his theory is based on idealized assumptions. Howe's model links noise reduction to the rate of vorticity intersecting streamlines, but it does not fully capture the complex vortex dynamics near the airfoil trailing edge. The interactions between flow and sound fields, influenced by serration size, can significantly impact both acoustic characteristics and aerodynamic performance. Limited in this paper, further research is needed in order to investigate the impact of serration aspect ratio on noise reduction.

\subsection{Sound Field Directivity}

\begin{figure}[H]
	\centering 
	\subfigure[400 Hz]{ 
		\label{634421_1000_directivity_400Hz}
		\includegraphics[width=.22\textwidth]{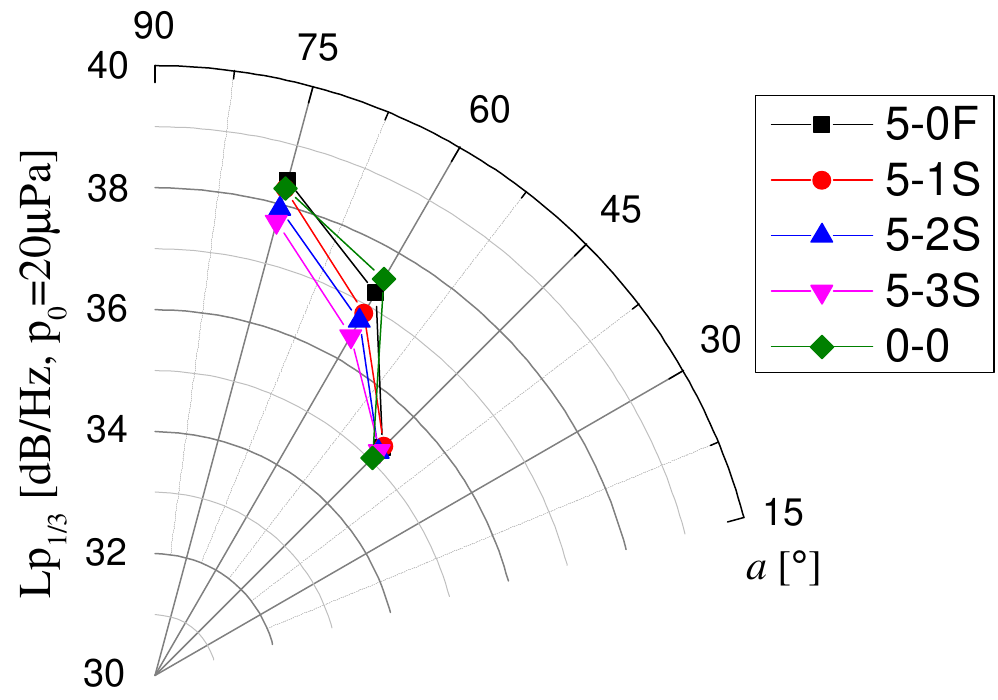} 
	} 
	\subfigure[1000 Hz]{ 
		\label{634421_1000_directivity_1000Hz}
		\includegraphics[width=.22\textwidth]{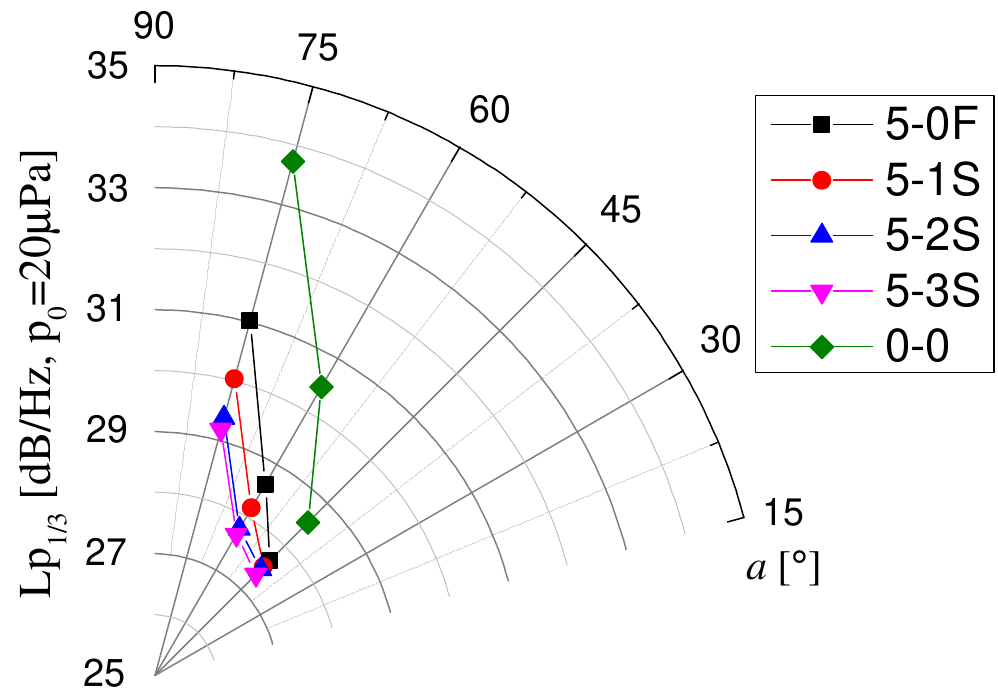} 
	}
	\subfigure[3150 Hz]{ 
		\label{634421_1000_directivity_3150Hz}
		\includegraphics[width=.22\textwidth]{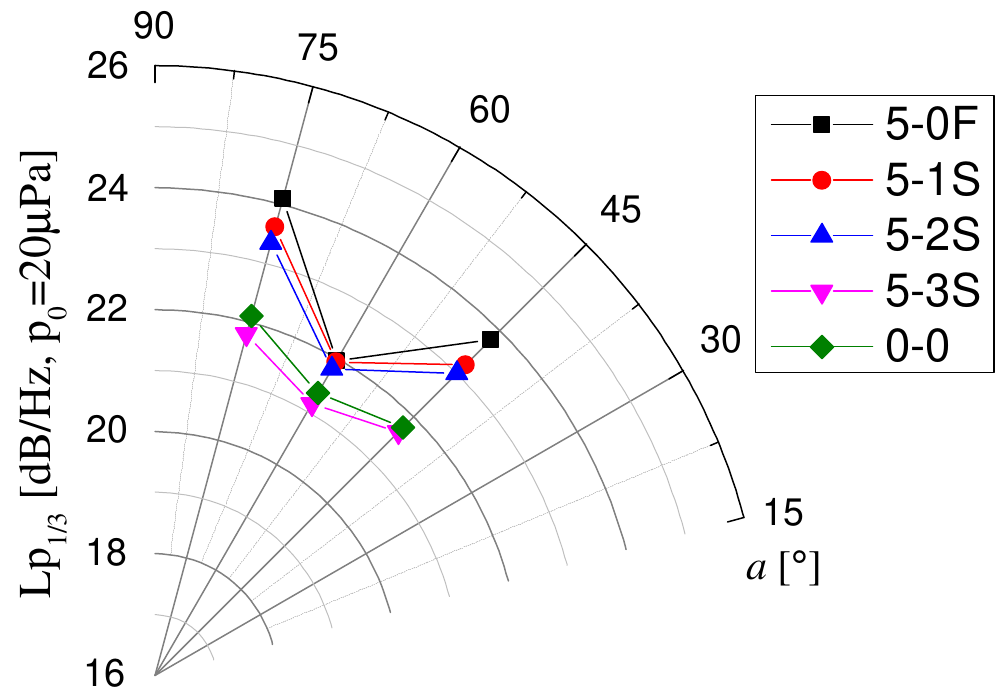} 
	}
	\subfigure[10000 Hz]{ 
    	\label{634421_1000_directivity_10000Hz}
    	\includegraphics[width=.22\textwidth]{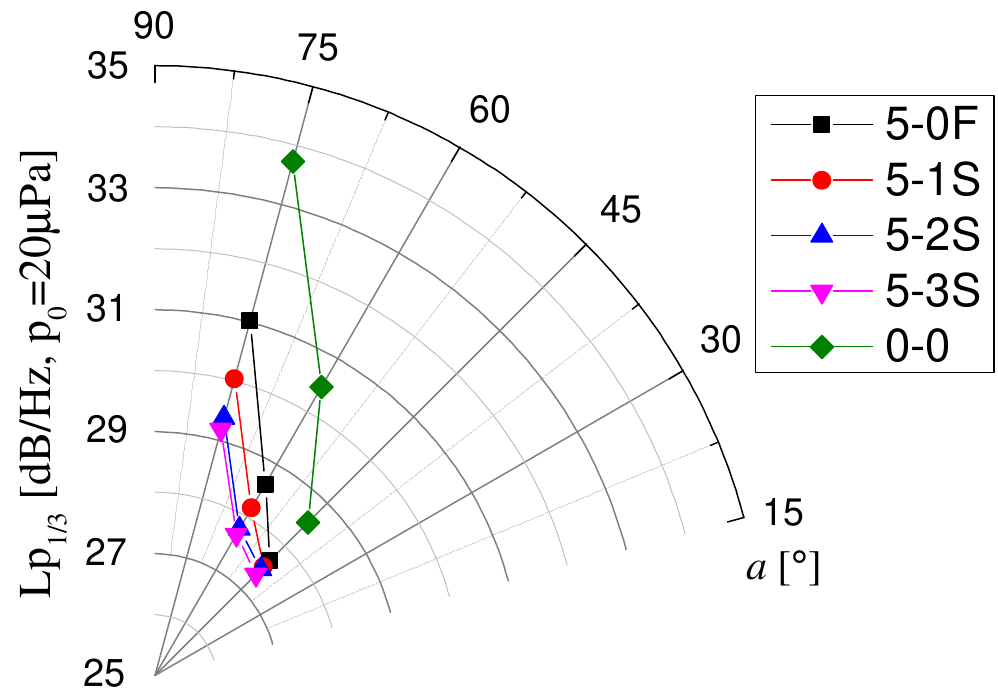} 
    }    
	\caption{Sound directivity of NACA 63(4)-421 airfoil at $Re=0.7\times10^5$} 
	\label{634421_1000_directivity}
\end{figure}

\begin{figure}[H]
	\centering 
	\subfigure[400 Hz]{ 
		\label{634421_1500_directivity_400Hz}
		\includegraphics[width=.22\textwidth]{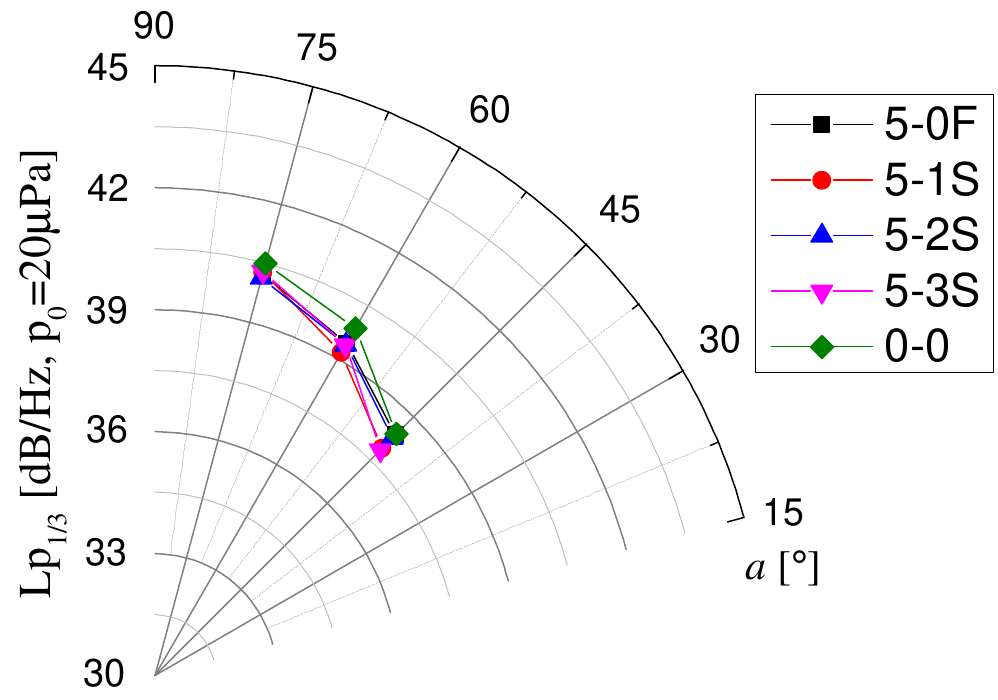} 
	} 
	\subfigure[1000 Hz]{ 
		\label{634421_1500_directivity_1000Hz}
		\includegraphics[width=.22\textwidth]{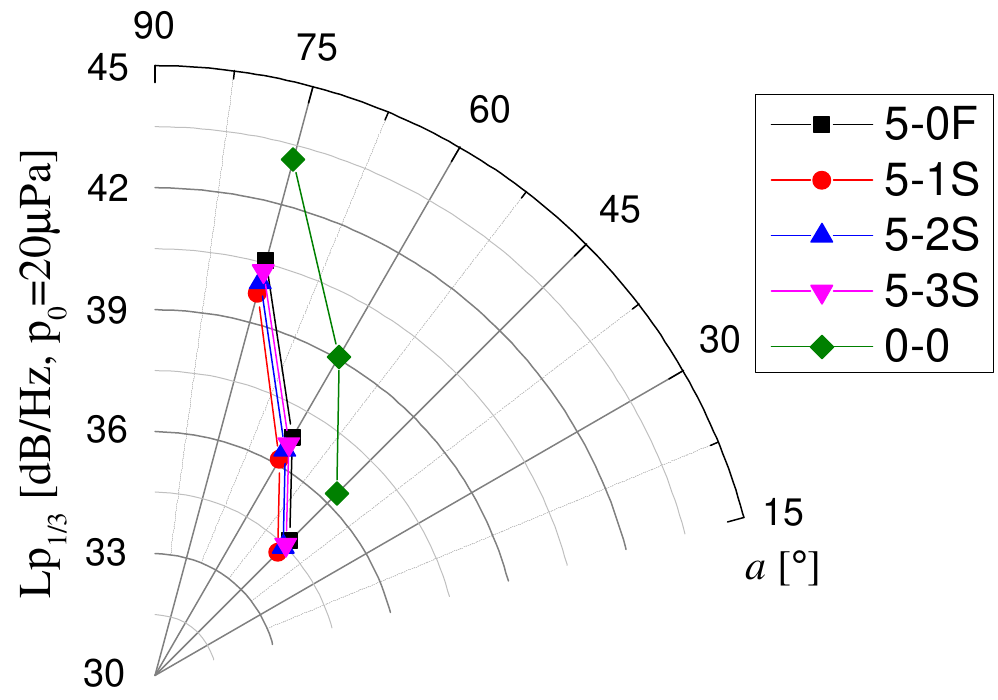} 
	}
	\subfigure[3150 Hz]{ 
		\label{634421_1500_directivity_3150Hz}
		\includegraphics[width=.22\textwidth]{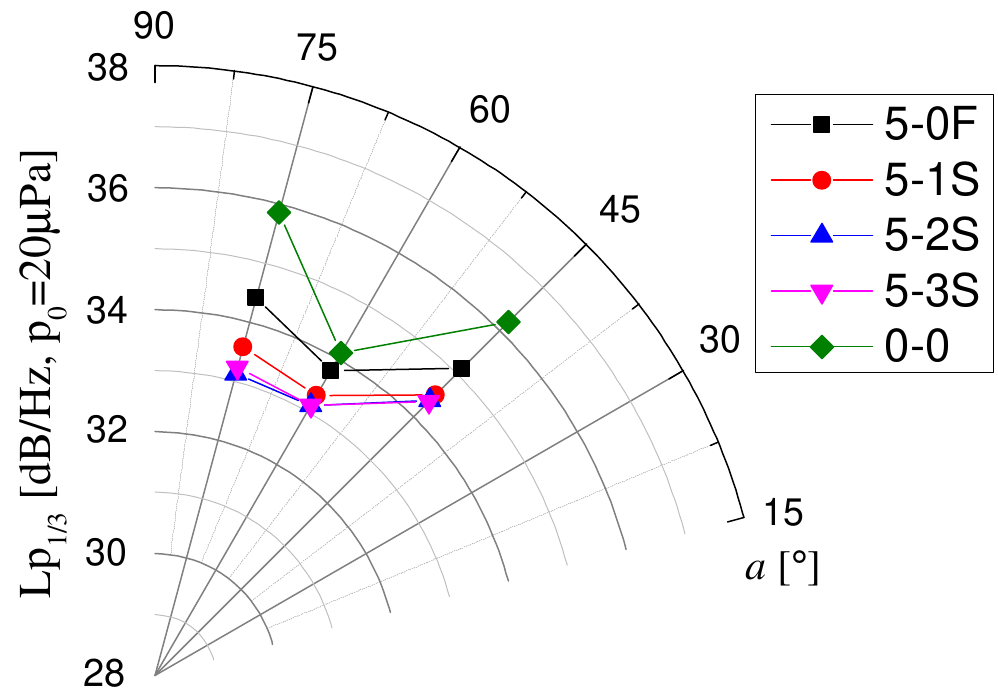} 
	}
	\subfigure[10000 Hz]{ 
    	\label{634421_1500_directivity_10000Hz}
    	\includegraphics[width=.22\textwidth]{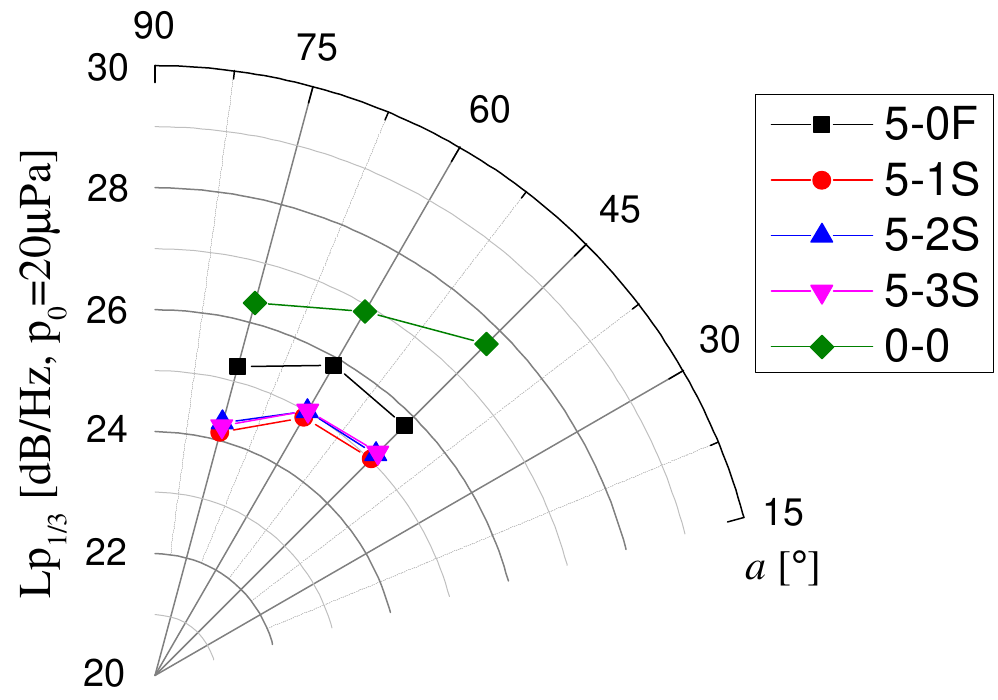} 
    } 
	\caption{Sound directivity of NACA 63(4)-421 airfoil at $Re=1.2\times10^5$} 
	\label{634421_1500_directivity}
\end{figure}

\begin{figure}[H]
	\centering 
	\subfigure[400 Hz]{ 
		\label{634421_2000_directivity_400Hz}
		\includegraphics[width=.22\textwidth]{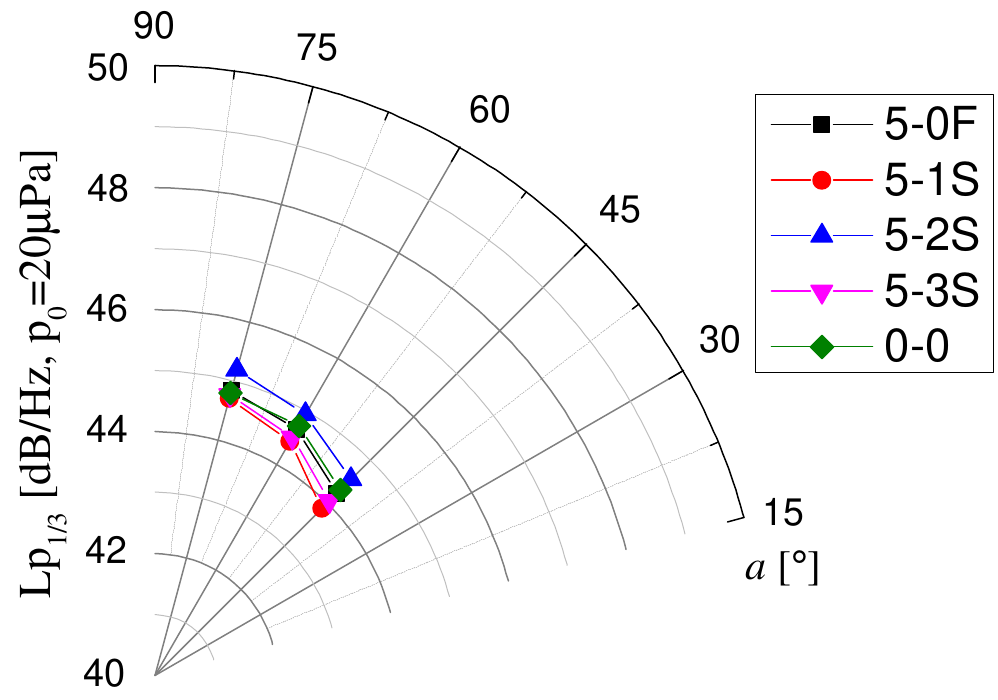} 
	} 
	\subfigure[1000 Hz]{ 
		\label{634421_2000_directivity_1000Hz}
		\includegraphics[width=.22\textwidth]{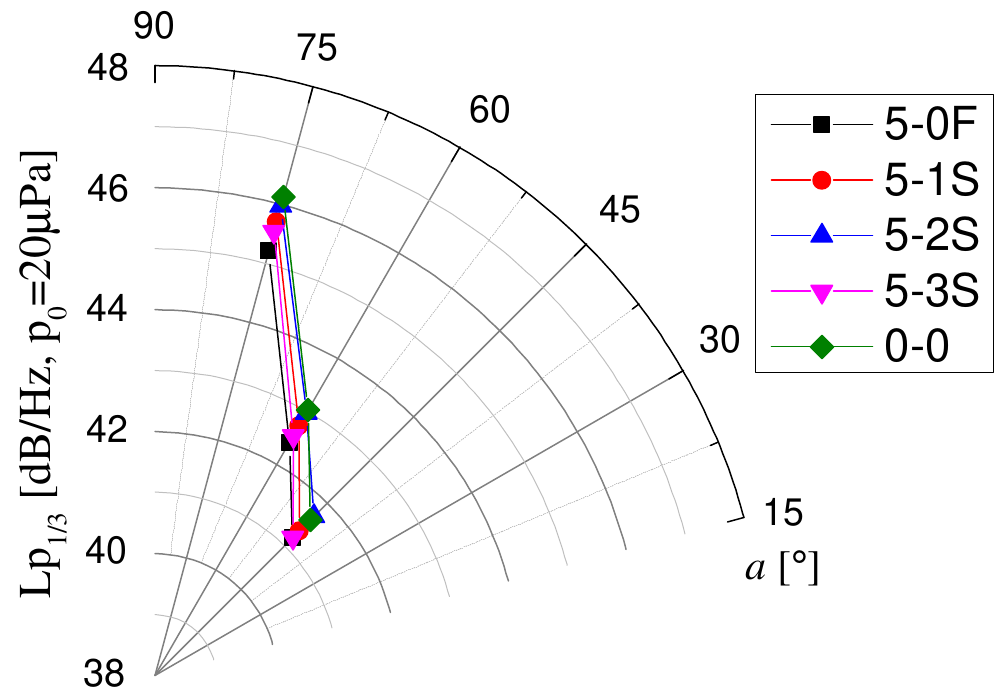} 
	}
	\subfigure[3150 Hz]{ 
		\label{634421_2000_directivity_3150Hz}
		\includegraphics[width=.22\textwidth]{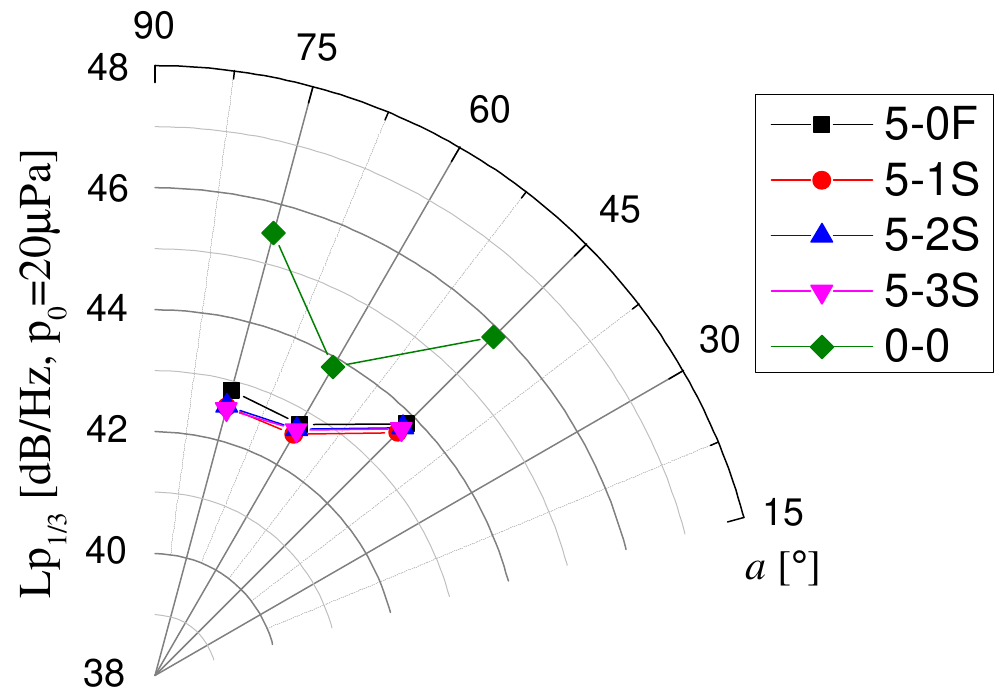} 
	}
	\subfigure[10000 Hz]{ 
    	\label{634421_2000_directivity_10000Hz}
    	\includegraphics[width=.22\textwidth]{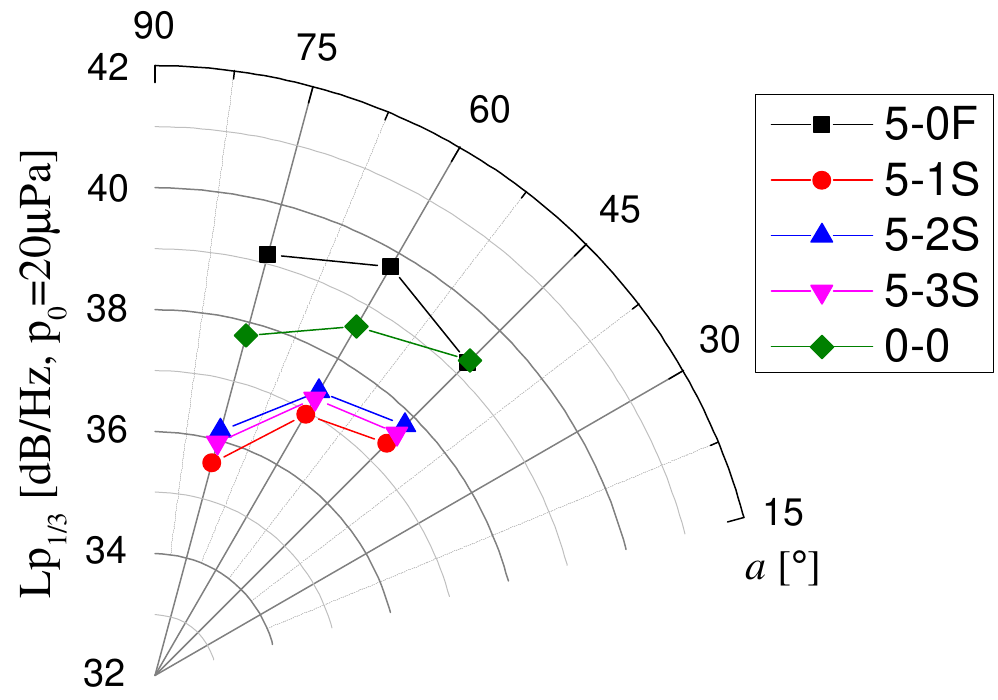} 
    } 
	\caption{Sound directivity of NACA 63(4)-421 airfoil at $Re=1.6\times10^5$} 
	\label{634421_2000_directivity}
\end{figure}

The sound pressure levels shown in Fig.~\ref{634421_1000_directivity}, Fig.~\ref{634421_1500_directivity}, and Fig.~\ref{634421_2000_directivity} represent the original intensities recorded by the microphones at their respective positions for the NACA 63(4)-421 airfoil at Reynolds numbers of $0.7 \times 10^5$, $1.2 \times 10^5$, and $1.6 \times 10^5$, respectively. The directivity patterns in these figures are presented in terms of 1/3-octave band sound pressure levels, which provide a smoothed representation of the acoustic data, making it easier to observe trends in frequency-dependent noise behavior.

Several key observations about noise propagation characteristics and the effectiveness of different trailing edge modifications can be revealed. At low to medium frequencies, around 400 Hz, both straight-edge and serrated trailing edges show minimal impact on noise reduction and do not significantly alter the acoustic directivity patterns of the airfoil noise. As the frequency increases to 1000 Hz and beyond, both straight-edge and serrated trailing edges demonstrate a noticeable reduction in sound intensity. However, these modifications do not alter the inherent acoustic directivity characteristics of the airfoil, which contrasts with the findings of Tian et al.~\cite{tian2022prediction}, who demonstrated that trailing edge serrations can significantly modify directivity patterns. For instance, at 1000 Hz, the highest noise intensity is observed in directions approximately perpendicular to the airfoil surface, and this pattern remains unchanged with trailing edge modifications. Serrated trailing edges notably outperform straight-edge modifications in noise reduction, particularly at frequencies above 3150 Hz. The serrated configurations consistently show lower noise levels compared to straight edges, indicating their superior efficacy in mitigating noise at higher frequencies. This observation is consistent with the findings of Lyu et al.~\cite{lyu2016prediction}, which emphasized that primary noise reduction arises from interference effects near the trailing edge but also highlighted substantial alterations in directivity characteristics at high frequencies.

The directivity results for the NACA 63(3)-418 airfoil exhibit similar trends and are not explicitly presented here, as the overall patterns and observations closely align with those of the analyzed configurations. Overall, the findings suggest that while trailing edge modifications can significantly reduce noise levels, especially at higher frequencies, they do not fundamentally alter the directivity characteristics of the airfoil's acoustic emissions.

\section{Wake Flow Measurement}

\subsection{Mean and Fluctuation Streamwise Velocity}

\begin{figure}[H]
	\centering 
	\subfigure[Mean velocity]{ 
		\label{634421_average_velocity}
		\includegraphics[width=.45\textwidth]{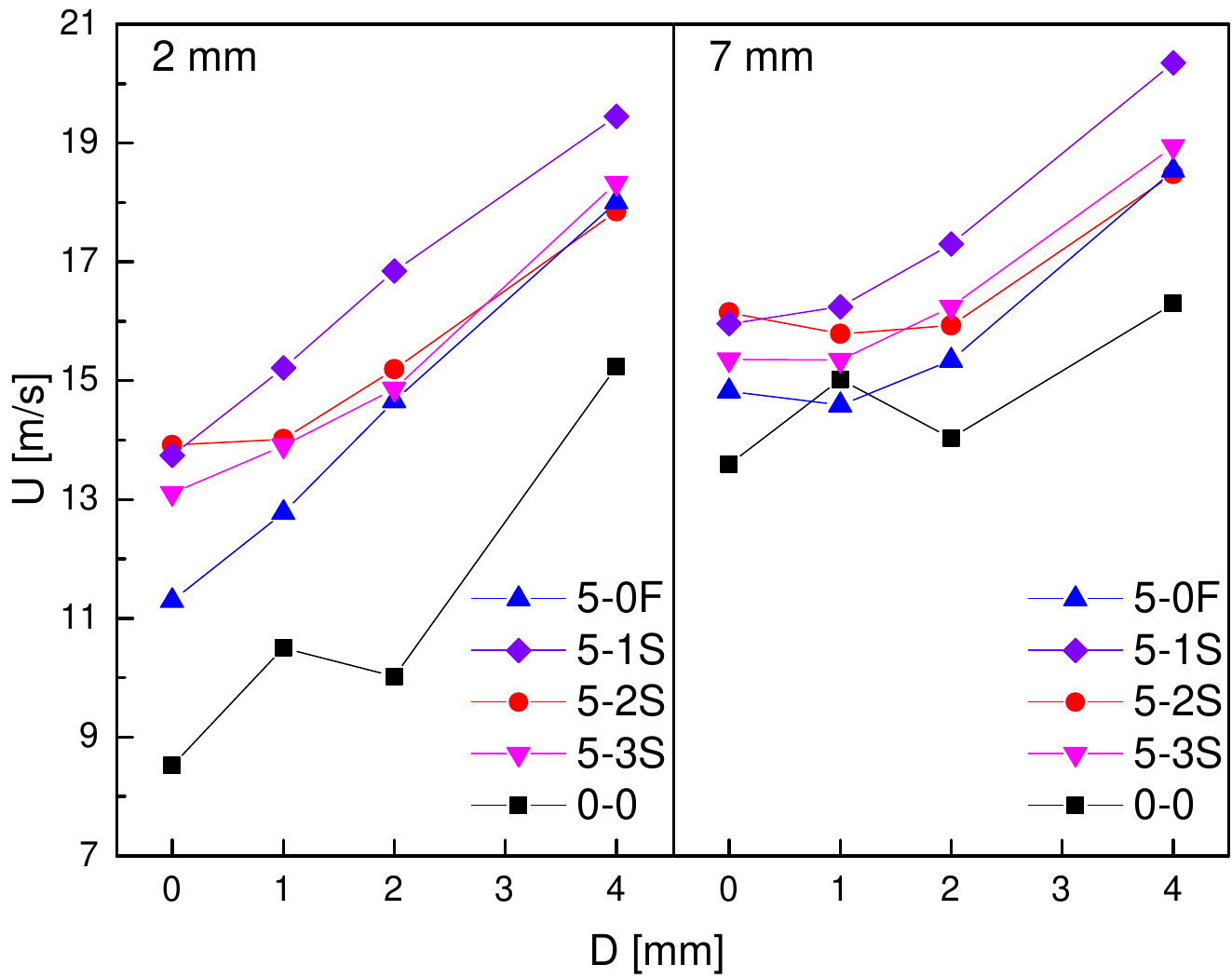} 
	} 
	\subfigure[Fluctuation velocity]{ 
		\label{634421_fluctuation_velocity}
		\includegraphics[width=.45\textwidth]{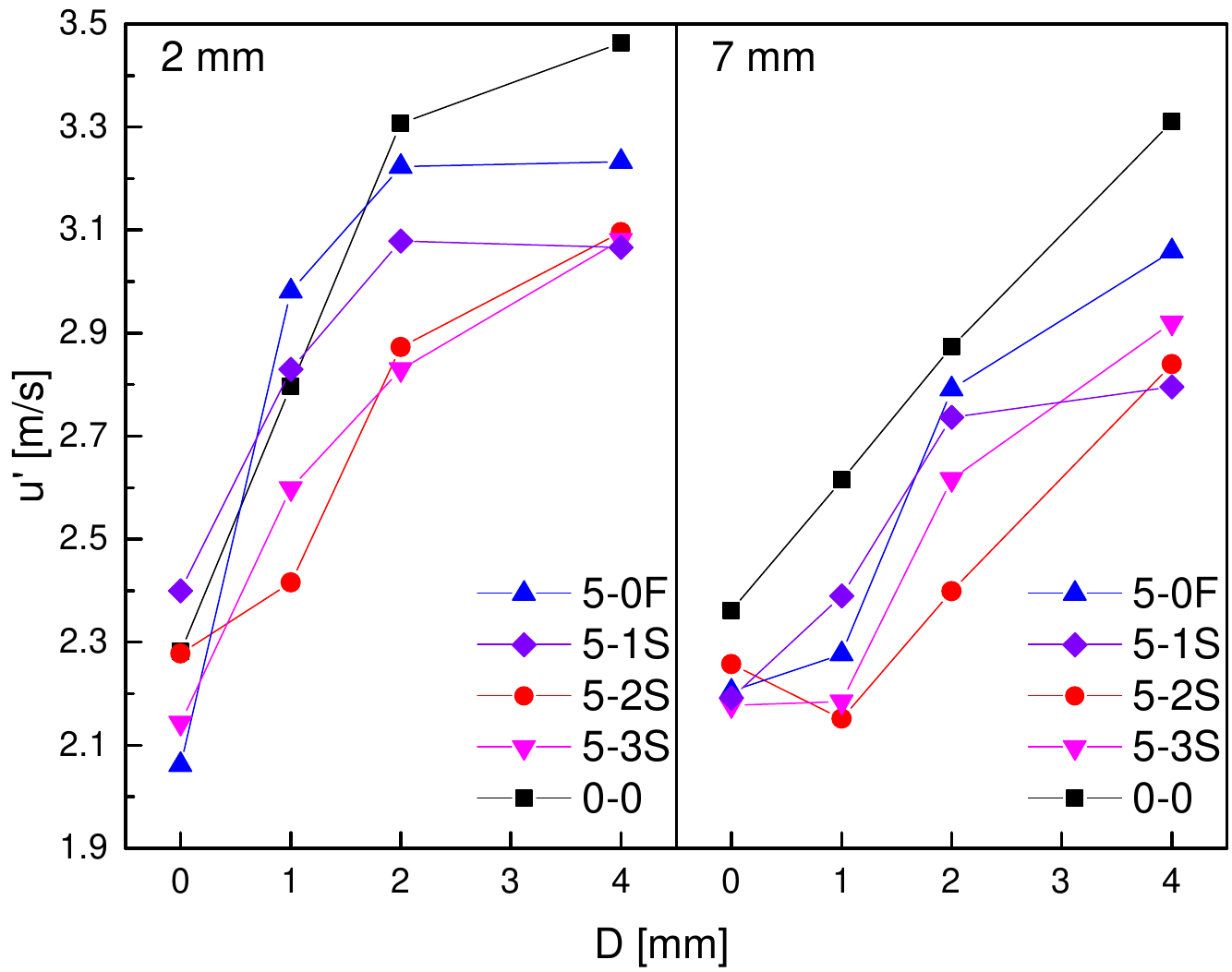} 
	} 
	\caption{Velocity measurement for the NACA 63(4)-421 airfoil ($Re=1.2\times10^5$)} 
	\label{634421_velocity}
\end{figure}

\begin{figure}[H]
	\centering 
	\subfigure[Mean velocity]{ 
		\label{634421_average_velocity_2000}
		\includegraphics[width=.45\textwidth]{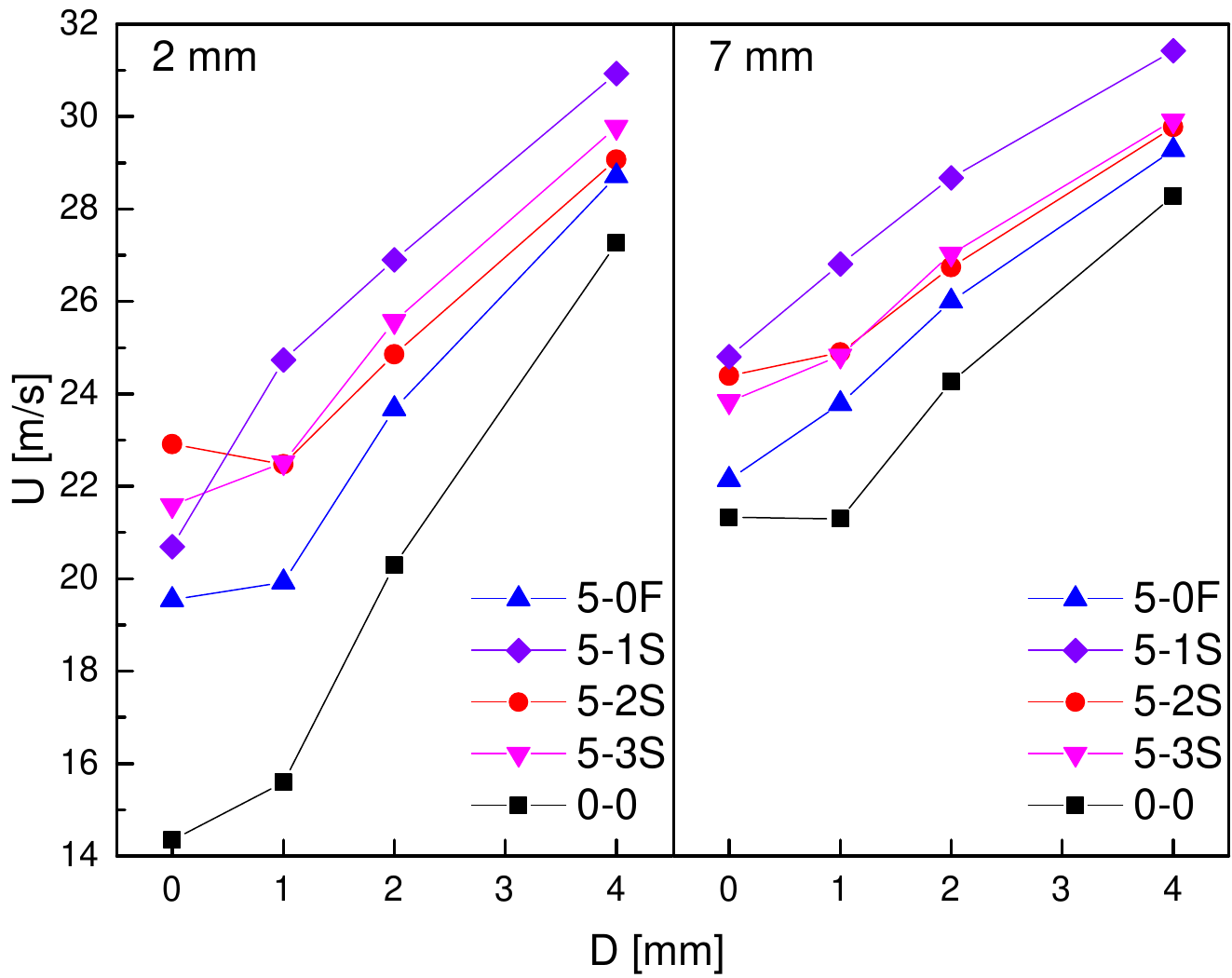} 
	} 
	\subfigure[Fluctuation velocity]{ 
		\label{634421_fluctuation_velocity_2000}
		\includegraphics[width=.45\textwidth]{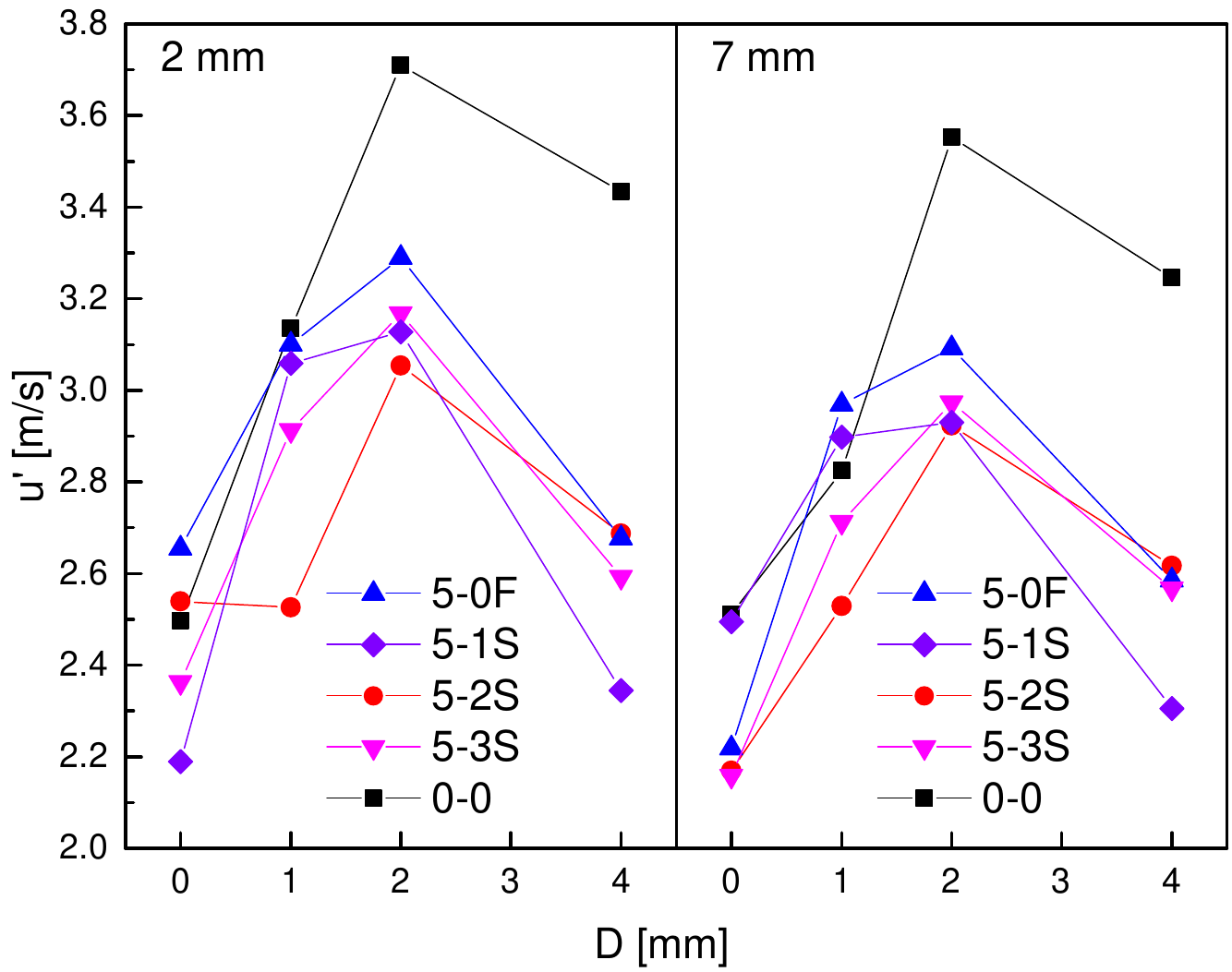} 
	} 
	\caption{Velocity measurement for the NACA 63(4)-421 airfoil ($Re=1.6\times10^5$)} 
	\label{634421_velocity_2000}
\end{figure}

Fig.~\ref{634421_velocity} and Fig.~\ref{634421_velocity_2000} show the mean velocity and fluctuation velocity measurements for the NACA 63(4)-421 airfoil with various trailing edge configurations at a Reynolds number of $1.2 \times 10^5$ and $1.6 \times 10^5$, respectively. Each figure includes the mean and fluctuation velocities at distances of 2 mm and 7 mm downstream. The horizontal axis indicates the offset distance from the trailing edge apex, as depicted in Fig.~\ref{measurement_locations}.

For both Reynolds numbers, the original and straight-edge airfoils exhibit the lowest mean velocities, while the serrated airfoils, particularly with the largest serration, show higher mean velocities. Fluctuation velocity generally decreases with distance from the trailing edge, and serrations effectively reduce fluctuation intensity at multiple locations. These results indicate that serrated trailing edges enhance streamwise turbulence reduction.

The observed correlation between reduced wake fluctuating velocities and noise reduction may be explained by the interaction between the serrations and the trailing edge flow. Serrations likely disrupt coherent vortex structures in the wake, breaking them into smaller, less energetic eddies that generate less noise. This disruption reduces the strength of the velocity fluctuations in the wake, which is directly linked to the sound generated by the trailing edge. The reduction in fluctuating velocities near the trailing edge, therefore, leads to a decrease in noise levels.

\begin{figure}[H]
	\centering 
	\subfigure[Point "10"]{ 
		\label{634421_1500_10_turbulence_spectrum}
		\includegraphics[width=.45\textwidth]{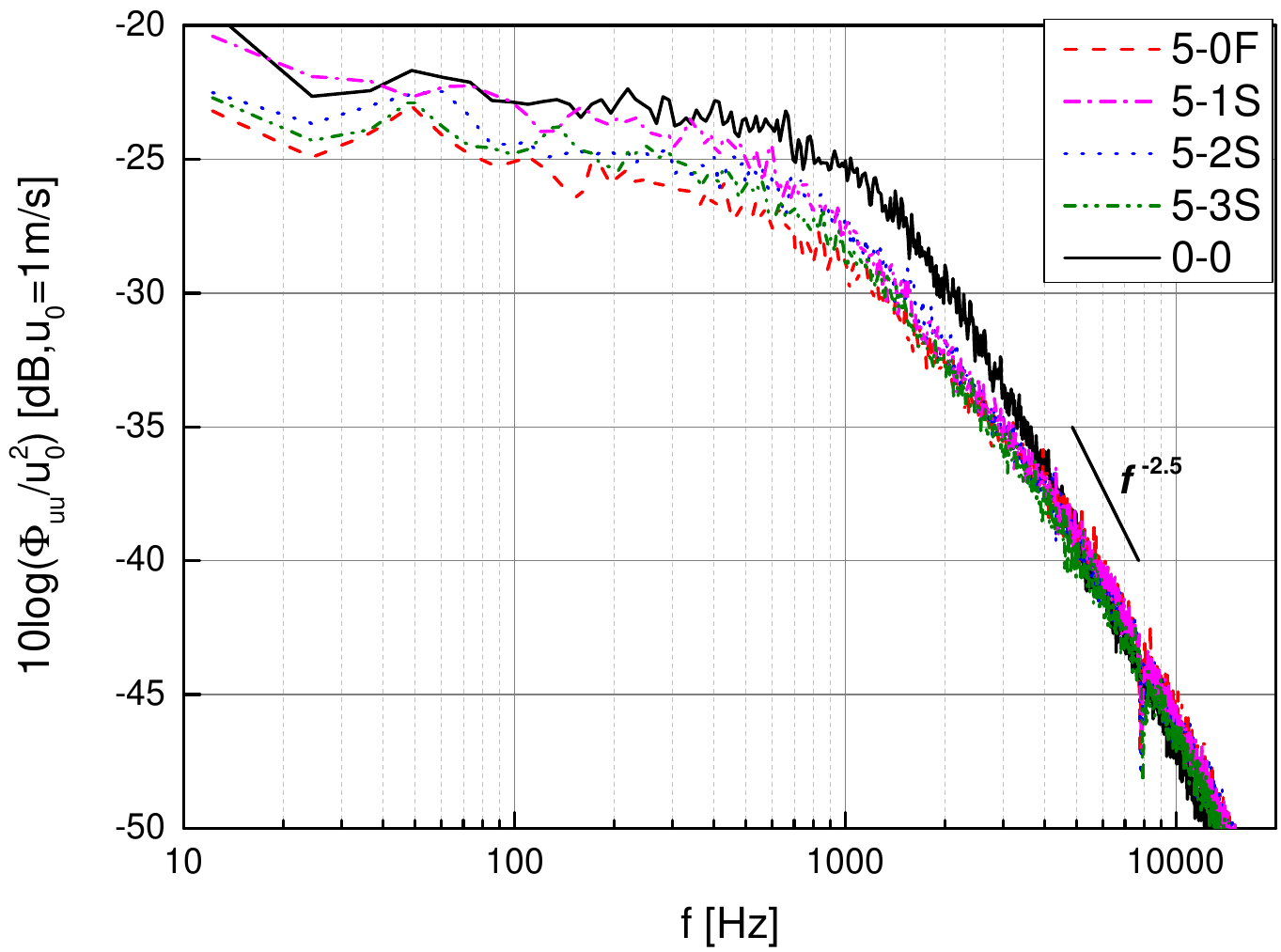} 
	} 
	\subfigure[Point "20"]{ 
		\label{634421_1500_20_turbulence_spectrum}
		\includegraphics[width=.45\textwidth]{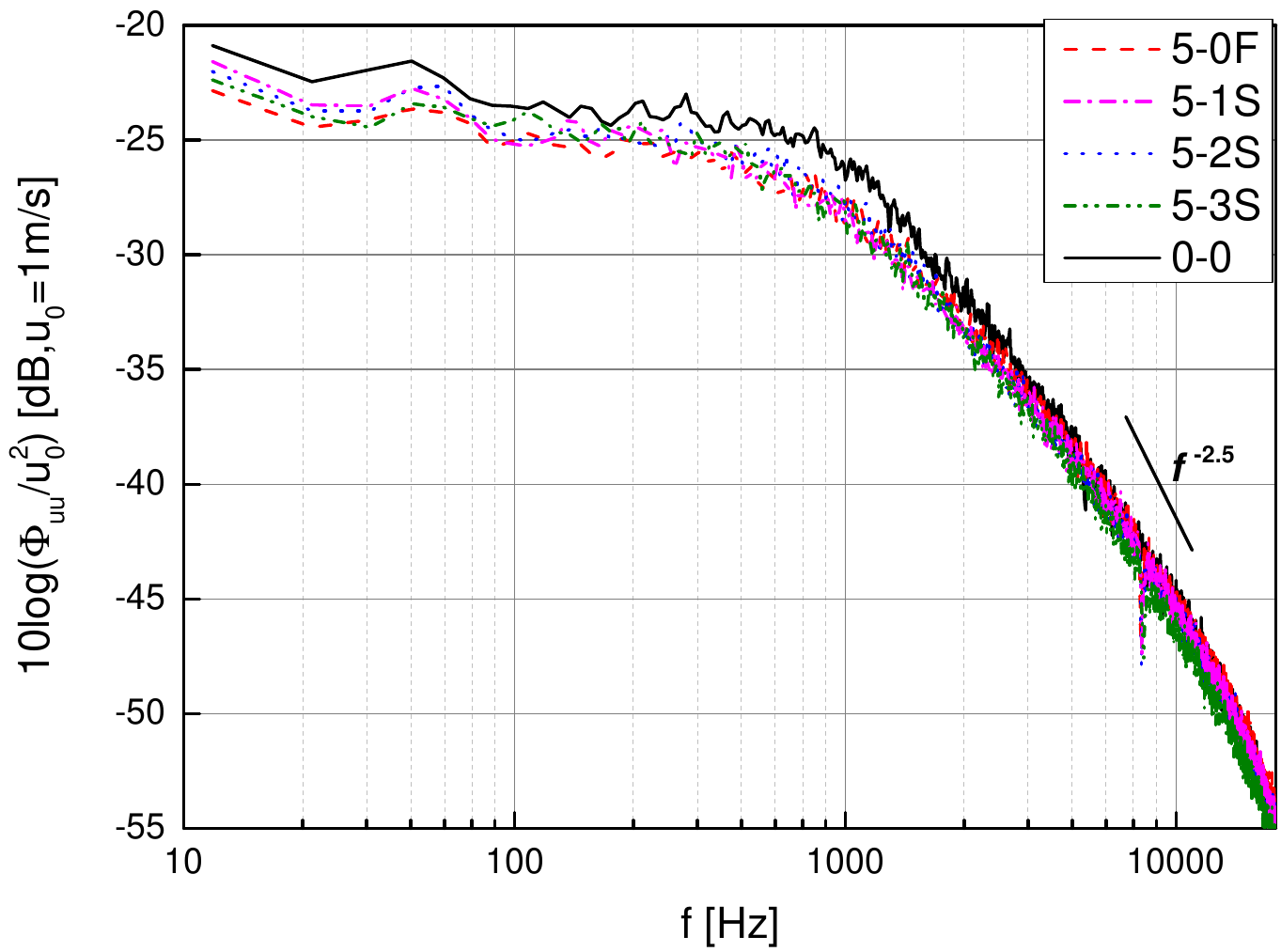} 
	} 
	\caption{PSD of turbulent velocity fluctuations for NACA 63(4)-421 at $Re=1.2\times10^5$} 
	\label{634421_1500_turbulence_spectrum}
\end{figure}

Fig.~\ref{634421_1500_turbulence_spectrum} displays the power spectral density (PSD) of turbulent velocity fluctuations at measurement points "10" and "20" (notations shown in Fig.~\ref{measurement_locations}) for $Re=1.2\times10^5$. These points, located downstream of the trailing edge, are highly influenced by the flow field. The original airfoil exhibits the highest PSD in the mid-low frequency range (300$\sim$4000 Hz), with differences up to 4 dB, correlating with noise reduction ranges. This indicates that reducing streamwise fluctuations near the trailing edge contributes to noise reduction. Serrations disrupt larger vortex structures, with finer serrations proving more effective. High-frequency turbulence remains unaffected, following a -2.5 power law, consistent with Ref~\cite{gruber2010experimental}.

\begin{figure}[H]
	\centering 
	\subfigure[Measurement point "10"]{ 
		\label{633418_1500_10_turbulence_spectrum}
		\includegraphics[width=.45\textwidth]{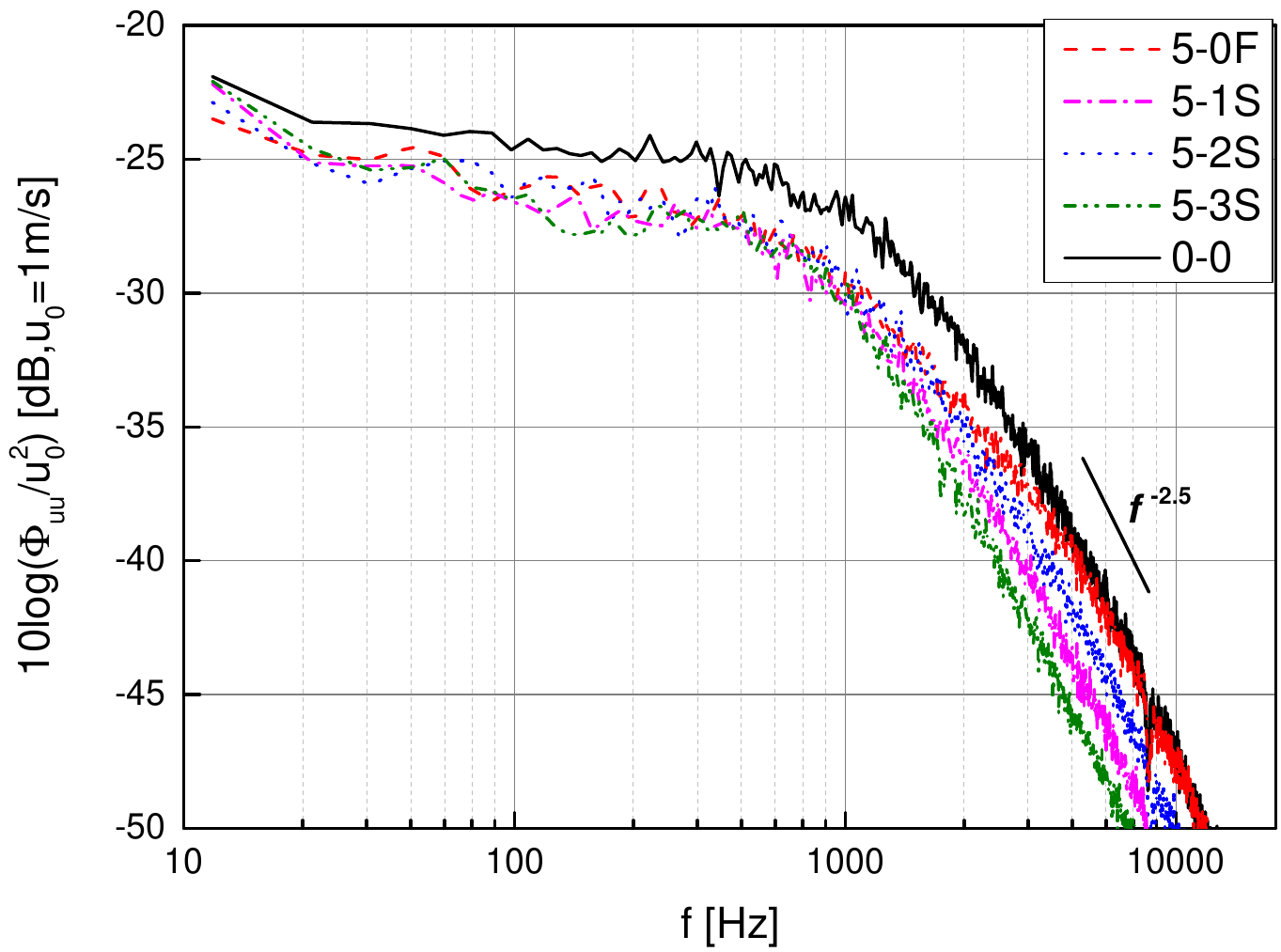} 
	} 
	\subfigure[Measurement point "20"]{ 
		\label{633418_1500_20_turbulence_spectrum}
		\includegraphics[width=.45\textwidth]{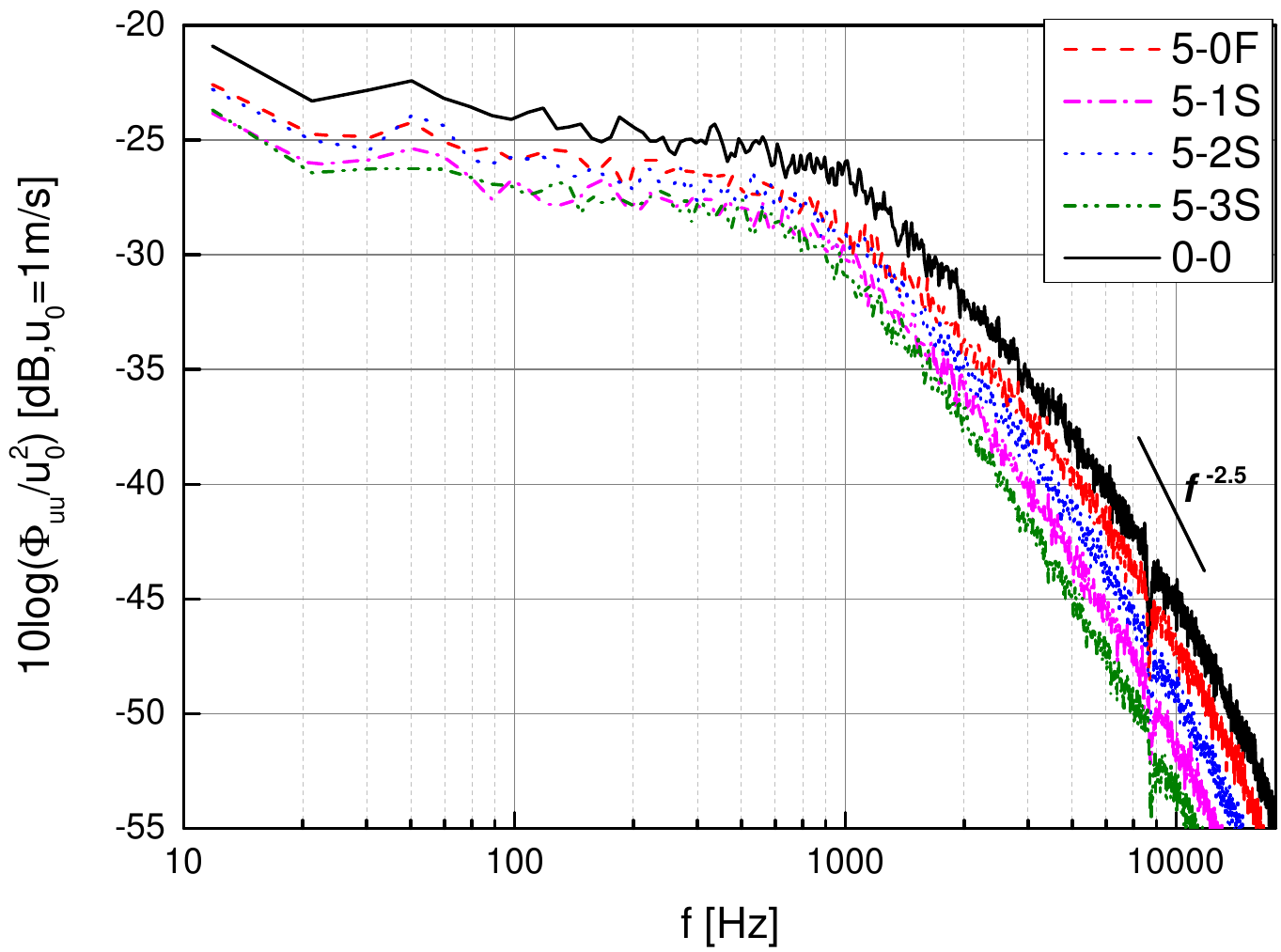} 
	} 
	\caption{PSD of turbulent velocity fluctuations for NACA 63(3)-418 at $Re=1.2\times10^5$} 
	\label{633418_1500_turbulence_spectrum}
\end{figure}

Fig.~\ref{633418_1500_turbulence_spectrum} shows the PSD of turbulent velocity fluctuations at measurement points "10" and "20" for the NACA 63(3)-418 airfoil at $Re=1.2\times10^5$. The results indicate that the PSD for the serrated trailing edge model is significantly lower than that for both the original airfoil and the airfoil with flat plate serrations. This suggests that the serrations are effective in reducing turbulence intensity. The difference in the wake flow behavior, when compared to the NACA 63(4)-421 airfoil, can be attributed to differences in the airfoil profiles, which affect the wake flow characteristics. Further measurements are required in future studies to explore these differences in more detail.

While these results suggest a causal relationship between reduced wake fluctuations and noise reduction, the empirical data primarily indicate a correlation. Further experimentation, particularly involving direct measurements of the vortex dynamics and their contribution to noise generation, would be beneficial to validate this connection and fully understand the underlying mechanisms.

\subsection{Autocorrelation}

The autocorrelation function of fluctuation velocity, denoted as \( R_{uu}(x, \tau) \), is a method to analyze turbulence structures and quantify the persistence of turbulent eddies in the flow field. The autocorrelation function at a given location \( x \) is defined in Eq.~\ref{autocorr}:

\begin{equation}
R_{uu}(x, \tau) = \langle u'(x, t) u'(x, t + \tau) \rangle
\label{autocorr}
\end{equation}
where \( u'(x, t) \) represents the streamwise fluctuation velocity component at time \( t \) and position \( x \), and \( \tau \) is the time lag. The notation \( \langle \cdot \rangle \) denotes the ensemble average, which provides a statistical measure of the correlation between velocity fluctuations at different time instances. The autocorrelation function, therefore, provides a measure of how the velocity at a particular point remains correlated with itself over time. 

The characteristic eddy scale, \( L_u(x) \), can be estimated by finding the time lag \( \tau_0 \) at which the autocorrelation function decays to half of its maximum value, as shown in Eq.~\eqref{half_autocorr}:

\begin{equation}
\frac{R_{uu}(x, \tau_0)}{R_{uu}(x, 0)} = 0.5
\label{half_autocorr}
\end{equation}

The eddy scale is then calculated using the convection velocity \( U_c \), typically approximated as \( U_c = 0.7 U \), with \( U \) representing the freestream velocity:

\begin{equation}
L_u(x) = U_c \tau_0
\label{eddy_scale}
\end{equation}

\begin{figure}[H]
	\centering 
	\subfigure[0-0 model]{ 
		\label{634421_1500_2mm_autocorelation_0-0}
		\includegraphics[width=.45\textwidth]{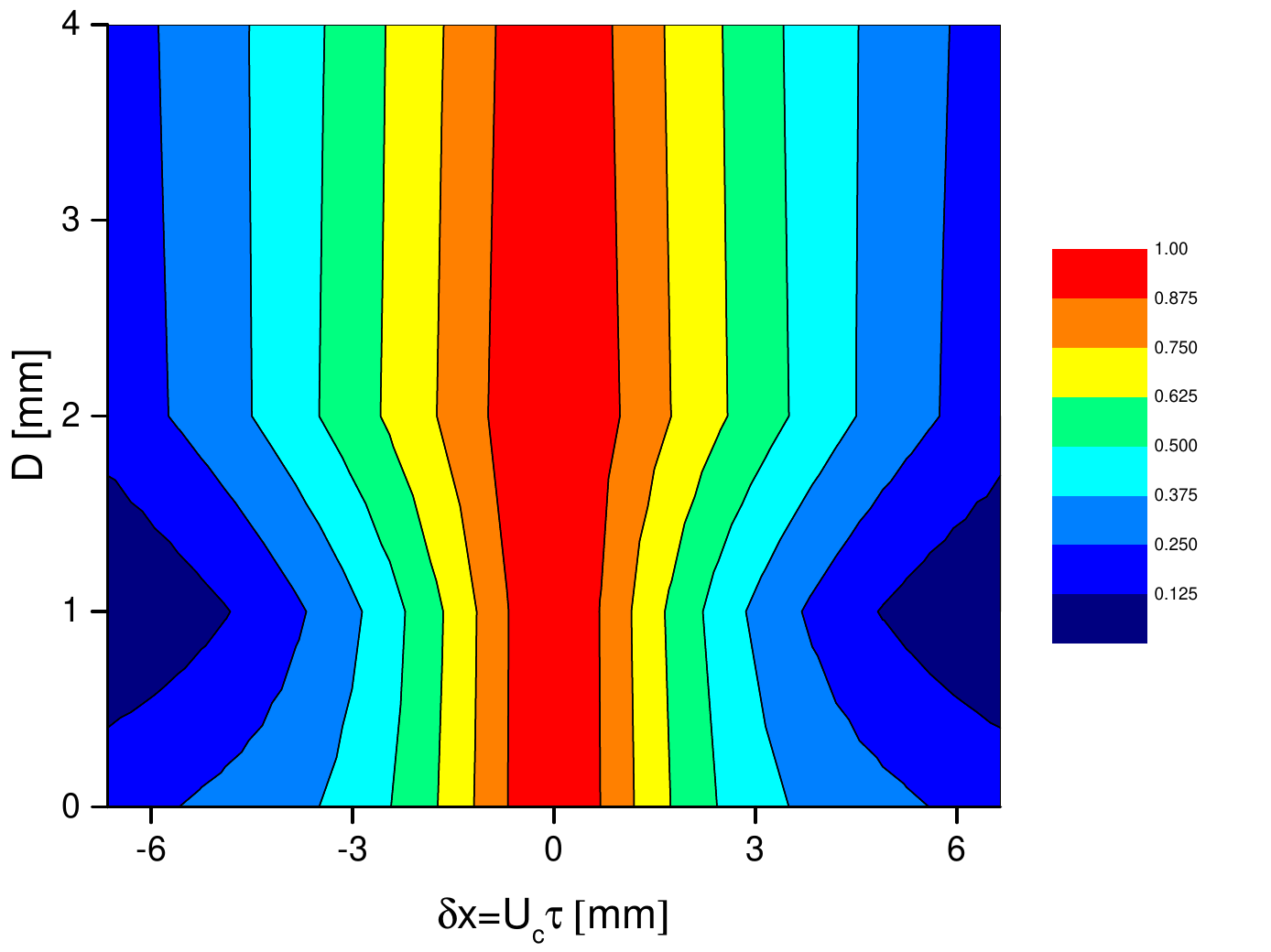} 
	} 
	\subfigure[5-2S model]{ 
		\label{634421_1500_2mm_autocorelation_5-2S}
		\includegraphics[width=.45\textwidth]{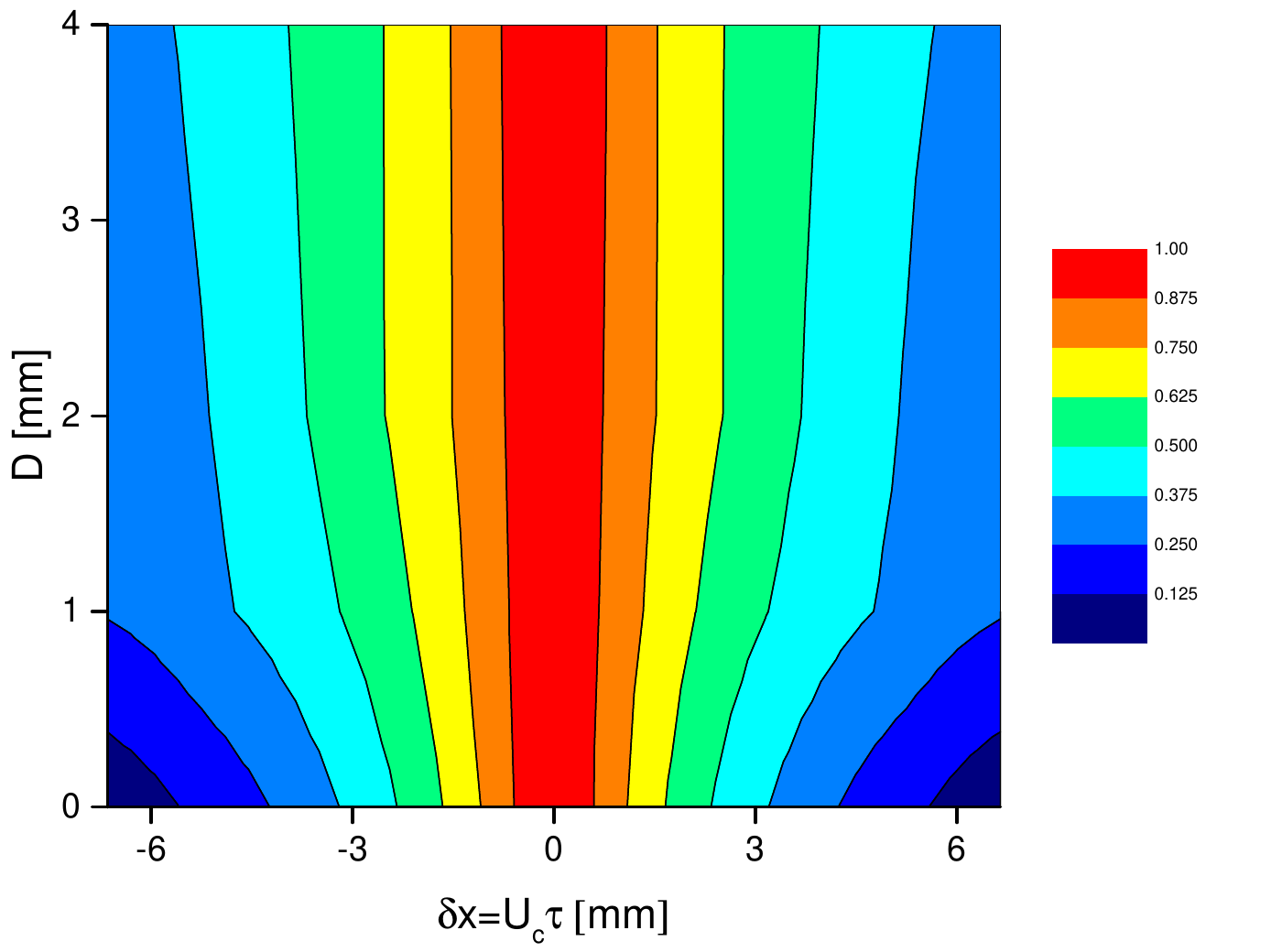} 
	} 
	\caption{Autocorrelation of turbulent velocity fluctuations at 2 mm downstream of the trailing edge for NACA 63(4)-421} 
	\label{634421_1500_2mm_autocorelation}
\end{figure}

\begin{figure}[H]
	\centering 
	\subfigure[0-0 model]{ 
		\label{634421_1500_7mm_autocorelation_0-0}
		\includegraphics[width=.45\textwidth]{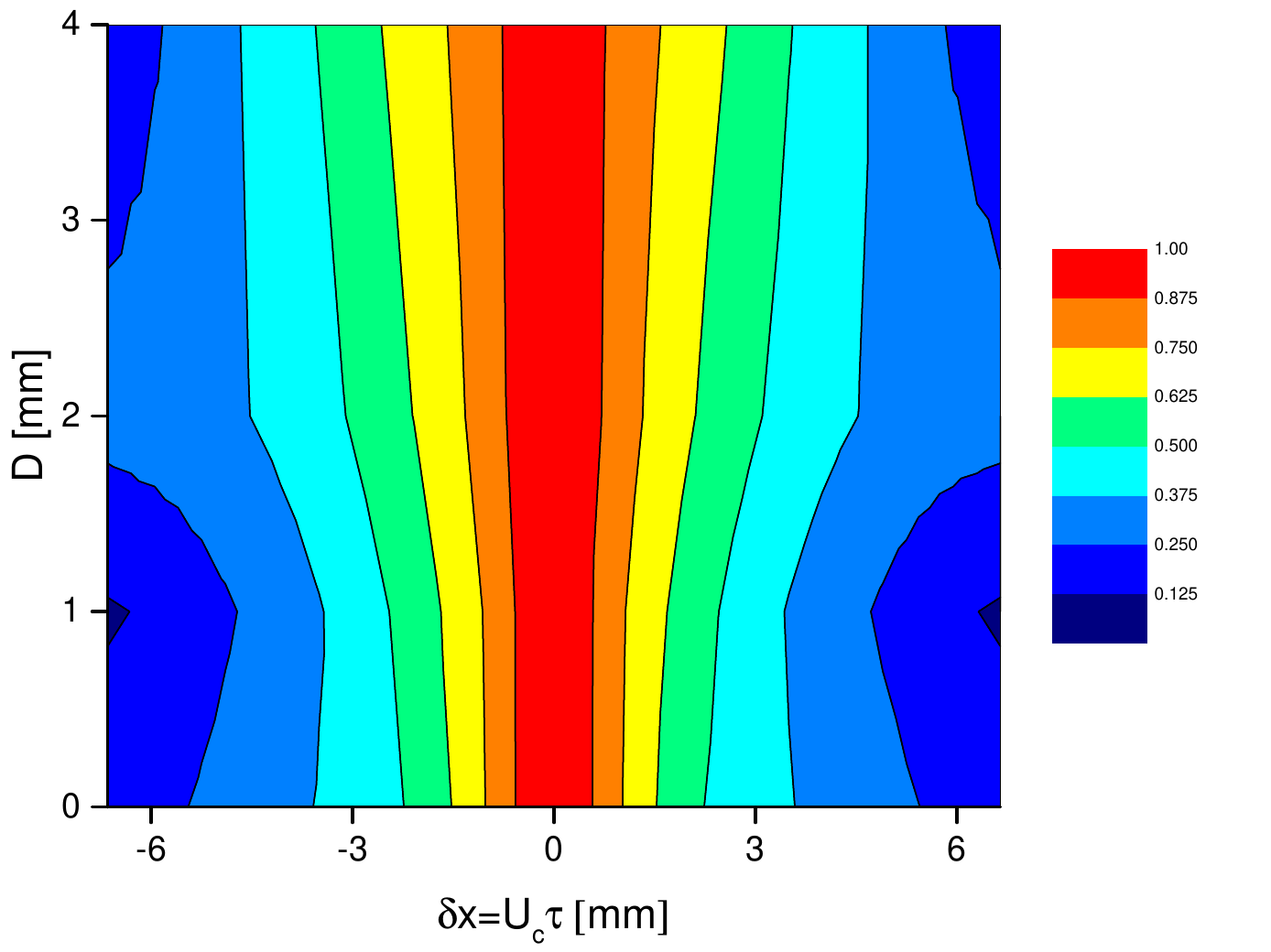} 
	} 
	\subfigure[5-2S model]{ 
		\label{634421_1500_7mm_autocorelation_5-2S}
		\includegraphics[width=.45\textwidth]{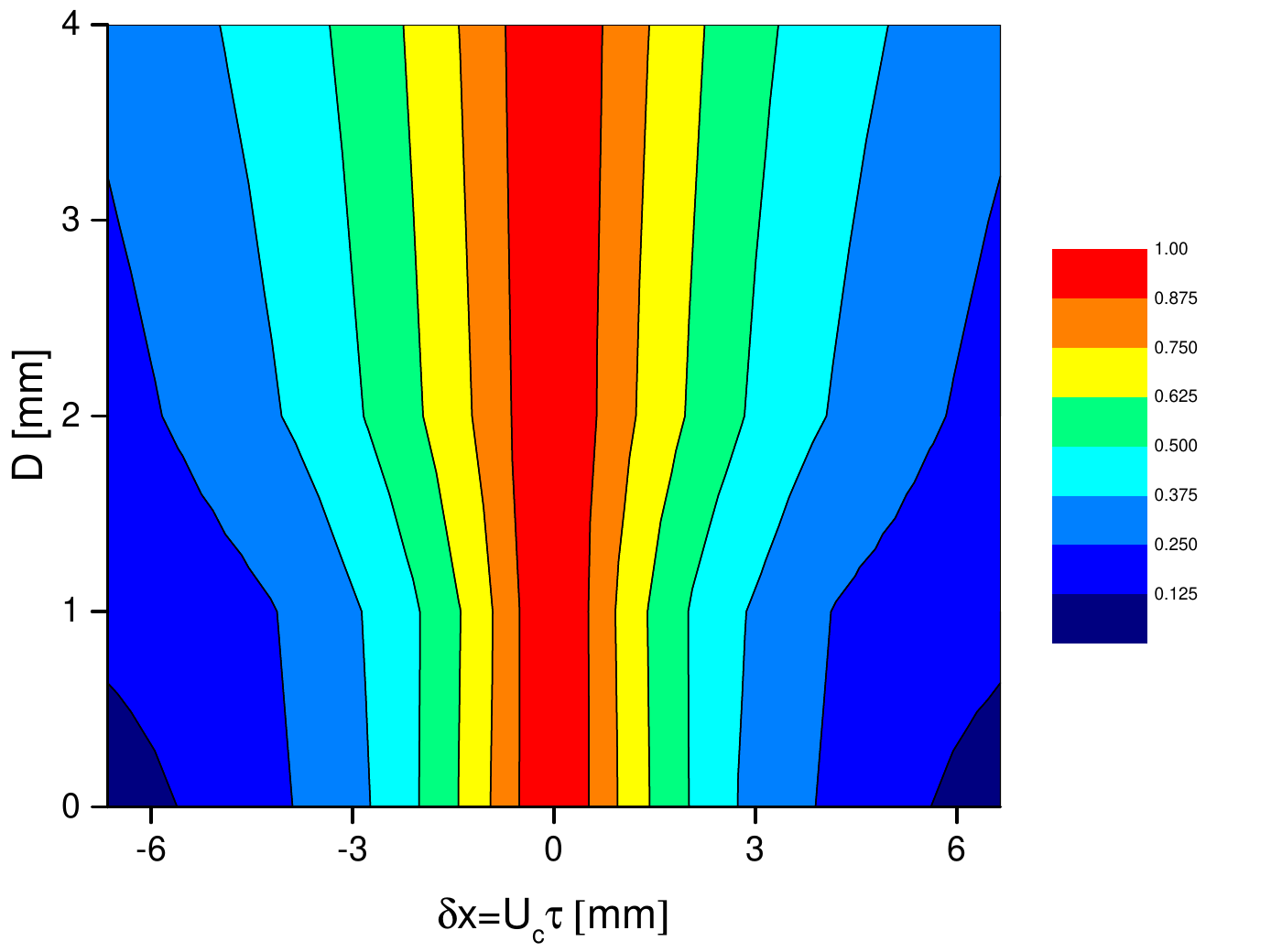} 
	} 
	\caption{Autocorrelation of turbulent velocity fluctuations at 7 mm downstream of the trailing edge for NACA 63(4)-421} 
	\label{634421_1500_7mm_autocorelation}
\end{figure}

Fig.~\ref{634421_1500_2mm_autocorelation} and Fig.~\ref{634421_1500_7mm_autocorelation} show the autocorrelation maps of turbulent velocity fluctuations for the original NACA 63(4)-421 airfoil (0-0) and the 5-2S model at 2 mm and 7 mm downstream, respectively, for $Re=1.2\times10^5$. The autocorrelation map for the 0-0 model reveals a broader high-value region and a narrower mid-low-value region at both locations, indicating more persistent turbulence structures compared to the 5-2S model. Additionally, for the same model, the high-value region at 2 mm is noticeably broader than at 7 mm, suggesting a larger characteristic eddy scale closer to the trailing edge. This observation suggests that serrations effectively reduce the correlation of flow near the sound source, thereby diminishing the efficiency of the sound source and contributing to noise reduction.

\begin{figure}[H]
	\centering 
	\subfigure[Measurement point "10"]{ 
		\label{634421_1500_turbulent_length_2mm}
		\includegraphics[width=.45\textwidth]{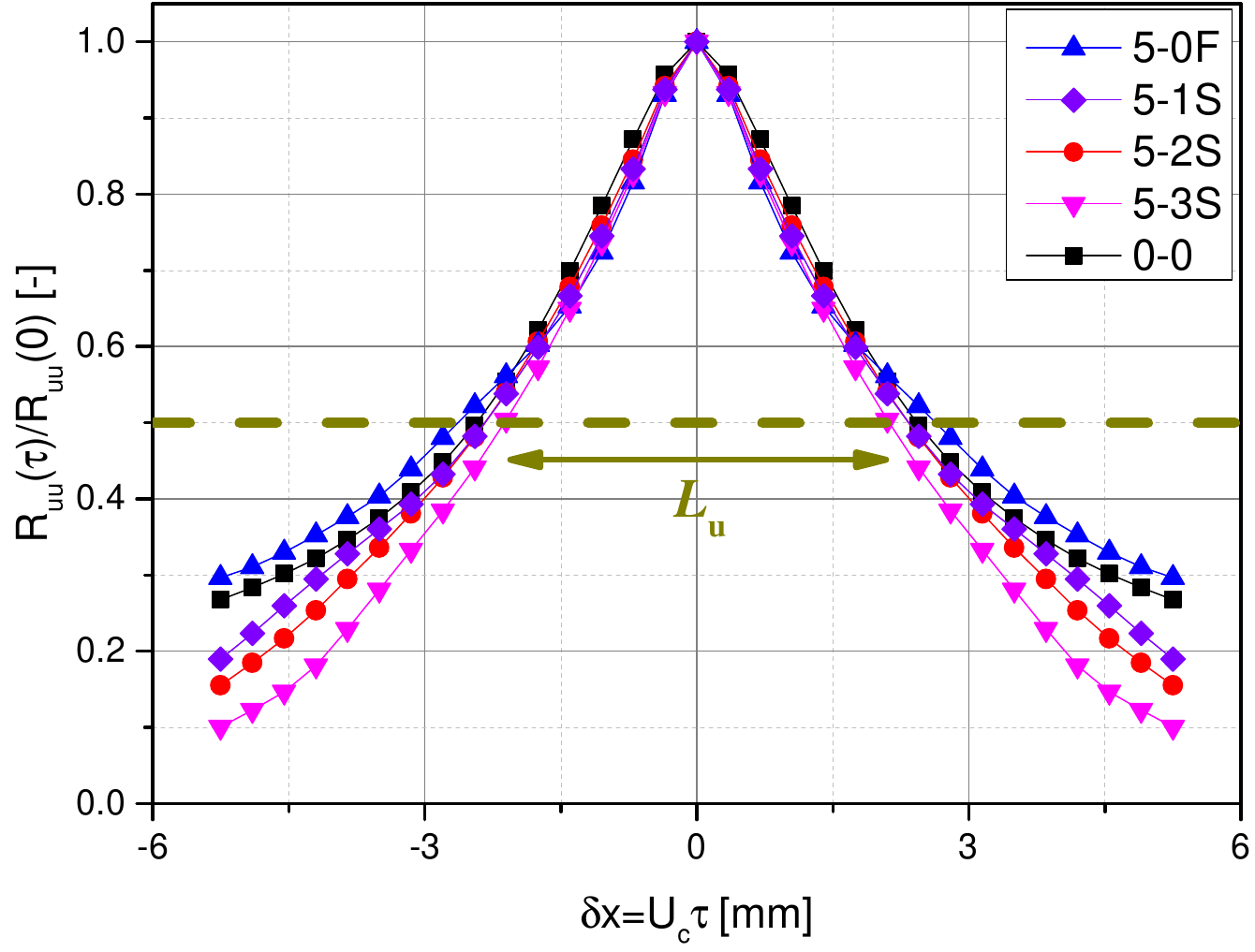} 
	} 
	\subfigure[Measurement point "20"]{ 
		\label{634421_1500_turbulent_length_7mm}
		\includegraphics[width=.45\textwidth]{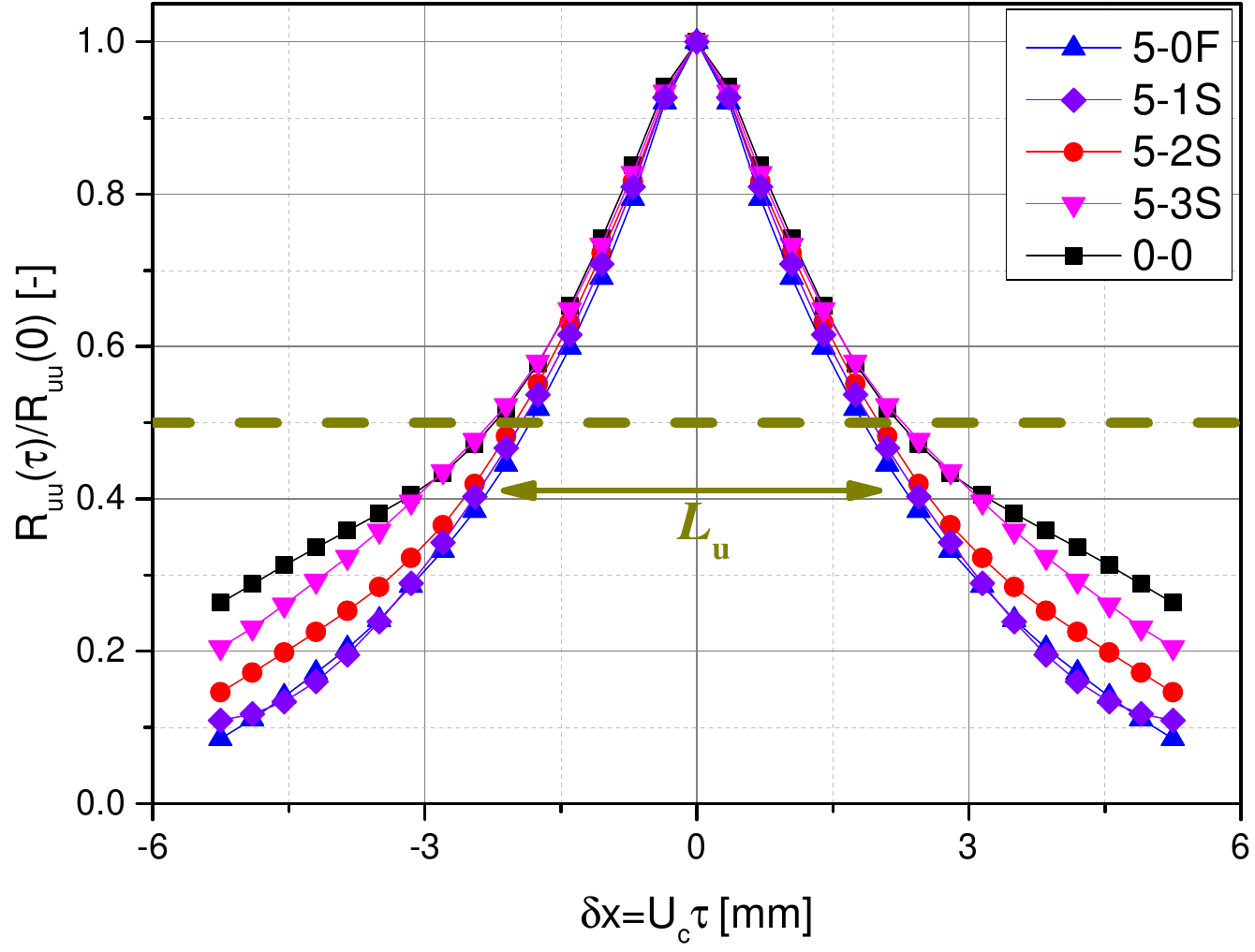} 
	} 
	\caption{Autocorrelation of different trailing edges for $Re=1.2\times10^5$}
	\label{634421_1500_turbulent_length}
\end{figure}

To facilitate a comprehensive comparison of the autocorrelation functions and characteristic eddy scales across various trailing edge configurations, these metrics are presented together in a single graph, as shown in Fig.~\ref{634421_1500_turbulent_length}. The labels $L_{u}$ indicate the characteristic eddy scales for each type of trailing edge.

Fig.~\ref{634421_1500_turbulent_length} shows that the largest turbulent length scale is observed for the 0-0 trailing edge. In contrast, serrated trailing edges, particularly those with larger serrations, show a reduction in turbulent length scales near the trailing edge. This decrease in the characteristic eddy scale suggests that serrations effectively disrupt larger vortex structures, thereby reducing aeroacoustic noise intensity in this region. Additionally, longer and larger serrations have a more significant effect on altering the vortex scale in the flow field near the trailing edge, further enhancing their noise reduction performance.

These measurements indicate that the primary reason for noise reduction by serrated trailing edges may be the alteration of the flow field near the source location, rather than changes in the radiation efficiency of the source~\cite{jones2010numerical, sandberg2010reprint, lyu2016prediction, mayer2019semi, ayton2018analytic, gelot2020effect}. Another possible noise reduction mechanism is destructive interference of the scattered surface pressure~\cite{lyu2016prediction}, but cannot be validated in this study. There are significant observed discrepancies between numerical and experimental results, underscoring the need for further investigation and a deeper understanding of the underlying mechanisms involved in the noise reduction process.

\section{Conclusions}
\label{conclusions}

This study offers an experimental measurement of noise reduction achieved with airfoils featuring serrated trailing edges in a low turbulence wind tunnel. The measurement can be divided into two main segments: acoustic spectral characteristics and wake flow field measurements. The first segment evaluates the influence of several factors on sound power level, including Reynolds number, angle of attack, serration parameters, and model type, along with far-field sound radiation patterns. The second segment investigates the relationship between noise reduction and flow field characteristics, focusing on fluctuation velocities at specific measurement points downstream. This part examines how changes in the wake flow field correlate with variations in noise reduction performance.

This study provides new insights into the effects of serrated trailing edges on noise reduction, particularly across a broader frequency range than previously reported in the literature. While past research primarily associates serrated trailing edges with noise reduction at low to mid frequencies, our findings demonstrate effectiveness across both low-to-mid and high frequencies, with the most pronounced reduction occurring in the mid-to-high frequency range when compared to straight trailing edge airfoils. Contrary to common assumptions, the geometry of the serrations does not play a critical role in noise reduction under the experimental conditions in this study. The serrated trailing edges show variable noise reduction effectiveness depending on the angle of attack and airfoil profile, with noise reduction observed across all four tested angles of attack within a specific frequency range. Moreover, while serration modifications substantially reduce noise, particularly at higher frequencies, they do not significantly alter the directivity pattern of the airfoil’s acoustic emissions.

Wake flow velocity spectra measurements reveal a clear correlation between reduced wake turbulence and noise reduction. Serrated trailing edges effectively lower the power spectral density of turbulent velocity fluctuations and reduce the turbulent length scales near the sound source, effectively disrupting larger vortex structures and potentially decreasing the efficiency of the noise-generating mechanisms. While this relationship is promising, further experimental validation is required to strengthen these findings.

Future research should expand upon these findings by integrating simultaneous sound and flow field measurements of serrated trailing edge noise reduction in larger anechoic wind tunnels (diameter $>$ 400 mm) or real-world wind turbine settings. Employing flow visualization techniques with larger models and serrated trailing edges could provide deeper insights into the underlying mechanisms of noise reduction. Additionally, high-fidelity numerical simulations are essential for capturing accurate results, facilitating a more comprehensive understanding of the noise reduction phenomena. Incorporating analytical models may further elucidate the mechanisms involved in noise reduction.

\section*{Data availability}

The data that support the findings of this study are available from the corresponding author upon reasonable request.




\bibliographystyle{elsarticle-num-names} 






\bibliography{mybib}

\end{document}